\documentclass[preprint,3p,12pt]{elsarticle}
\usepackage{mathrsfs}
\usepackage{amsmath}
\usepackage{stmaryrd}
\usepackage{bbding}
\usepackage{dcolumn}
\usepackage{graphicx}
\usepackage{amsfonts}
\usepackage{amssymb}
\usepackage{psfrag}
\usepackage{wrapfig}
\usepackage{subfigure}
\usepackage{makeidx}
\usepackage{bm}
\usepackage{epsf}
\usepackage{epsfig}
\usepackage{setspace}
\usepackage{graphicx}
\usepackage{epstopdf}
\usepackage{psfrag}
\usepackage{subfigure}
\usepackage{multirow}
\usepackage{diagbox}
\usepackage{verbatim}

\usepackage{makecell}

\begin{document}

\title{A Family of High-order Gas-kinetic Schemes and Its Comparison with Riemann Solver Based High-order Methods}

\author[hkust1]{Xing Ji}
\ead{xjiad@connect.ust.hk}
\author[iapcm]{Fengxiang Zhao}
\ead{kobezhao@126.com}
\author[hkust2]{Wei Shyy }
\ead{weishyy@ust.hk}
\author[hkust1,hkust2]{Kun Xu\corref{cor}}
\ead{makxu@ust.hk} \cortext[cor]{Corresponding author}
\address[hkust1]{Department of mathematics, Hong Kong University of Science and Technology, Clear Water Bay, Kowloon, Hong Kong}
\address[hkust2]{Department of Mechanical and Aerospace Engineering, Hong Kong University of
Science and Technology, Clear Water Bay, Kowloon, Hong Kong}

\address[iapcm]{Institute of Applied Physics and Computational Mathematics, Beijing, China}
\begin{abstract}
Most high order computational fluid dynamics (CFD) methods for compressible flows are based on Riemann solver for the flux evaluation and Runge-Kutta (RK) time stepping technique for temporal accuracy.
The main advantage of this kind of approach is the easy implementation and stability enhancement
by introducing more middle stages.
However, the nth-order time accuracy needs no less than n stages for the RK method,
which is very time and memory consuming for a high order method.
On the other hand, the multi-stage multi-derivative (MSMD) method
can be used to achieve the same order of time accuracy  using less middle stages, once the time derivatives of the flux function is used.
For the traditional Riemann solver based CFD methods, the lack of time derivatives in the flux function
prevents its direct implementation of the MSMD method.
However, the gas kinetic scheme (GKS) provides such a time accurate evolution model.
By combining the second-order or third-order GKS flux functions with the MSMD technique,
a family of high order gas kinetic methods can be constructed.
As an extension of the previous 2-stage 4th-order GKS, the 5th-order schemes with 2 and 3 stages will be developed in this paper.
Based on the same 5th-order WENO reconstruction, the performance of gas kinetic schemes from the 2nd- to
the 5th-order time accurate methods will be evaluated.
The results show that the 5th-order scheme can achieve the theoretical order of accuracy for the Euler equations,
and present accurate Navier-Stokes solutions as well due to the coupling of inviscid and viscous terms in the GKS formulation.
In comparison with Riemann solver based 5th-order RK method,
the high order GKS has advantages in terms of efficiency, accuracy, and robustness, for all test cases.
The 4th-order and 5th-order GKS have the same robustness as the 2nd-order scheme for the
capturing of discontinuous solutions.
The current high order MSMD GKS is a multidimensional scheme with incorporation of both normal and tangential spatial derivatives
of flow variables at a cell interface in the flux evaluation.
The scheme can be extended straightforwardly to viscous flow computation in unstructured mesh.
It provides a promising direction for the development of high-order
CFD methods for the computation of complex flows, such as turbulence and acoustics with shock interactions.

\end{abstract}

\maketitle

\section{Introduction}

The flux of the gas kinetic scheme (GKS) is based on the time-dependent evolution solution of the kinetic equation, such as
the Bhatnagar-Gross-Krook (BGK) model \cite{BGK}. It targets on the Euler and NS solutions \cite{originalGKS,implicitGKS}.
In comparison with traditional Riemann solver based CFD methods, the distinguishable points of GKS include the following.
Firstly, the time evolving gas distribution function provides a multiple scale flow physics from the kinetic
particle transport to the hydrodynamic wave propagation \cite{xu-liu}. The particle transport supplies numerical dissipation in the
discontinuous region and the wave propagating solution provides accurate solution in the smooth region.
The multiple scale nature of the flux function bridges the evolution from the upwind flux vector splitting to the
central difference Lax-Wendroff type discretization, where
both inviscid and viscous fluxes are obtained from the moments of a single gas distribution function \cite{originalGKS,implicitGKS,multi-tem-GKS}.
Secondly, the GKS is intrinsically a multi-dimensional scheme, where both normal and tangential derivatives of flow variables
around a cell interface participate the time evolution of the gas distribution function \cite{implicitGKS}.
The hyperbolic kinetic equation with local relaxation provides a compact physical domain for the design of the numerical schemes in comparison with
the direct solvers of the Navier-Stokes equations. The numerical solutions in GKS are not sensitive to the mesh distributions
due to the absence of direct evaluation of dissipative terms.
Thirdly, the time dependent flux function achieves higher order time accuracy than that in the Riemann flux.
A one-step 3rd-order scheme can be
directly constructed \cite{3rdGKS-Li}. Fourthly, a unified GKS (UGKS) can be developed for multi-scale gas
dynamics \cite{UGKS,xu-liu}.

The second order schemes were mostly developed in 1980s and they are the main numerical schemes used in engineering applications.
Great efforts have been paid on the development of high order (3rd order or higher) methods in the past decades,
which are expected to provide more accurate solutions with less computational cost than second order methods \cite{high-order-review}.
It is still early to point out the appropriate approaches for high order schemes,
especially for high speed compressible flow with discontinuities.
There are many attempts in the construction of high order schemes, such as  the k-exact \cite{k-exact},
weighted essentially non-oscillatory (WENO) \cite{weno,wenoz,wenoz+},
multi-moment constrained (MCV)  methods \cite{mcv}, discontinuous Galerkin (DG) \cite{DG},
flux reconstruction approach \cite{CPR}, and many others.
Most approaches are focusing on high order initial reconstruction, which achieves high order  accuracy in smooth region and
avoids oscillation in the discontinuous region.
In terms of evolution model, same as the second order schemes,
the Riemann solver  is commonly used for the flux evaluation \cite{rm-book}.
In order to improve the time accuracy, the traditional Runge-Kutta (RK) time-stepping method is used \cite{RK-Jameson}.
The RK methods separate the spatial and temporal discretization, which could improve the stability for hyperbolic problems
in comparison with the single stage or the Adams family of methods under the same reconstruction \cite{RK-advantage1,RK-advantage2}.
The RK method can be used in GKS as well under FV or DG frameworks \cite{HGKS-Magnet,RK-DG-Ren1,RK-DG-Ren2}.
However, the nth-order accuracy in RK method requires no less than n stages.
For a classical 4th-order RK method, we need 4 stages. While for a 5th-order RK method,
usually  6 stages are needed \cite{fifth-RK}. For a high order scheme, the initial reconstruction for the middle stages may take most of the
computational time.
All those are related to the use of Riemann solver, which may be the real barrier for the development of efficient high order schemes.

On the other hand, the GKS is based on the high order gas evolution model, which provides
time-dependent flux function. The second and third order schemes can be developed through
a single updating step without middle stages \cite{3rdGKS-Li}. With the 5th-order WENO reconstruction,
a 3rd order GKS has been developed for both 2D and 3D inviscid and viscous flow computations \cite{3rdGKS-Luo,3rdGKS-3D-Pan}.
A compact 3rd-order scheme  in structured  and unstructured meshes has been
developed as well \cite{structured-compact-gks,unstructured-compact-gks}.
Although the 3rd-order GKS flux function takes approximately  $4$ to $6$ times of the computational time of a 2nd order GKS flux,
the one step method still shows high efficiency against a 3 stages RK method,
since the spatial reconstructions take a significant amount of CPU time.
The 3rd-order GKS flux function depends on the initial reconstructions of derivatives, such as 1st, 2nd, and 3rd-order ones, which
become  more and more unreliable numerically, especially close to the discontinuous regions.
The one step 3rd-order GKS becomes less robust than the second order one.
Instead of continuously constructing high order one step GKS methods, it may become a good choice to combine
the advantages of both time accuracy of the GKS flux function and the RK technique for the robustness.

Starting from 1940s,  methods with multiple stages and  multiple derivatives (MSMD) have been used for numerical solution of ODEs \cite{MMMD1}.
This group of methods was reviewed and defined  by Hairer and Wanner \cite{MMMD2}, where
the MSMD including 2nd-order derivatives was studied up to 7th-order accuracy method and was compared with RK methods.
However, this technique has hardly been applied to CFD methods since most schemes here are based on the 1st-order Riemann solver.
After realizing the benefits from the MSMD, a DG method with MSMD has been proposed recently \cite{multi-derivative}.
Almost at the same time, a 4th-order 2 stages scheme based on the generalized Riemann problem (GRP) for the  Euler equations
has been proposed \cite{4th2stage-Li}. Similarly, a 4th-order 2 stages GKS has been designed for the Navier-Stokes solutions \cite {4th2stage-Pan}.
Benefiting from the 2nd-order GKS flux function and two stages strategy,
the 4th-order 2 stages GKS shows great accuracy and outstanding robustness due to the absence two reconstructions,
and the efficiency of the scheme is also superior in comparison with
other Riemann solver based 4th-order schemes with 4 stages RK technique.

In this paper, two kinds of 5th-order GKS will be proposed by using MSMD technique and
taking the advantages of high order time-accurate GKS flux function.
One of the 5th-order schemes has 3 stages GKS with the use of a 2nd-order GKS flux function.
Another 5th-order scheme has 2 stages only, where a 3rd-order flux function is used.
To further improve the efficiency of the 5th-order 2 stages GKS, a simplified 3rd-order GKS flux function is adopted \cite{3rdGKS-simplified}. Meanwhile, in order to present a complete picture about the high order gas kinetic schemes and get a comparison among GKS methods,
the 3rd-order 1 stage multi-dimensional GKS \cite{3rdGKS-Luo}  and a
4th-order 2 stage GKS  \cite{4th2stage-Pan} will be summarized in this paper as well.
Thus a family and a variation of high order GKS from 3rd-order up to 5th-order will be presented.
For spatial reconstruction, the WENO technique  has achieved great success, especially on structured mesh.
Since we are focusing on the improvement of time accuracy for the schemes,
the same 5th-order WENOZ reconstruction \cite{wenoz} based on characteristic variables will be used for all schemes,
including the 5th-order Godunov method with RK technique.

This paper is organized as follows. Section 2 introduces the multi-stage time integrating techniques.
Section 3 gives a brief review of GKS flux solvers and introduces the numerical algorithm for MSMD GKS.
Section 4 presents the numerical results from different schemes and their comparison with other standard high order methods with
exact Riemann solver in terms of accuracy, efficiency, and robustness. Finally we end up with some concluding remarks.

\section{Multi-stage multi-derivative Mathods}

The conservation laws
\begin{equation}\label{ms-1}
\begin{split}
\textbf{w}_t+ \nabla \cdot \textbf{F}(\textbf{w})=0,\textbf{w}(0,\textbf{x})=\textbf{w}_0(\textbf{x}),\textbf{x}\in \Omega \subseteq \mathbb{R}^d,
\end{split}
\end{equation}
for conserved mass, momentum, and energy  $\textbf{w}$ can be
written as
$$\textbf{w}_t=-\nabla \cdot {\textbf{F}}(\textbf{w}) .$$
With the spatial discretization $\textbf{w}^h$ and appropriate evaluation $-\nabla \cdot {\textbf{F}}(\textbf{w})$,
the original PDEs become a system of  ordinary differential equation (ODE)
\begin{equation}\label{ms-2}
\begin{split}
\textbf{w}^h_t=L(\textbf{w}^h),t=t_n.
\end{split}
\end{equation}
The well established numerical scheme for ODE can be used to solve this initial value problem.

For a smooth function $L$,
the solution $\textbf{w}(\Delta t)$ around ($t=t_n=0$) becomes
\begin{equation}\label{ms-3}
\begin{split}
\textbf{w}(\Delta t)&=\textbf{w}(0)+\Delta t \textbf{w}^{(1)}(0)+\frac{{\Delta t}^2}{2} \textbf{w}^{(2)}(0)+\frac{{\Delta t}^3}{6} \textbf{w}^{(3)}(0)
\\&+\frac{{\Delta t}^4}{24} \textbf{w}^{(4)}(0)+\frac{{\Delta t}^5}{120} \textbf{w}^{(5)}(0)
+\cdots+\frac{{\Delta t}^n}{n!} \textbf{w}^{(n)}(0)+\mathcal{O}(\Delta t^{n+1}),
\end{split}
\end{equation}
where $\textbf{w}^{(n)}(t)(n=1,2,3,...)$ refers to
\begin{equation}\label{ms-4}
\begin{split}
\textbf{w}^{(n)}(t)=\frac{d^n\textbf{w}(t)}{dt^n}=\frac{d^{n-1}L(\textbf{w}(t))}{dt^{n-1}}.
\end{split}
\end{equation}
For simplicity of presentation, we define $L = L(w(t))$ and $L^{(n)} = \frac{d^nL(t)}{dt^n}$.
A $n$th-order time marching scheme can be constructed straightforwardly if the  time derivatives of $L^{(n)}$ up to $(n-1)$th-order are provided.
However, under most circumstances, we can easily evaluate  low order derivatives, such as $L$ for the
approximate Remiann solver, $L^{(1)}$ for the generalized Riemann problem (GRP) and the 2nd-order GKS flux function,
and $L^{(2)}$ for the 3rd-order GKS flux function.
A continuous construction of higher order derivatives becomes prohibited, such as the extremely complicated 4th-order GKS flux function \cite{liu-tang}.
Another approach, similar to RK method, is to introduce the middle stages.
The update at $t^{n+1}$ becomes a linear combination of $L$ and their derivatives in the multiple stages.
If $L$ is used only, the traditional RK method is recovered. But, with the inclusion of $L^{(n)}$, the multi-stage multi-derivative (MSMD)
method can be constructed.

\subsection{Multi-stage Multi-derivative High Order Method }
 \textbf{Definition 1} According to \cite{multi-derivative}, given a collection of real numbers ${a_{ij}^{(1)},a_{ij}^{(2)},a_{ij}^{(3)},b_{i}^{(1)},b_{i}^{(2)},b_{i}^{(3)}}$,
 a multi-derivative (up to 3), s-stage method can be defined as the following
\begin{equation}\label{multiequation1}
\begin{split}
\textbf{w}_{n+1}=\textbf{w}_{n}+\Delta{t}\sum_{i=1}^sb_i^{(1)}L(\textbf{w}^{i})+{\Delta{t}}^2\sum_{i=1}^sb_{i}^{(2)}L^{(1)}(\textbf{w}^{i})
+{\Delta{t}}^3\sum_{i=1}^sb_{i}^{(3)}L^{(2)}(\textbf{w}^{i}),
\end{split}
\end{equation}
where intermediate stage values are given by
\begin{equation}\label{multiequation2}
\begin{split}
\textbf{w}^{i}=\textbf{w}_{n}+\Delta{t}\sum_{j=1}^{i-1}a_{ij}^{(1)}L(\textbf{w}^{j})+{\Delta{t}}^2\sum_{j=1}^{i-1}a_{ij}^{(2)}L^{(1)}
(\textbf{w}^{j})+{\Delta{t}}^3\sum_{j=1}^{i-1}a_{ij}^{(3)}L^{(2)}(\textbf{w}^{j}).
\end{split}
\end{equation}
Since MSMD is an explicit method, at every intermediate stage the state only depends on the states and  derivatives of previous ones.
The Butcher tableau, which is widely used to list all coefficients in Runge-Kutta and MSMD method \cite{multi-derivative}, is shown in Table. \ref{buther_tableau}, where $c_i=\sum_{j=1}^s a_{ij}$. Note that the explicit method makes all coefficient $a_{ij}=0$ if $i<=j$.

\begin{table}[!h]
\begin{center}
\begin{tabular}{c|ccc|ccc|ccc}

$c_1$&$a_{11}^{(1)}$&$\cdots$&$a_{1s}^{(1)}$&$a_{11}^{(2)}$&$\cdots$&$a_{1s}^{(2)}$&$a_{11}^{(3)}$&$\cdots$&$a_{1s}^{(3)}$\\
$\vdots$&$\vdots$&$\ddots$&$\vdots$&$\vdots$&$\ddots$&$\vdots$&$\vdots$&$\ddots$&$\vdots$\\
$c_s$&$a_{s1}^{(1)}$&$\cdots$&$a_{ss}^{(1)}$&$a_{s1}^{(2)}$&$\cdots$&$a_{ss}^{(2)}$&$a_{s1}^{(3)}$&$\cdots$&$a_{ss}^{(3)}$\\
\hline

~&$b_{1}^{(1)}$&$\cdots$&$b_{s}^{(1)}$&$b_{1}^{(2)}$&$\cdots$&$b_{s}^{(2)}$&$b_{1}^{(3)}$&$\cdots$&$b_{s}^{(3)}$\\

\end{tabular}
\vspace{-1mm} \caption{\label{buther_tableau} Butcher tableau for a multi-derivative (up to 3) multi-stage method.}
\end{center}
\end{table}

In the following, a few cases which are related high order CFD methods are presented.

\subsection{Tradition Runge-Kutta Methods:  RK4 and RK5}

The Butcher tableau for classical 4th order, four-stages Runge-Kutta (RK4) \cite{weno} and 5th-order, six-stages Runge-Kutta (RK5) methods \cite{RK56} are given in Table \ref{4th4stage} and Table \ref{5th6stage}.
The computational time and robustness of the above RK5 with Riemann solvers for the flux evaluation will be compared
with our newly proposed 5th-order MSMD GKS methods.
Since most high order schemes with Riemann solver use 3rd-order or 4th-order time accurate RK methods, in this paper
many comparison will be done with 4th-order RK (RK4) scheme.

\begin{table}[!h]
\begin{center}
\begin{tabular}{c|cccc}

0&0&0&0&0\\
1/2&1/2&0&0&0\\
1/2&0&1/2&0&0\\
1&0&0&1&0\\
\hline
~&1/6&1/3&1/3&1/6\\

\end{tabular}
\vspace{-1mm} \caption{\label{4th4stage} Butcher tableau for RK4.}
\end{center}
\end{table}

\begin{table}[!h]
\begin{center}
\begin{tabular}{c|cccccc}

0&0&0&0&0&0&0\\
1/4&1/4&0&0&0&0&0\\
3/8&3/32&9/32&0&0&0&0\\
12/13&1932/2197&-7200/2197&7296/2197&0&0&0\\
1&439/216&-8&3680/513&-845/4104&0&0\\
1/2&-8/27&2&-3544/2565&1859/4104&-11/40&0\\
\hline
~&16/135&0&6656/12825&28561/56430&-9/50&2/55\\

\end{tabular}
\vspace{-1mm} \caption{\label{5th6stage} Butcher tableau for RK5.}
\end{center}
\end{table}

\subsection{A 3rd-Order 1 stage Method: S1O3}
A 3rd-order accurate GKS flux function provides
up to 3rd-order time derivatives $\textbf{w}^{(3)}=L^{(2)}$. A direct Taylor expansion method can be used to update the solution,
$$\textbf{w}_{n+1}=\textbf{w}_{n}+\Delta{t}L+\frac{1}{2}\Delta{t}^2L^{(1)}+\frac{1}{6}\Delta{t}^3L^{(2)}.$$
This is the 3rd-order one stage time-accuracy scheme \cite{3rdGKS-Li,3rdGKS-Luo,3rdGKS-simplified}.
The corresponding Butcher tableau is given in Table \ref{3rd1stage}.

\begin{table}[!h]
\begin{center}
\begin{tabular}{c|c|c|c}

0&0&0&0\\
\hline
~&1&1/2&1/6\\

\end{tabular}
\vspace{-1mm} \caption{\label{3rd1stage} Butcher tableau for one stage 3rd-order method (S1O3).}
\end{center}
\end{table}

\subsection{A 4th-Order 2 Stages Method: S2O4}

 There is uniqueness for the 2 stages (S2) 4th-order (O4) method \cite{MMMD3}. This method has been used in CFD applications \cite{multi-derivative,4th2stage-Li,4th2stage-Pan}, which show good accuracy, high efficiency, and robustness.
 It can be written as

\begin{equation}\label{4th2stage1}
\begin{split}
\textbf{w}^{1}&=\textbf{w}_{n}+\frac{1}{2}\Delta{t}L(\textbf{w}_n)+\frac{1}{8}\Delta{t}^2L^{(1)}(\textbf{w}_n),\\
\textbf{w}_{n+1}&=\textbf{w}_{n}+\Delta{t}L(\textbf{w}_n)+\frac{1}{2}\Delta{t}^2[\frac{1}{3}L^{(1)}(\textbf{w}_n)+\frac{2}{3}L^{(1)}(\textbf{w}^{1})].
\end{split}
\end{equation}

The Butcher tableau for the two-stage fourth-order (S2O4) method is given in Table \ref{4th}.
The region of A-stability is plotted in Fig. \ref{A-stability}.

\begin{table}[!h]
\begin{center}
\begin{tabular}{c|cc|cc}

0&0&0&0&0\\
1/2&1/2&0&1/8&0\\
\hline
~&1&0&1/6&1/3\\

\end{tabular}
\vspace{-1mm} \caption{\label{4th} Butcher tableau for S2O4.}
\end{center}
\end{table}

\subsection{5th-Order 3 Stage Methods: S3O5}

For the 5th-order MSMD methods, the coefficients are not uniquely defined.
With the constraints of $a^{(1)}_{ij}=0, j\neq 1$, several choices are given in \cite{MMMD3}.
The choices of the coefficients have been studied in \cite{ssp}.
However, the real performance from different schemes for the Euler and N-S equations haven't been reported yet.
Here we will construct two kinds of 5th-order 3 stages (S3O5) methods.
The first choice is given by:
\begin{equation}\label{s3o51}
\begin{split}
\textbf{w}^{1}&=\textbf{w}_{n}+\frac{2}{5}\Delta{t}L(\textbf{w}_n)+\frac{2}{25}\Delta{t}^2L^{(1)}(\textbf{w}_n),\\
\textbf{w}^{2}&=\textbf{w}_{n}+\Delta{t}L(\textbf{w}_n)+\frac{1}{2}\Delta{t}^2[-\frac{1}{2}L^{(1)}(\textbf{w}_n)+\frac{3}{2}L^{(1)}(\textbf{w}^{1})],\\
\textbf{w}_{n+1}&=\textbf{w}_{n}+\Delta{t}L(\textbf{w}_n)+\frac{1}{2}\Delta{t}^2[\frac{1}{4}L^{(1)}(\textbf{w}_n)+\frac{25}{36}L^{(1)}(\textbf{w}^{1})+\frac{1}{18}L^{(1)}(\textbf{w}^{2})],
\end{split}
\end{equation}
which is denoted by S3O5 and the Butcher tableau is given in Table \ref{s3o511}.
Note that $a^{(2)}_{31}=-1/4 <0$.

\begin{table}[!h]
\begin{center}
\begin{tabular}{c|ccc|ccc}

0&0&0&0&0&0&0\\
2/5&2/5&0&0&2/25&0&0\\
1&1&0&0&-1/4&3/4&0\\
\hline
~&1&0&0&1/8&25/72&1/36\\

\end{tabular}
\vspace{-1mm} \caption{\label{s3o511} Butcher tableau for S3O5.}
\end{center}
\end{table}

The second choice is given by:
\begin{equation}\label{s3o52}
\begin{split}
\textbf{w}^{1}&=\textbf{w}_{n}+\frac{3}{10}\Delta{t}L(\textbf{w}_n)+\frac{9}{200}\Delta{t}^2L^{(1)}(\textbf{w}_n),\\
\textbf{w}^{2}&=\textbf{w}_{n}+\frac{3}{4}\Delta{t}L(\textbf{w}_n)+\frac{9}{32}\Delta{t}^2L^{(1)}(\textbf{w}^{1}),\\
\textbf{w}_{n+1}&=\textbf{w}_{n}+\Delta{t}L(\textbf{w}_n)+\frac{1}{2}\Delta{t}^2[\frac{5}{27}L^{(1)}(\textbf{w}_n)+\frac{50}{81}L^{(1)}(\textbf{w}^{1})+\frac{16}{81}L^{(1)}(\textbf{w}^{2})],
\end{split}
\end{equation}
which is named S3O5+ and the Butcher tableau is given in Table \ref{s3o522}.
\begin{table}[!h]
\begin{center}
\begin{tabular}{c|ccc|ccc}

0&0&0&0&0&0&0\\
3/10&3/10&0&0&9/200&0&0\\
3/4&3/4&0&0&0&9/32&0\\
\hline
~&1&0&0&5/54&25/81&8/81\\

\end{tabular}
\vspace{-1mm} \caption{\label{s3o522} Butcher tableau for S3O5+.}
\end{center}
\end{table}

In comparison with S305, S305+ keeps all coefficients positive,
which may have better stability property in numerical simulations.
The numerical performance will be conducted in this paper, and S3O5+ does have a better robustness.

\subsection{A 5th-Order 2 Stages Methods: S2O5}

To achieve fifth-order time accuracy, we may construct a scheme with 3rd-order derivatives and two stages \cite{4th2stage-Pan}.
The scheme is given by,
\begin{equation}\label{s2o51}
\begin{split}
\textbf{w}^{1}&=\textbf{w}_{n}+\frac{2}{5}\Delta{t}L(\textbf{w}_n)+\frac{2}{25}\Delta{t}^2L^{(1)}(\textbf{w}_n),\\
\textbf{w}_{n+1}&=\textbf{w}_{n}+\Delta{t}L(\textbf{w}_n)+\frac{1}{2}\Delta{t}^2L^{(1)}(\textbf{w}_n)+\frac{1}{6}\Delta{t}^3[\frac{3}{8}L^{(2)}(\textbf{w}_n)+\frac{5}{8}L^{(2)}(\textbf{w}^{1})].
\end{split}
\end{equation}

\begin{figure}[!h]
\centering
\includegraphics[width=0.444\textwidth]{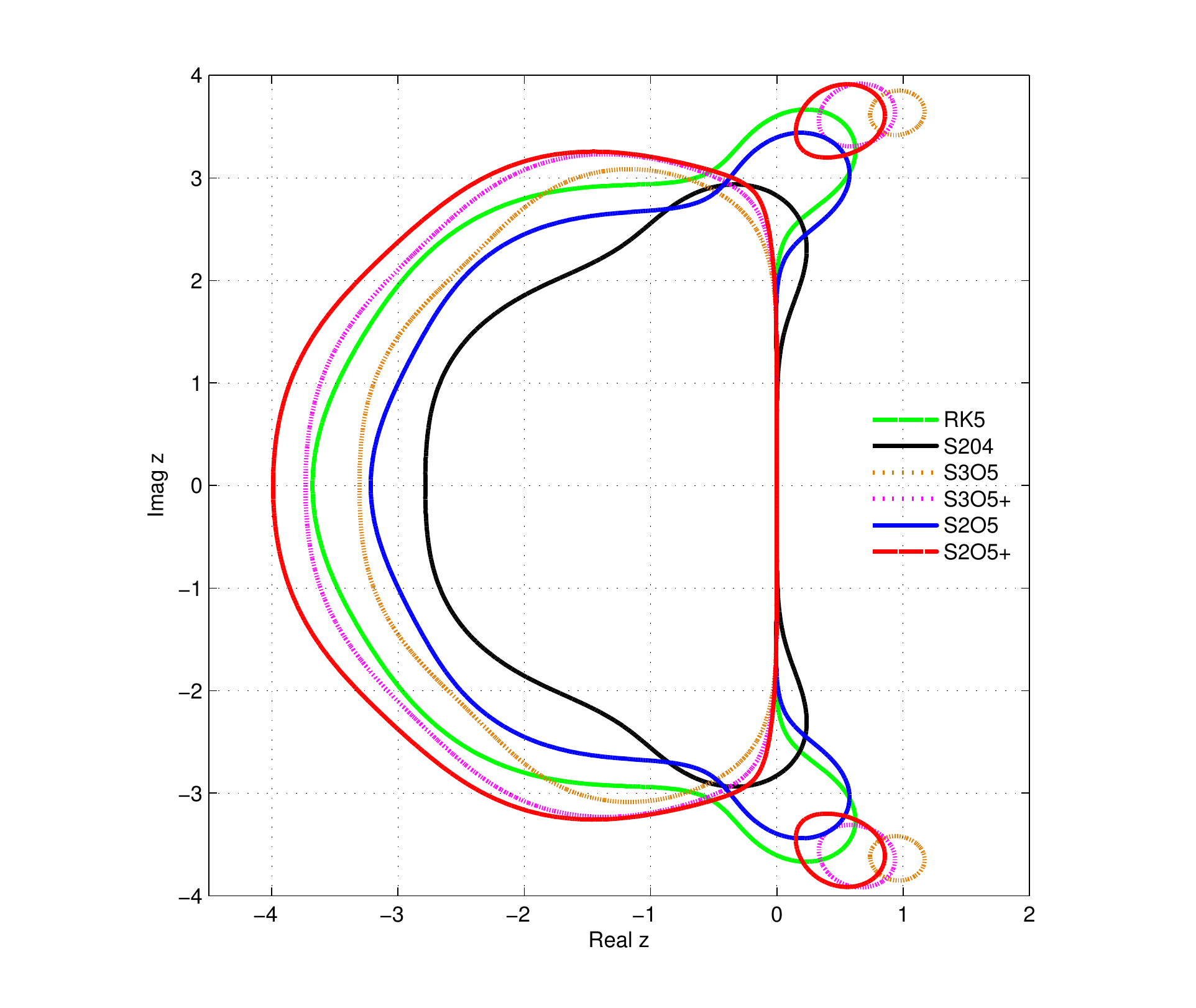}\includegraphics[width=0.444\textwidth]{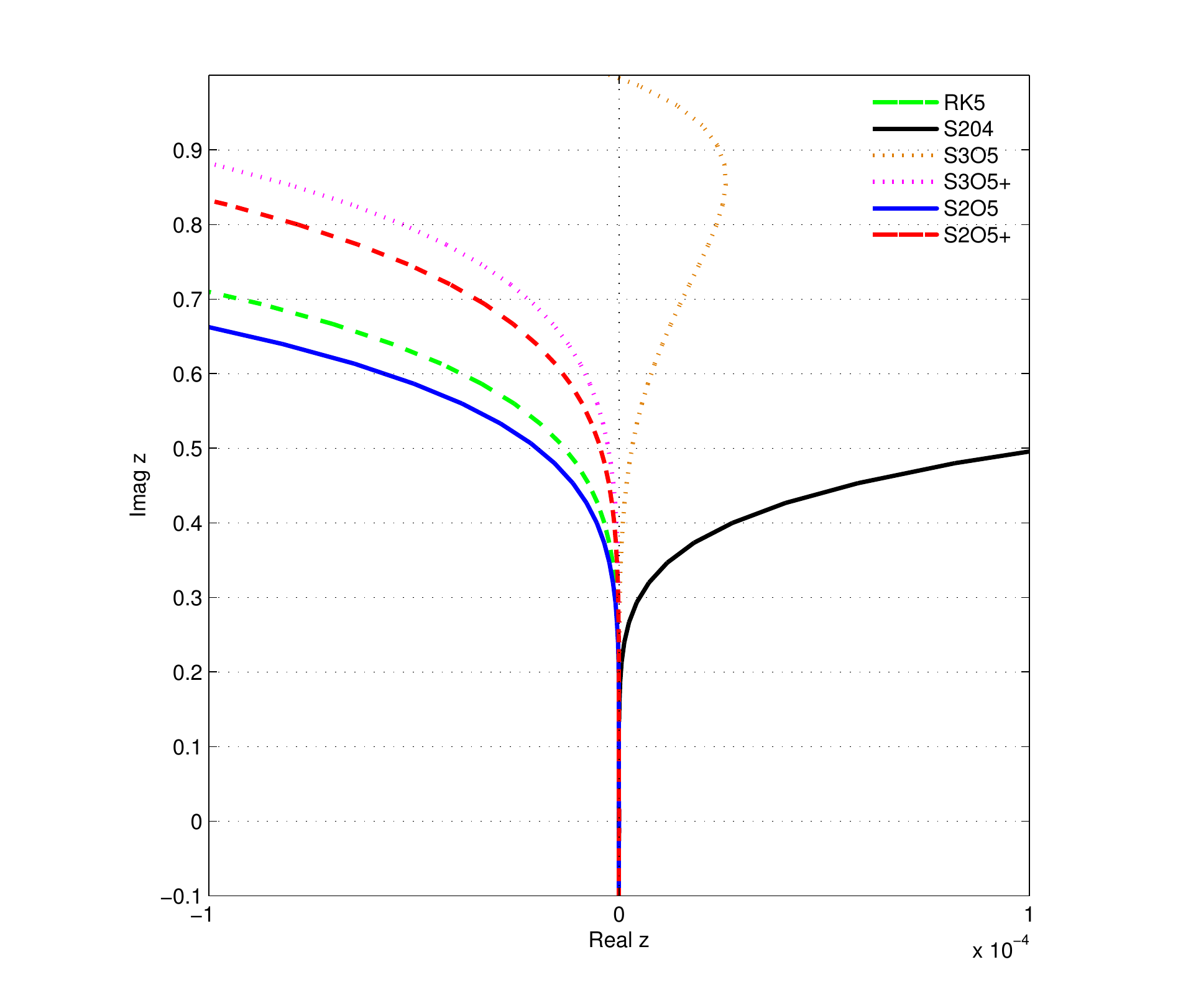}

\caption{\label{A-stability} Left figure: regions of A-stability for RK5, S2O4, S2O5, S2O5+, S3O5, and S3O5+ schemes.
Right figure:  local enlarged region of left figure. }
\end{figure}

The Butcher tableau for the 5th order two-stages (S2O5) scheme with up to 3rd-order derivatives is given in Table \ref{5th2stage}.
The region of stability is plotted in Fig. \ref{A-stability}.

\begin{table}[!h]
\begin{center}
\begin{tabular}{c|cc|cc|cc}

0&0&0&0&0&0&0\\
2/5&2/5&0&2/25&0&0&0\\

\hline
1&1&0&1/2&0&1/16&5/48\\

\end{tabular}
\vspace{-1mm} \caption{\label{5th2stage} Butcher tableau for S2O5.}
\end{center}
\end{table}

For RK method, the range covered in the imaginary axis indicates the stability region.
As shown in Fig. \ref{A-stability}, the S2O4 method has the largest imaginary axis covered by the stability contour
among all schemes. This is confirmed by the numerical results, which shows that S2O4 has best robustness.
On the other hand, the S2O5 method in Table. \ref{5th2stage} shows the weakest A-stability.
However, for the S2O5 method the coefficient $a^{(3)}_{21}$ is a free parameter, which can be chosen to
increase its stability without losing its accuracy.
How to optimize this coefficient is still not clear. But, with the modified coefficients in Table. \ref{5th2stage_modified},
which is named S2O5+, the scheme improves the A-stability, as shown in Fig. \ref{A-stability}.
The S2O5+ method contains the largest A-stable area on the negative left plane in Fig. \ref{A-stability}.
The improvement of robustness from S2O5+ will be shown in the numerical tests.

\begin{figure}[!h]
\centering
\includegraphics[width=0.5\textwidth]{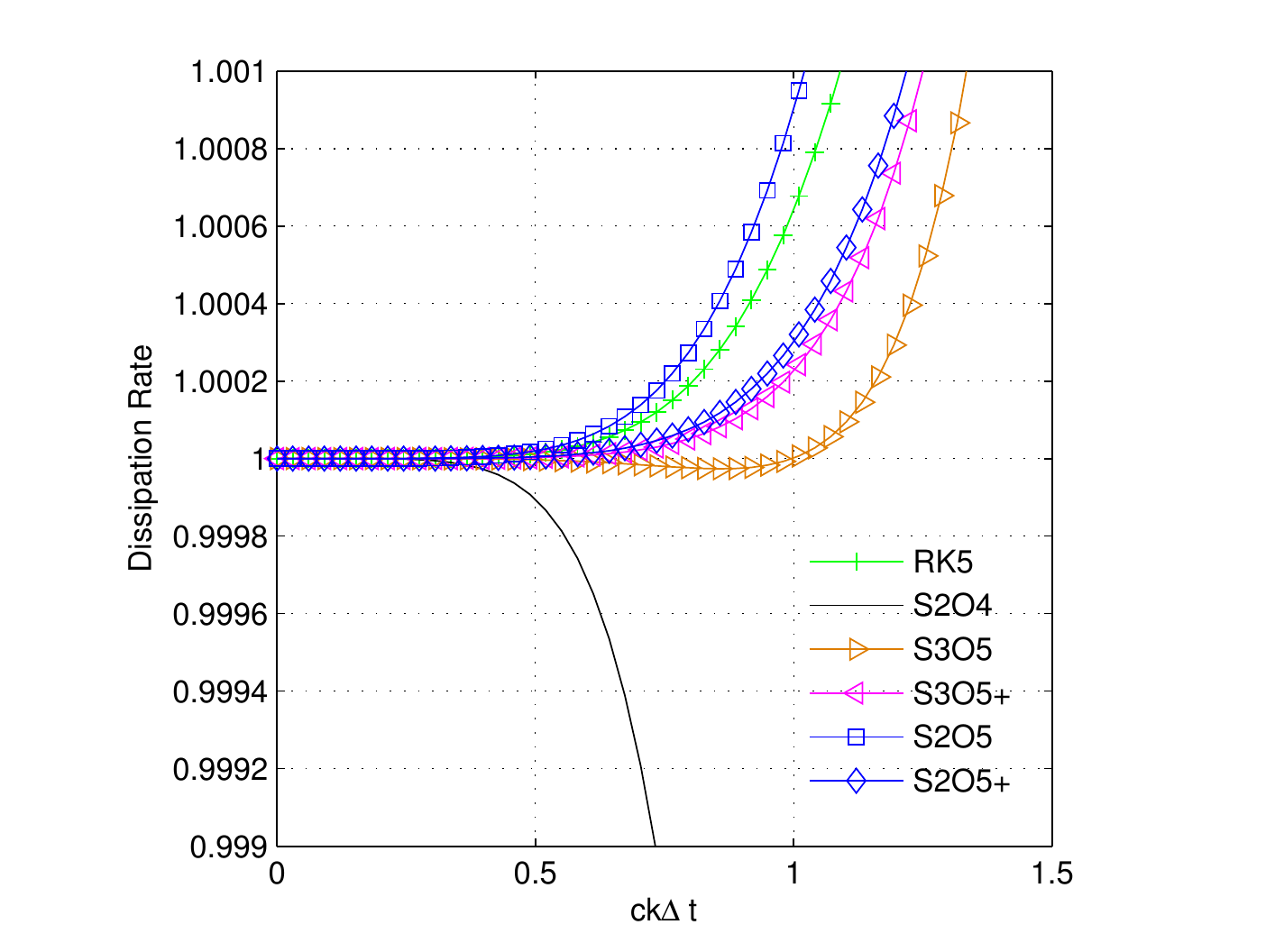}\includegraphics[width=0.5\textwidth]{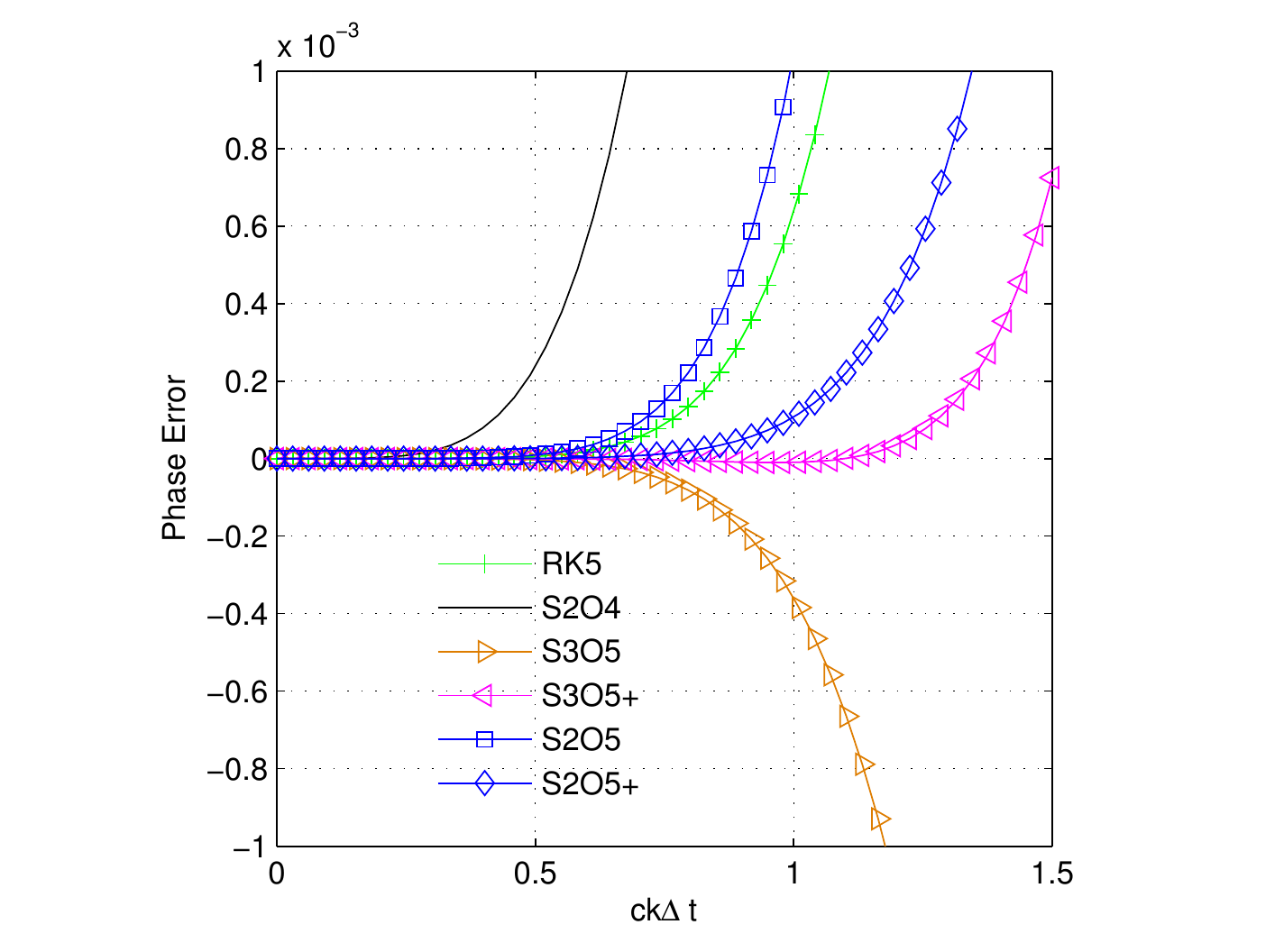}

\caption{\label{Dissipation} The dissipation (Left) and dispersion (Right) properties for different schemes.
Dissipation rate $||r|-1|<0.001$ and phase error $|\delta|<0.001$ are set as accuracy limit in \cite{RK-Acoustic}.}
\end{figure}

In the RK method, lot of efforts have been paid for minimizing the dissipation and dispersion error of the schemes \cite{RK-Acoustic},
which is important in turbulence and acoustic computations.
For different schemes studied in this paper, the dissipation rate and phase error defined in \cite{RK-Acoustic} are plotted
in Fig. \ref{Dissipation}, where $c$ is the wave speed for linear advection equation and $k$ is the wave number.
It shows that with the same time step both S3O5+ and S2O5+ have less dissipation and dispersion error than that of RK5.
This indicates that high order MSMD methods have potential to present more accurate solutions than those with traditional RK technique.

\begin{table}[!h]
\begin{center}
\begin{tabular}{c|cc|cc|cc}

0&0&0&0&0&0&0\\
2/5&2/5&0&2/25&0&4/375&0\\

\hline
1&1&0&1/2&0&1/16&5/48\\

\end{tabular}
\vspace{-1mm} \caption{\label{5th2stage_modified} Butcher tableau for S2O5+.}
\end{center}
\end{table}

\section{High order MSMD Gas Kinetic Schemes}

\subsection{General finite volume framework and kinetic model equation}

In a 2D rectangular mesh, Eq.(\ref{ms-2}) can be written in a semi-discrete form
\begin{equation}\label{finite_volume1}
\begin{split}
\frac{d\bar{W}_{ij}^n}{dt}=L_{ij}(W):=-\frac{1}{\Delta x_{ij}}({F}_{i+1/2,j}^n-{F}_{i-1/2,j}^n)-\frac{1}{\Delta y_{ij}}({G}_{i,j+1/2}^n-{G}_{i,j-1/2}^n),
\end{split}
\end{equation}
where $W=(\rho ,\rho U,\rho V,\rho E)$ are the conservative flow variables,
$\bar{W}$ are the cell average values. Here $F(W(t))=(F_{\rho},F_{\rho U},F_{\rho V},F_{\rho E})$
are the corresponding fluxes across the cell interface in the x-direction, similarly for $G(W(t))$ in the y-direction.
The key point for constructing MSMD method is to obtain a time dependent flux function $F(t)$ and $G(t)$.
Most Godunov type schemes solve the Riemann problem with time-independent flux. But,
for the GKS solver and generalized Riemann problem (GRP), time-dependent fluxes are provided.

The gas kinetic scheme is to solve the kinetic equation
\begin{equation}\label{finite_volume2}
\begin{split}
f_t+\textbf{u}\cdot\nabla f=\frac{g-f}{\tau},
\end{split}
\end{equation}
where $\textbf{u}$ is particle velocity.
Here $f$ is the gas distribution function, which is a function in
the physical and velocity space, i.e., $f(x,y,t,u,v,\xi)$ as a function of particle velocity $(u,v)$,
space and time coordinates $(x,y,t)$, and internal variable $\xi$. Here $g$ is the corresponding equilibrium state of $f$,
and $\tau$ is the relaxation time from $f$ to $g$.
The collision term satisfies the compatibility condition
\begin{equation}\label{finite_volume3}
\begin{split}
\int \frac{g-f}{\tau}\psi d\Xi=0,
\end{split}
\end{equation}
where $\psi =(\psi_1, \psi_2, \psi_3, \psi_4)^T = (1,u,v,\frac{1}{2}(u^2+v^2+\xi ^2))^T$,
$d\Xi=dudvd\xi$, $d\xi = d\xi_1d\xi_2...d\xi_K$,
$K$ is the degrees of internal variable $\xi$ with the relationship $K=(4-2\gamma)/(\gamma-1)$ fora 2D problem,
and  $\gamma$ is the specific heat ratio.

The connections between macroscopic mass $\rho$, momentum ($\rho U, \rho V$), and energy $\rho E$ with the distribution function $f$ are
\begin{equation}\label{finite_volume4}
\left(
\begin{array}{c}
\rho\\
\rho U\\
\rho V\\
\rho E\\
\end{array}
\right)
=\int \psi fd\Xi.
\end{equation}
The time-dependent numerical fluxes across a cell interface, for example in the x-direction, can be evaluated by
\begin{equation}\label{finite_volume5}
F(t)=\int_{-\frac{1}{2}\Delta y}^{\frac{1}{2}\Delta y}\int u\psi f(0,y,t,u,v,\xi)d\Xi dy,
\end{equation}
where the construction of the time-dependent cell interface distribution function $f$ is the core of the gas kinetic scheme.

\subsection{GKS flux solver}

The integral solution of kinetic model equation is
\begin{equation}\label{flux_solver1}
\begin{split}
f(0,y,t,u,v,\xi)=&\frac{1}{\tau} \int_0^t g(-u(t-t'),y-v(t-t'),t',u,v,\xi)e^{-(t-t')/\tau}dt'
\\&+e^{-t/\tau}f_0(-ut,y-vt,u,v,\xi).
\end{split}
\end{equation}\\
The initial term $f_0$ in the above integral solution is defined as,
\begin{equation}\label{flux_solver2}
\begin{split}
f=f_0^l(x,y,u,v,\xi)(H(x))+f_0^r(x,y,u,v,\xi)(1-H(x)),
\end{split}
\end{equation}
where $H(x)$ is the Heaviside function, $f_0^l$ and $f_0^r$ are the initial gas distribution functions on both sides of a cell interface.
To keep  a third-order accuracy, the gas distribution function in space around $(x,y)=(0,0)$ can be expanded as
\begin{equation}\label{3rd-left-right}
\begin{split}
f_0^{l,r}(x,y)=f_0^{l,r}(0,0)+\frac{\partial{f_0^{l,r}}}{\partial{x}}x+\frac{\partial{f_0^{l,r}}}{\partial{y}}y
+\frac{1}{2}\frac{\partial^2{f_0^{l,r}}}{\partial{x^2}}x^2
+\frac{\partial^2{f_0^{l,r}}}{\partial{xy}}xy
+\frac{1}{2}\frac{\partial^2{f_0^{l,r}}}{\partial{y^2}}y^2.
\end{split}
\end{equation}
\\
According to the Chapman-Enskog theory, for the Euler equations $f_0^{l,r}(0,0)$ are the equilibrium states
\begin{equation}\label{flux_solver4}
\begin{split}
&f_0^{l,r}(0,0)=g_0^{l,r}.
\end{split}
\end{equation}
For the Navier-Stokes equations, they are given by
\begin{equation}\label{flux_solver4}
\begin{split}
&f_0^{l,r}(0,0)=g_0^{l,r}-\tau(\frac{\partial{g_0^{l,r}}}{\partial{x}}u+\frac{\partial{g_0^{l,r}}}{\partial{y}}v+\frac{\partial{g_0^{l,r}}}{\partial{t}}),
\end{split}
\end{equation}
where  the Maxwellian distribution function $g_0^{l,r}$ are written as
\begin{equation}\label{flux_solver4}
\begin{split}
g_0=\rho(\frac{\lambda}{\pi})^{\frac{K+2}{2}}e^{\lambda((u-U)^2+(v-V)^2+\xi^2)},
\end{split}
\end{equation}
where $\lambda =m/2kT $, and $m, k, T$ represent the molecular mass, the Boltzmann constant, and temperature.
$g_0^{l,r}$ are the equilibrium states corresponding to the macroscopic flow variables $W_l,W_r$ at the left and right hand sides of a cell interface.

After determining the non-equilibrium part $f_0$, the equilibrium part $g$ in the integral solution can be expanded in space and time as follows
\begin{equation}\label{3rd-equ}
\begin{split}
g&=\bar{g}
+\frac{\partial{\bar{g}}}{\partial{x}}x+\frac{\partial{\bar{g}}}{\partial{y}}y+\frac{\partial{\bar{g}}}{\partial{t}}t
\\&+\frac{1}{2}\frac{\partial^2{\bar{g}}}{\partial{x^2}}x^2
+\frac{\partial^2{\bar{g}}}{\partial{xy}}xy
+\frac{1}{2}\frac{\partial^2{\bar{g}}}{\partial{y^2}}y^2
+\frac{1}{2}\frac{\partial^2{\bar{g}}}{\partial{t^2}}t^2
+\frac{\partial^2{\bar{g}}}{\partial{xt}}xt
+\frac{\partial^2{\bar{g}}}{\partial{yt}}yt,
\end{split}
\end{equation}
where $\bar{g}$ can be obtained using the compatibility condition in Eq.(\ref{finite_volume3}),
\begin{equation}\label{g0-collision}
\begin{split}
\int \psi \bar{g}d\Xi=\bar{W}=\int_{u>0} \psi g_ld\Xi+\int_{u<0} \psi g_rd\Xi.
\end{split}
\end{equation}
Before calculating all derivatives, let's introduce the following notations
\begin{equation}\label{coe-relation}
\begin{split}
&a_1=g_x/g,a_2=g_y/g,A=g_t/g,d_{11}=\frac{\partial{a_1}}{\partial{x}},
\\&d_{12}=\frac{\partial{a_1}}{\partial{y}}=\frac{\partial{a_2}}{\partial{x}},
d_{22}=\frac{\partial{a_2}}{\partial{y}},
b_1=\frac{\partial{a_1}}{\partial{t}}=\frac{\partial{A}}{\partial{x}},
b_2=\frac{\partial{a_2}}{\partial{t}}=\frac{\partial{A}}{\partial{y}},
B=\frac{\partial{A}}{\partial{t}}.
\end{split}
\end{equation}

All coefficients $a_1,a_2,A,...$ are determined by conservative flow variables and their gradients.
Each coefficient can be written as $\Lambda=\Lambda_1 \psi_1+\Lambda_2 \psi_2+\Lambda_3 \psi_3+\Lambda_4 \psi_4$,
which is determined in the following. \\
First order derivatives:
\begin{equation}\label{coe-determine-1st}
\begin{split}
&\left\langle a_1\right\rangle=\frac{\partial{W}}{\partial{x}},
\left\langle a_2\right\rangle=\frac{\partial{W}}{\partial{y}},
\left\langle A+a_1u+a_2v\right\rangle=0.
\end{split}
\end{equation}
Second order derivatives:
\begin{equation}\label{coe-determine-2nd}
\begin{split}
&\left\langle a_1^2+d_{11}\right\rangle=\frac{\partial^2{W}}{\partial{x^2}},
\left\langle a_2^2+d_{22}\right\rangle=\frac{\partial^2{W}}{\partial{y^2}},
\left\langle a_1a_2+d_{12}\right\rangle=\frac{\partial^2{W}}{\partial{xy}},
\\
&\left\langle (a_1^2+d_{11})u+(a_1a_2+d_{12})v+(Aa_1+b_1)\right\rangle=0,
\\
&\left\langle (a_1a_2+d_{12})u+(a_2^2+d_{22})v+(Aa_2+b_2)\right\rangle=0,
\\
&\left\langle (Aa_1+b_1)u+(Aa_2+b_2)v+(A^2+B)\right\rangle=0,
\end{split}
\end{equation}
where $\left\langle ... \right\rangle$ are the moments of a gas distribution function defined by
\begin{equation}\label{flux_solver10}
\begin{split}
\left\langle ...\right\rangle=\int \psi g(...)d\Xi.
\end{split}
\end{equation}

In the following subsections the final expressions for the 3rd-order and 2nd-order GKS flux functions are listed.
The detailed consideration for the GKS flux construction is given in
\cite{originalGKS} for the 2nd-order flux and \cite{3rdGKS-Luo} for the 3rd-order one.

\subsubsection{Full 3rd-order GKS flux in 2D}

With the definition of a physical particle collision time $\tau$ and a numerical one $\tau_n$ for proving additional dissipation
in unresolved region, the integral solution becomes
\begin{equation}\label{tau-n}
\begin{split}
f(0,y,t,u,v,\xi)=&\frac{1}{\tau_n} \int_0^t g(-u(t-t'),y-v(t-t'),t',u,v,\xi)e^{-(t-t')/\tau_n}dt'
\\&+e^{-t/\tau_n}f_0(-ut,y-vt,u,v,\xi).
\end{split}
\end{equation}
Substituting Eq.(\ref{3rd-left-right}) and (\ref{3rd-equ}) with coefficients in Eq.(\ref{coe-relation})
into the above equation, we get
\begin{equation}\label{3rdsolver_equ1}
\begin{split}
&\frac{1}{\tau_n} \int_0^t g(-u(t-t'),y-v(t-t'),t',u,v,\xi)e^{-(t-t')/\tau_n}dt'
\\&=C_1 \bar{g}+C_2 \bar{g}\bar{a_1}u+C_1 \bar{g}\bar{a_2}y+C_2 \bar{g}\bar{a_2}v+C_3 \bar{g}\bar{A}+\frac{1}{2}C_4 \bar{g}(\bar{a_1}^2+\bar{d_{11}})u^2
\\+&\frac{1}{2}C_1 \bar{g}(\bar{a_2}^2+\bar{d_{22}})y^2+C_2 \bar{g}(\bar{a_2}^2+\bar{d_{22}})vy+\frac{1}{2}C_4 \bar{g}(\bar{a_2}^2+\bar{d_{22}})v^2
\\+&C_2 \bar{g}(\bar{a_1}\bar{a_2}+\bar{d_{12}})uy+C_4 \bar{g}(\bar{a_1}\bar{a_2}+\bar{d_{12}})uv+\frac{1}{2}C_5 \bar{g}(\bar{A}^2+\bar{B})
\\+&C_6 \bar{g}(\bar{A}\bar{a_1}+\bar{b_{1}})u+C_3 \bar{g}(\bar{A}\bar{a_2}+\bar{b_{2}})y+C_6 \bar{g}(\bar{A}\bar{a_2}+\bar{b_{2}})v,
\end{split}
\end{equation}
and
\begin{equation}\label{3rdsolver_equ2}
\begin{split}
e^{-t/\tau_n}f_0(-ut,y-vt,u,v,\xi)=
\begin{cases}
e^{-t/\tau_n}f_0^l(-ut,y-vt,u,v,\xi),&u>0,\\
e^{-t/\tau_n}f_0^r(-ut,y-vt,u,v,\xi),&u<0,
\end{cases}
\end{split}
\end{equation}
where
\begin{equation}\label{3rdsolver_equ3}
\begin{split}
&e^{-t/\tau_n}f_0^{l,r}(-ut,y-vt,u,v,\xi)
\\&=C_7 g_0^{l,r}[1-\tau(a_1^{l,r}u+a_2^{l,r}v+A^{l,r})]
\\&+C_8 g_0^{l,r}[a_1^{l,r}u-\tau (((a_1^{l,r})^2+d_{11}^{l,r})u^2+(a_1^{l,r}a_2^{l,r}+d_{12}^{l,r})uv+(A^{l,r}a_1^{l,r}+b_1^{l,r})u)]
\\&+C_7g_0^{l,r}[a_2^{l,r}-\tau ((a_1^{l,r}a_2^{l,r}+d_{12}^{l,r})u+((a_2^{l,r})^2+d_{22}^{l,r})v+A^{l,r}a_2^{l,r}+b_2^{l,r})]y
\\&+C_8 g_0^{l,r}[a_2^{l,r}v-\tau (((a_1^{l,r}a_2^{l,r}+d_{12}^{l,r})uv+((a_2^{l,r})^2+d_{22}^{l,r})v^2+(A^{l,r}a_2^{l,r}+b_2^{l,r})v)]
\\&+\frac{1}{2}C_9 g_0^{l,r}((a_1^{l,r})^2+d_{11}^{l,r})u^2+\frac{1}{2}C_7g_0^{l,r}((a_2^{l,r})^2+d_{22}^{l,r})y^2
\\&+C_8 g_0^{l,r}((a_2^{l,r})^2+d_{22}^{l,r})vy+\frac{1}{2}C_9 g_0^{l,r}((a_2^{l,r})^2+d_{22}^{l,r})v^2
\\&+C_8 g_0^{l,r}(a_1^{l,r}a_2^{l,r}+d_{12}^{l,r})uy+C_9 g_0^{l,r}(a_1^{l,r}a_2^{l,r}+d_{12}^{l,r})uv.
\end{split}
\end{equation}
The time integral coefficients are given by
\begin{equation}\label{3rdsolver_equ4}
\begin{split}
&C_1=1-e^{-t/ \tau _n},C_2=(t+\tau)e^{-t/ \tau _n}- \tau,C_3=t-\tau+\tau e^{-t/ \tau _n},C_4=(-t^2-2\tau t)e^{-t/\tau _n},
\\&C_5=t^2-2\tau t,C_6=-\tau t(1+e^{-t/ \tau _n}),C_7=e^{-t/ \tau _n},C_8=-te^{-t/ \tau _n},C_9=t^2e^{-t/ \tau _n}.
\end{split}
\end{equation}

\subsubsection{Simplified 3rd-order GKS flux in 2D}

The full 3rd-order flux function is very complicated. A simplified version has been proposed by Zhou et al. \cite{3rdGKS-simplified}.
The new set of coefficients is introduced as
\begin{equation}\label{flux_solver11}
\begin{split}
&a_x=a_1=g_x/g,a_y=a_2=g_y/g,a_t=A=g_t/g,
\\
&a_{xx}=g_{xx}/g,a_{xy}=g_{xy}/g,a_{yy}=g_{yy}/g,
\\
&a_{xt}=g_{xt}/g,a_{yt}=g_{yt}/g,a_{tt}=g_{tt}/g.
\end{split}
\end{equation}
And Eq.(\ref{coe-determine-1st}) and (\ref{coe-determine-2nd}) are replaced by
\begin{equation}\label{flux_solver12}
\begin{split}
&\left\langle a_x\right\rangle=\frac{\partial{W}}{\partial{x}},
\left\langle a_y\right\rangle=\frac{\partial{W}}{\partial{y}},
\left\langle a_t+a_xu+a_xv\right\rangle=0,
\\
&\left\langle a_{xx}\right\rangle=\frac{\partial^2{W}}{\partial{x^2}},
\left\langle a_{xy}\right\rangle=\frac{\partial^2{W}}{\partial{xy}},
\left\langle a_{yy}\right\rangle=\frac{\partial^2{W}}{\partial{y^2}},
\\
&\left\langle a_{xx}u+a_{xy}v+a_{xt}\right\rangle=0,
\\
&\left\langle a_{xy}u+a_{yy}v+a_{yt}\right\rangle=0,
\\
&\left\langle a_{xt}u+a_{yt}v+a_{tt}\right\rangle=0.
\end{split}
\end{equation}
The final distribution function becomes
\begin{equation}\label{3rd-simplify-flux}
\begin{split}
f(0,y,t,u,v,\xi)=&\bar{g}+\frac{1}{2}\bar{g}_{yy}y^2+\bar{g}_tt+\frac{1}{2}\bar{g}_{tt}t^2-\tau[(\bar{g}_t+u\bar{g}_x+v\bar{g}_y)+(\bar{g}_{tt}+u\bar{g}_{xt}+v\bar{g}_{yt})t]
\\&-e^{-t/\tau_n}[\bar{g}-(u\bar{g}_x+v\bar{g}_y)t]
\\&+e^{-t/\tau_n}[g^l-(ug^l_x+vg_y^l)t]H(u)+e^{-t/\tau_n}[g^r-(ug^r_x+vg^r_y)t](1-H(u)).
\end{split}
\end{equation}

Both the full and simplified 3rd-order GKS fluxes could achieve the theoretical accuracies.
The reason may come from the insensitivity of macroscopic flux function to the microscopic particle distribution function once the
conservation is fully imposed in the evolution process.
Since the simplified 3rd-order flux has about 4 times speed-up in comparison with the complete 3rd-order one in the 2D case,
the simplified flux function will be used in all test cases in this paper.

\subsubsection{2nd-order GKS flux in 2D}

If we drop all second-order derivative in Eq.(\ref{3rdsolver_equ1}) and Eq.(\ref{3rdsolver_equ3}),
the traditional 2nd-order GKS flux solver can be recovered
\begin{equation}\label{2ndgks-1}
\begin{split}
f(0,y,t,u,v,\xi)&=(1-e^{-t/\tau_n})\bar{g}+((t-\tau)e^{-t/ \tau_n}-\tau)(u\bar{a}_1+v\bar{a}_2)+(t-\tau +\tau e^{-t/ \tau_n})\bar{A}g_0
\\&+e^{-t/\tau_n}[g^l-(ug^l_x+vg_y^l)(\tau+t)-\tau A_r]H(u)
\\&+e^{-t/\tau_n}[g^r-(ug^r_x+vg^r_y)(\tau+t)-\tau A_l](1-H(u)).
\end{split}
\end{equation}

\subsection{Numerical Algorithm for MSMD GKS}

The numerical flux of the GKS is a complicated function of time in the non-smooth region.
In order to construct MSMD GKS, the 1st-order and 2nd-order time derivatives of the flux function have to be
properly evaluated. Since the main contribution of a flux function in a numerical scheme is about the total transport within a time
step between cells, the time derivatives of a flux function are evaluated on the average of a time step.
Denote the total transport of flux at the cell interface $i+1/2$ within a time interval $\delta$,
\begin{align*}
\mathbb{F}_{i+1/2}(W^n,\delta)
=\int_{t_n}^{t_n+\delta}F_{i+1/2}(W^n,t)dt&=\int_{t_n}^{t_n+\delta}\int
u \psi f(x_{i+1/2},t,u, v,\xi)d\Xi dt.
\end{align*}
For a 2nd-order GKS flux, the flux can be approximated to be a linear function within a time step,
\begin{align}\label{na1}
F_{i+1/2}(W^n,t)=F_{i+1/2}^n+ \partial_t F_{j+1/2}^nt.
\end{align}
The coefficients $F_{j+1/2}^n$ and $\partial_tF_{j+1/2}^n$ can be
determined as follows
\begin{align*}
F_{i+1/2}^n\Delta t&+\frac{1}{2}\partial_t
F_{i+1/2}^n\Delta t^2 =\mathbb{F}_{i+1/2}(W^n,\Delta t)\\
\frac{1}{2}F_{i+1/2}^n\Delta t&+\frac{1}{8}\partial_t
F_{i+1/2}^n\Delta t^2 =\mathbb{F}_{i+1/2}(W^n,\Delta t/2)
\end{align*}
By solving the linear system, we have
\begin{align}\label{na2}
F_{i+1/2}^n&=(4\mathbb{F}_{i+1/2}(W^n,\Delta t/2)-\mathbb{F}_{i+1/2}(W^n,\Delta t))/\Delta t,\nonumber\\
\partial_t F_{i+1/2}^n&=4(\mathbb{F}_{i+1/2}(W^n,\Delta t)-2\mathbb{F}_{i+1/2}(W^n,\Delta t/2))/\Delta
t^2.
\end{align}
Similar formulation can be obtained for the flux in the y-direction.

For the 3rd-order GKS flux, $F(t)$ is approximated by a quadratic function of time with second-order time derivative,
\begin{align}\label{na3}
F_{i+1/2}(W^n,t)=F_{i+1/2}^n+ \partial_t
F_{i+1/2}^nt+\frac{1}{2}\partial_{tt}F_{i+1/2}^nt^2.
\end{align}
Three conditions
\begin{align*}
F_{i+1/2}^n\Delta t+\frac{1}{2}\partial_t F_{i+1/2}^n\Delta
t^2+\frac{1}{6}\partial_{tt}
F_{i+1/2}^n\Delta t^3&=\mathbb{F}_{i+1/2}(W^n,\Delta t),\\
\frac{2}{3}F_{i+1/2}^n\Delta t+\frac{2}{9}\partial_t F_{i+1/2}^n\Delta
t^2+\frac{4}{81}\partial_{tt}
F_{i+1/2}^n\Delta t^3&=\mathbb{F}_{i+1/2}(W^n,2\Delta t/3),\\
\frac{1}{3}F_{i+1/2}^n\Delta t+\frac{1}{18}\partial_t F_{i+1/2}^n\Delta
t^2+\frac{1}{162}\partial_{tt} F_{i+1/2}^n\Delta
t^3&=\mathbb{F}_{i+1/2}(W^n,\Delta t/3)
\end{align*}
can be used to determine these coefficients
\begin{align}\label{na4}
F_{i+1/2}^n&=\frac{1}{\Delta t}(\mathbb{F}_{i+1/2}(W^n,\Delta t)-\frac{9}{2}\mathbb{F}_{i+1/2}(W^n,2\Delta t/3)+9\mathbb{F}_{i+1/2}(W^n,\Delta t/3)) , \nonumber\\
\partial_t F_{i+1/2}^n&=-\frac{9}{\Delta t^2}(\mathbb{F}_{i+1/2}(W^n,\Delta t)-4\mathbb{F}_{i+1/2}(W^n,2\Delta t/3)+5\mathbb{F}_{i+1/2}(W^n,\Delta t/3)) , \nonumber\\
\partial_{tt} F_{i+1/2}^n&=\frac{9}{\Delta t^3}(3\mathbb{F}_{i+1/2}(W^n,\Delta t)-9\mathbb{F}_{i+1/2}(W^n,2\Delta t/3)+9\mathbb{F}_{i+1/2}(W^n,\Delta t/3)) .\nonumber\\
\end{align}


\subsection{Remarks on spatial reconstructions}

In 1-D case, the standard WENO5-Z reconstruction \cite{wenoz} based on characteristic variables
is applied to obtain the cell interface values $W^{l,r}$.

In 2-D case, the reconstruction is conducted direction by direction, and the multi dimensional effect is included
through the flux evaluation at Gaussian quadrature  points on each cell interface.
For example, at the cell interface $(i+1/2,j)$,
1-D WENO5-Z reconstruction is first applied to get interface averaged values
$\widetilde{W}^{l,r}_{i+1/2,j}$ by using the averaged values within the neighboring cells $W_{i-2,j}...W_{i+3,j}$.
Then, the tangential reconstruction based on $\widetilde{W}^{l,r}_{i+1/2,j-2}...\widetilde{W}^{l,r}_{i+1/2,j+2}$
is conducted by using the 1-D WENO5-Z again in the y-direction to obtain the values at the Gaussian points.
The flux transport in the x-direction through the three Gaussian points is evaluated by quadratures
\begin{align}
\frac{1}{\Delta
y}\int_{y_{j-1/2}}^{y_{j+1/2}}F(W(x_{i+1/2},y,t))dy=\sum_{l=1}^3\omega_lF(W(x_{i+1/2},y_l,t)),
\end{align}
where $y_l=0,\pm\frac{1}{2}\sqrt{\frac{3}{5}}\Delta y$ and $\omega_l=\frac{4}{9},\frac{5}{18},\frac{5}{18}$ accordingly.
This guarantees a fifth-order accuracy for the flux calculation in the tangential direction.
This reconstruction procedure is exactly the same as the Class B method  defined in \cite{accuracy-FVM}.

For GKS, besides the point-wise values at a cell interface, the slopes on both sides of a cell interface are also needed in the flux evaluations.
For the initial discontinuous non-equilibrium part $g_0^{l,r }|_{x_{i+1/2},y_l}$,
they have one to one correspondences to $W^{l,r}_{i+1/2,j_l}$ at each Gaussian point, which are reconstructed
in the same way as that for the Riemann solvers.
For the equilibrium part $\bar{g}$, the interface averaged value $\widetilde{\bar{W}}_{i+1/2,j}$
is obtained by Eq.(\ref{g0-collision}).
Since $\bar{g}$ represents a continuous equilibrium flow,
a 4th-order polynomial can be uniquely determined by using $\widetilde{\bar{W}}_{i+1/2,j-2}...\widetilde{\bar{W}}_{i+1/2,j+2}$ and the point-wise value $\widetilde{\bar{W}}_{i+1/2,j_l}$ can be obtained.

For the non-equilibrium state within each cell $i$,
based on the the reconstructed cell interface values $(W_{i-1/2}^r, W_{i+1/2}^l)$ and cell averaged $W_i$,
a 2nd order polynomial could be constructed within the cell.
For the equilibrium state reconstruction at the cell interface $i+1/2$,
the stencil is the four cell average values $W_{i-1}...W_{i+2}$ and the interface value $\bar{W}_{i+1/2}$ itself.
Then, a 4th-order polynomial could be obtained  without using any limiter.
This equilibrium construction has a 5th-order accuracy in smooth region.
The detail construction could be found in \cite{3rdGKS-Luo}.

In the 2-D case, at the cell interface $(i+1/2,j)$,
after obtaining the interface averaged derivatives $\partial_x \widetilde{W}^{l,r}, \partial_{xx} \widetilde{W}^{l,r}$,
the derivatives at each Gaussian point is constructed by the same WENO reconstruction in tangential direction as for the
reconstruction of $W^{l,r}$.
On the other hand, after obtaining $W^{l,r}$ at these three Gaussian points,
$\partial_y W^{l,r},\partial_{yy} W^{l,r}$ could be determined by a 2nd-order polynomial which passes through these three points.
And $\partial_{xy} W^{l,r}$ could be calculated in the same way from the data $\partial_x W^{l,r}$ at the Gaussian points.
For the equilibrium part, rather than the WENO reconstruction, a 4th-order polynomial is used to determine all derivatives
along the tangential direction, and a 5th-order accuracy can be achieved in smooth region.

Since he Gaussian points are used to evaluate the flux transport along the cell interface,
the $y$ and $y^2$ terms in Eq.(\ref{3rdsolver_equ1}), (\ref{3rdsolver_equ3}), and (\ref{3rd-simplify-flux}) can be ignored
in the flux evaluation.

\section{Numerical Tests}

The schemes tested in this section include many MSMD GKS methods.
The names of the schemes are defined as SnOr with the definition of n-stages and r-th-order accuracy, such as
S1O2 (single stage, 2nd-order accuracy),  S1O3 (single stage, 3rd-order accuracy), etc.
S3O5 (three stages, 5th-order) takes the time marching strategy in Table. \ref{s3o511},
while S3O5+ takes the strategy in Table. \ref{s3o522}.
For different kinds of S2O5 schemes, the suffix "c" and "s" indicates its usage of complete 3rd-order flux
or the simplified one. And the suffix "+" refers the coefficients in Table. \ref{5th2stage_modified},
while the absence of "+" refers the one in Table. \ref{5th2stage}.
If no special illustration, the numerical results in comparison, especially with the Godunov type schemes,
are based on the same spatial reconstruction.

For inviscid flow computations, the physical collision time $\tau = \mu /p = 0$, where $\mu$ is the dynamical viscosity coefficient and
$p$ is the pressure, the numerical collision time
\begin{align*}
\tau_{n}=C_1 \Delta t+C_2\displaystyle|\frac{p_l-p_r}{p_l+p_r}|\Delta
t,
\end{align*}
where $C_1,C_2$ are two constants. Generally, $C_1\ll 1$ and $C_2\sim \mathcal{O}(1)$ under current WENOZ reconstruction.
For viscous flow computation, the physical collision time is defined as $\tau=\mu/{p}$ and the numerical collision time is
$$\tau_{n}= \frac{\mu}{p} + C_2 \displaystyle|\frac{p_l-p_r}{p_l+p_r}|\Delta t .$$


\subsection{Accuracy tests}

For the Euler equations, the smooth density propagation is used for the accuracy evaluation.
In these cases, both the physical viscosity and the collision time related to the
numerical dissipation are set to zero.
The initial condition for the 1-D density advection is given by
\begin{align*}
\rho(x)=1+0.2\sin(\pi x), U(x)=1, p(x)=1, x\in[0,2].
\end{align*}
The exact solution under periodic boundary condition is
\begin{align*}
\rho(x,t)=1+0.2\sin(\pi(x-t)), U(x,t)=1, p(x,t)=1.
\end{align*}
The numerical solutions from different schemes after one period of propagation at time $t=2$
are obtained and compared with the exact solution.
Since the same fifth-order spatial reconstruction is used for all schemes,
the leading truncation error of a $r$th-order GKS is expected to be $\mathcal{O}(\Delta x^5 +\Delta t^r)$.
Assume that $\Delta t = c \Delta x$ with the CFL condition,
the truncation error is then proportional to  $\mathcal{O}(\Delta x^5 + c^r (\Delta x)^r)$.
If $c\ll 1$, the leading error of $\mathcal{O}(\Delta x^5) $ due to spatial discretization
might become dominant, and it is hard to evaluate the time accuracy.
So a rather large time step $\Delta t = 0.25 \Delta x$, which corresponds to $CFL \approx 0.5$,
is used to test the spatial and temporal accuracy together.
Based on $L^1$ error,  the orders of different schemes at $t=2$ are presented in Table. \ref{1d_accuracy2}.

All schemes from 2nd-order up to 5th-order ones achieve their theoretical accuracy. Among all schemes,
the 5th-order 2-stages method with the complete 3rd-order GKS flux function has the smallest absolute error,
while the same scheme with simplified 3rd-order GKS flux function has slightly larger absolute error.
But, the order of accuracy of the scheme with simplified flux keeps the theoretical value.

The test is extended to 2-D case, where
the density perturbation propagates in the diagonal direction,
\begin{align*}
&\rho(x,y)=1+0.2\sin(\pi x)sin(\pi y),
\\& U(x,y)=1,V(x,y)=1, p(x)=1,
\end{align*}
with the exact solution
\begin{align*}
&\rho(x,y)=1+0.2\sin(\pi (x-t))sin(\pi (y-t)),
\\&U(x,y)=1,V(x,y)=1, p(x)=1.
\end{align*}
The computation domain is $[-1,1]\times[-1,1]$ and the $N{\times}N$ uniform mesh points are used with periodic boundary condition.
The results are shown in Table. \ref{2d_accuracy2} and \ref{2d_accuracy3}, with validated theoretical accuracy.

\begin{table}[!h]
\small
\begin{center}
\def\temptablewidth{0.85\textwidth}
{\rule{\temptablewidth}{1pt}}
\begin{tabular*}{\temptablewidth}{@{\extracolsep{\fill}}c|cc|cc|cc}
~ &  S1O2 &~ &  S1O3 & ~&  S2O4 & ~ \\
\hline
mesh & $L^1$ error & Order & $L^1$ error & Order& $L^1$ error & Order  \\
\hline
160&1.6449e-005&~     &6.465536e-008&~    &1.762567e-009&~\\
320& 4.11231e-006 &2.000&7.934459e-009&3.027&5.558891e-011&4.954\\
640& 1.02808e-006  &2.000&9.871997e-010&3.001&1.793678e-012&5.166\\
1280& 2.57021e-007  &2.000  &1.232565e-010&3.001&6.391980e-014&4.811\\
\Xhline{1.2pt}
~ &  S3O5 & &  S3O5+ &~ &  S2O5c \\
\hline
mesh & $L^1$ error & Order & $L^1$ error & Order& $L^1$ error & Order  \\
\hline
160&1.72333e-09&~&1.72327e-09&~&1.55396e-09&~\\
320& 5.38507e-11&5.000&5.38492e-11&5.000&4.85575e-11 &5.000\\
640& 1.68291e-12&5.000&1.68297e-12&5.000&1.51832e-12&5.000\\
1280& 5.36963e-14&4.970&5.34834e-14&4.976&4.82038e-14&4.977\\
\Xhline{1.2pt}
~ &  S2O5c+ & &  S2O5s &~ &  S2O5s+ \\
\hline
mesh & $L^1$ error & Order & $L^1$ error & Order& $L^1$ error & Order  \\
\hline
160&1.55413e-09&~&1.55396e-09&~&1.578850e-009&~\\
320& 4.85622e-11&5.000&4.85568e-11&5.000&4.933414e-011&5.000\\
640& 1.51805e-12&5.000&1.51805e-12&5.000&1.541736e-012&5.000\\
1280& 4.83028e-14&4.974&4.90587e-14&4.952&4.924277e-014&4.969\\

\end{tabular*}
{\rule{\temptablewidth}{0.1pt}}
\end{center}
\vspace{-4mm} \caption{\label{1d_accuracy2} Accuracy test for the 1-D advection
of density perturbation by GKS with different temporal accuracy under same fifth order reconstruction. $\Delta t = 0.25\Delta x$.}
\end{table}

\begin{table}[!h]
\small
\begin{center}
\def\temptablewidth{1\textwidth}
{\rule{\temptablewidth}{1pt}}
\begin{tabular*}{\temptablewidth}{@{\extracolsep{\fill}}c|cc|cc|cc}

mesh & $L^1$ error & Order & $L^2$ error & Order& $L^{\infty}$ error & Order  \\
\hline
20*20&3.67810e-06&~ & 4.46724e-06 &~ &8.88218e-06&~\\
40*40&5.85160e-08&5.974 & 7.09236e-08 & 5.976&1.40828e-07&5.979\\
80*80&9.17888e-10 &5.994 &  1.11208e-09&5.995 &　 2.20522e-09&5.997\\
160*160&1.43537e-11&5.999 & 1.73878e-11& 5.999& 3.44513e-11&6.000\\
320*320&2.24360e-13 &5.999 &2.71745e-13 & 6.000&5.435787e-13&5.989\\
\end{tabular*}
{\rule{\temptablewidth}{0.1pt}}
\end{center}
\vspace{-4mm} \caption{\label{2d_accuracy2} Accuracy test for the 2D advection
of density perturbation for S3O5+ scheme, $\Delta t = 0.1\Delta x$.}
\end{table}

\begin{table}[!h]
\small
\begin{center}
\def\temptablewidth{1\textwidth}
{\rule{\temptablewidth}{1pt}}
\begin{tabular*}{\temptablewidth}{@{\extracolsep{\fill}}c|cc|cc|cc}

mesh & $L^1$ error & Order & $L^2$ error & Order& $L^{\infty}$ error & Order  \\
\hline
20*20&3.63607e-06&~ & 4.41462e-06 &~ &8.77532e-06&~\\
40*40&5.78414e-08&5.974 & 7.00846e-08 & 5.976&1.39129e-07&5.979\\
80*80&9.07345e-10 &5.994 &  1.09893e-09&5.995 &　 2.17865e-09&5.997\\
160*160&1.41889e-11&5.999 & 1.71825e-11& 5.999& 3.40366e-11&6.000\\
320*320&2.21805e-13 &5.999 &2.68546e-13 & 6.000&5.38458e-13&5.981\\
\end{tabular*}
{\rule{\temptablewidth}{0.1pt}}
\end{center}
\vspace{-4mm} \caption{\label{2d_accuracy3} Accuracy test for the 2D advection
of density perturbation for S2O5s+ scheme, $\Delta t = 0.1\Delta x$.}
\end{table}

\subsection{One dimensional test cases}

Three Riemann problems in 1-D are selected to validate the high order GKS.
At the same time, the results from the RK4 Godunov method with exact Riemann solver are also included.
All simulations are based on the same initial reconstruction.

\bigskip
\noindent{\sl{(a) Sod problem}}

\begin{figure}[!h]
\centering
\includegraphics[width=0.4\textwidth]{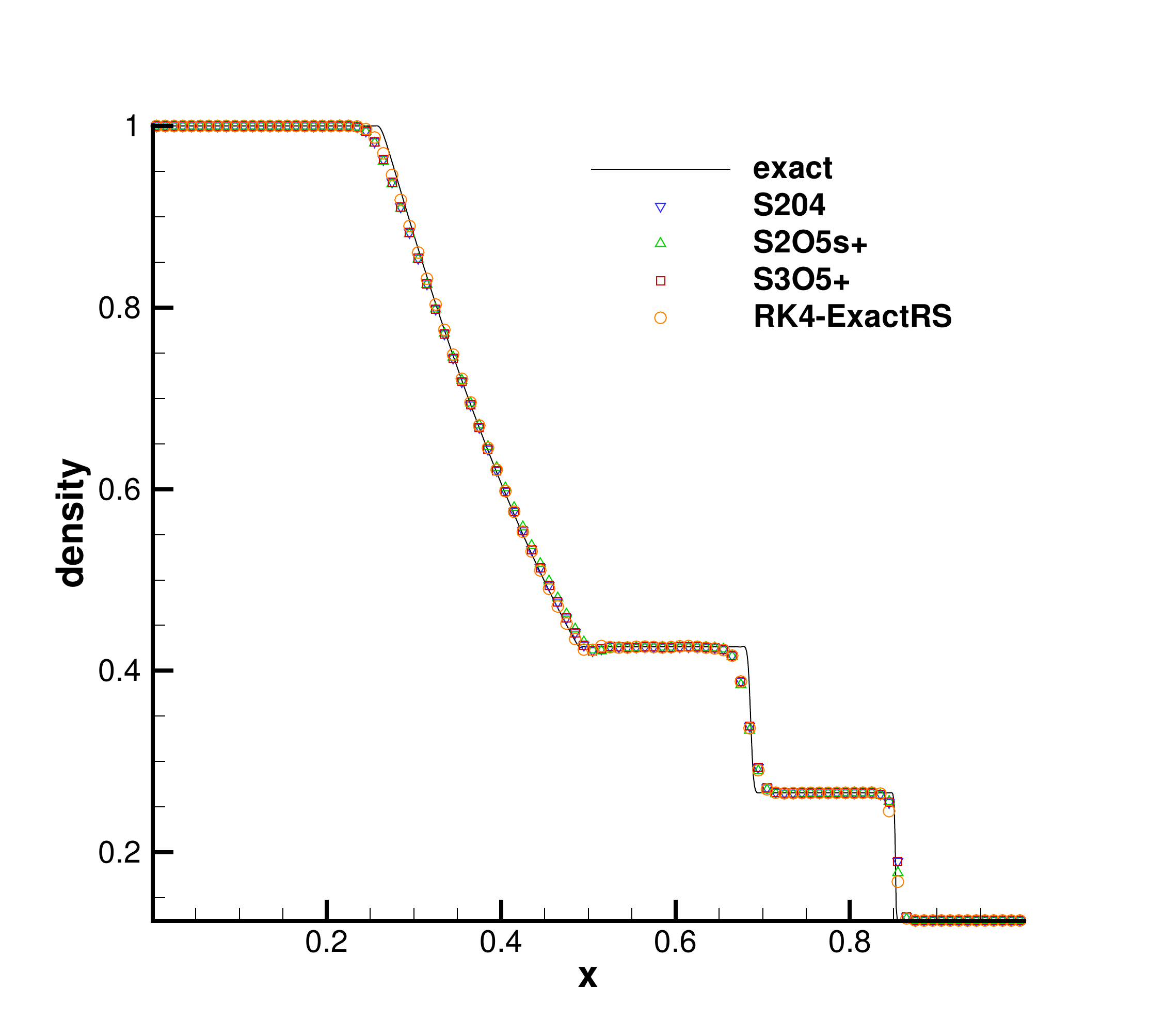}
\includegraphics[width=0.4\textwidth]{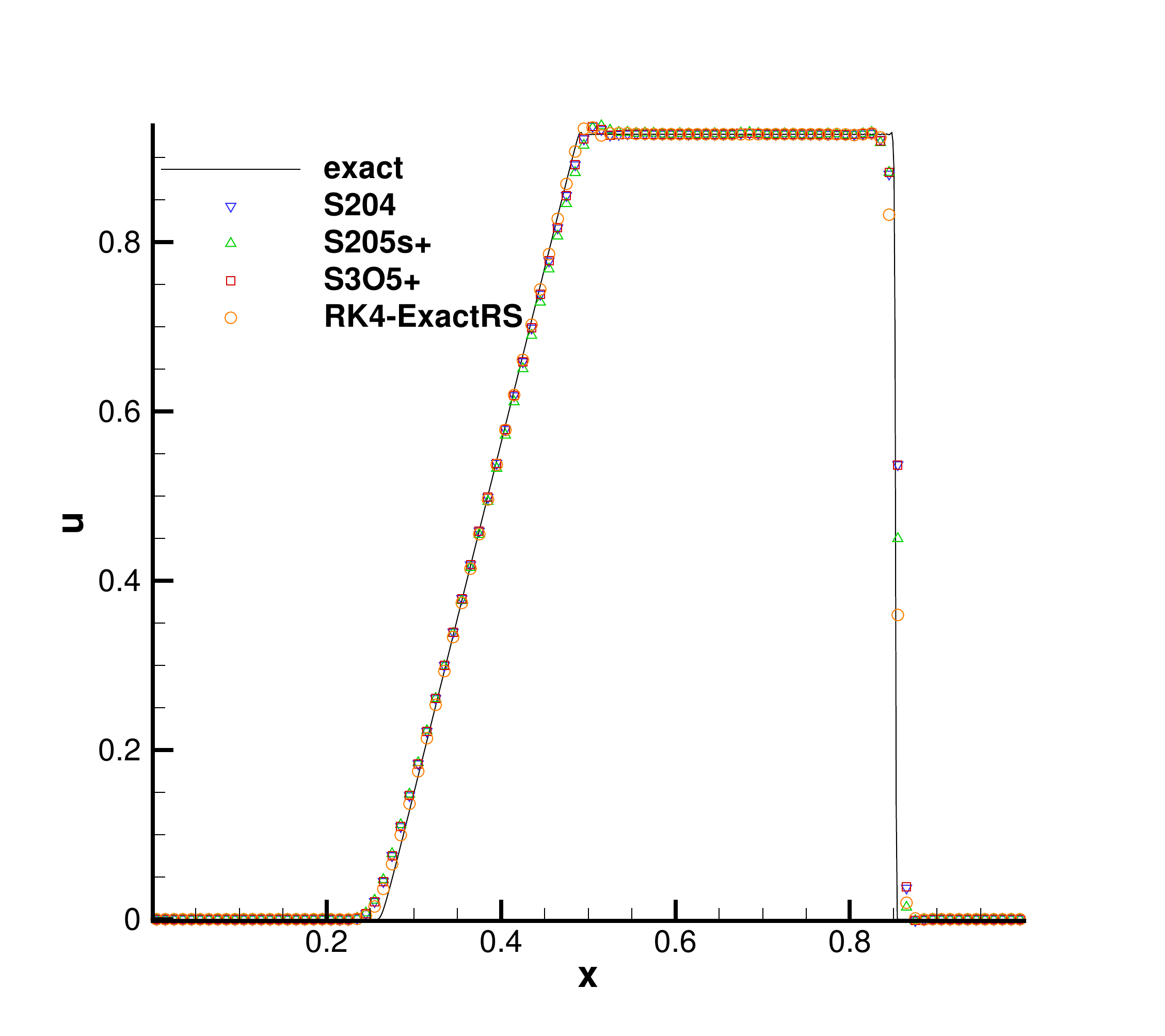}
\caption{\label{sod} The density and velocity distributions for 1-D sod problem at $t=0.2$ with $100$ cells. }
\end{figure}

The initial condition for the Sod problem is given by
\begin{equation*}
(\rho,u,p)=\left\{\begin{aligned}
&(1, 0, 1), 0<x<0.5,\\
&(0.125,0,0.1),  0.5 \leq x<1.
\end{aligned} \right.
\end{equation*}
The simulation domain is covered by $100$ uniform mesh points.
The solutions at $t=0.2$ are presented.
The WENO5-Z reconstruction is based on the characteristic variables.
The results from S2O4, S2O5s+, S3O5+, and the exact Riemann solver are shown in Fig. \ref{sod}.

\bigskip
\noindent{\sl{(b) Blast wave problem}}

\begin{figure}[!h]
\centering
\includegraphics[width=0.4\textwidth]{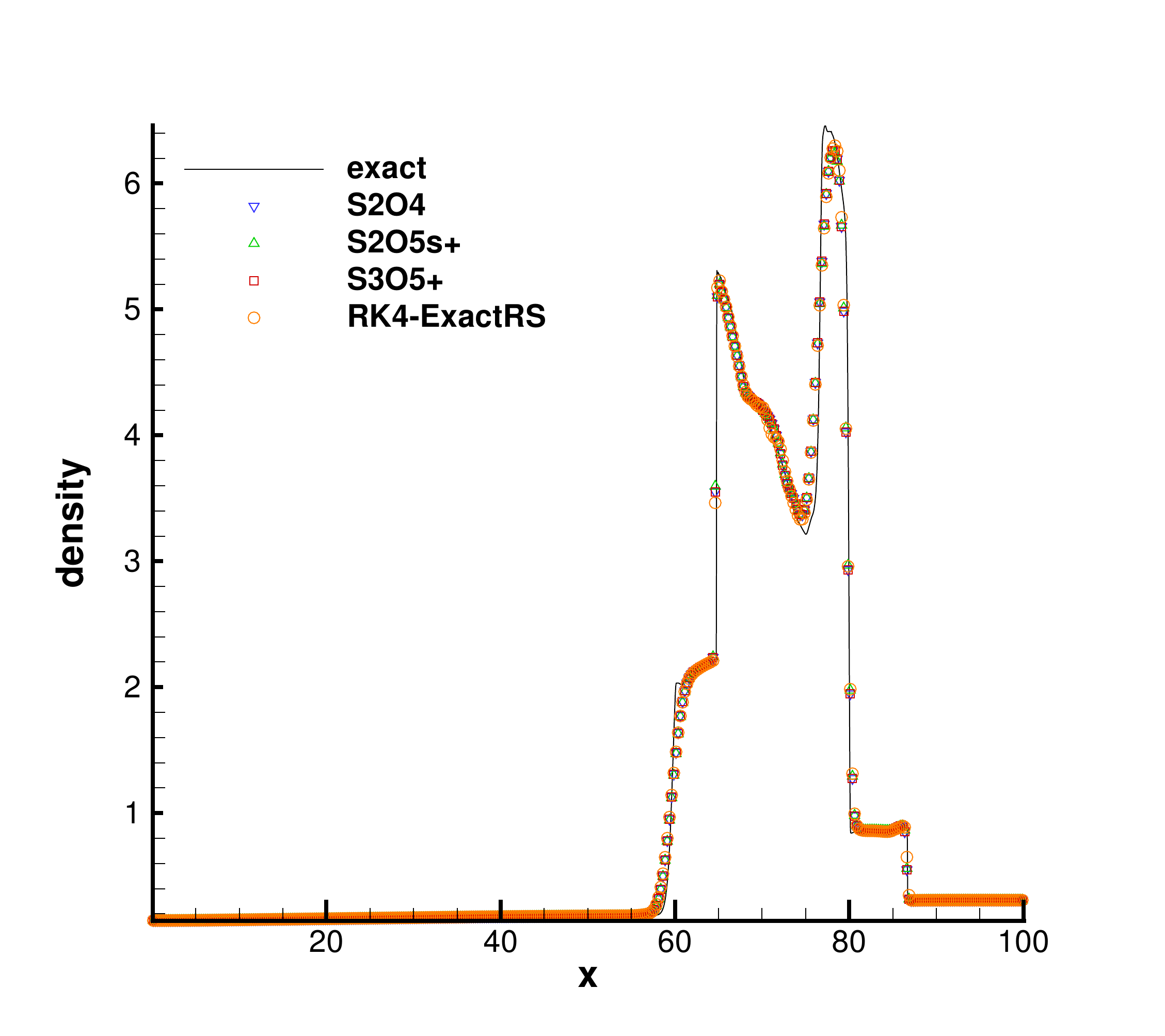}
\includegraphics[width=0.4\textwidth]{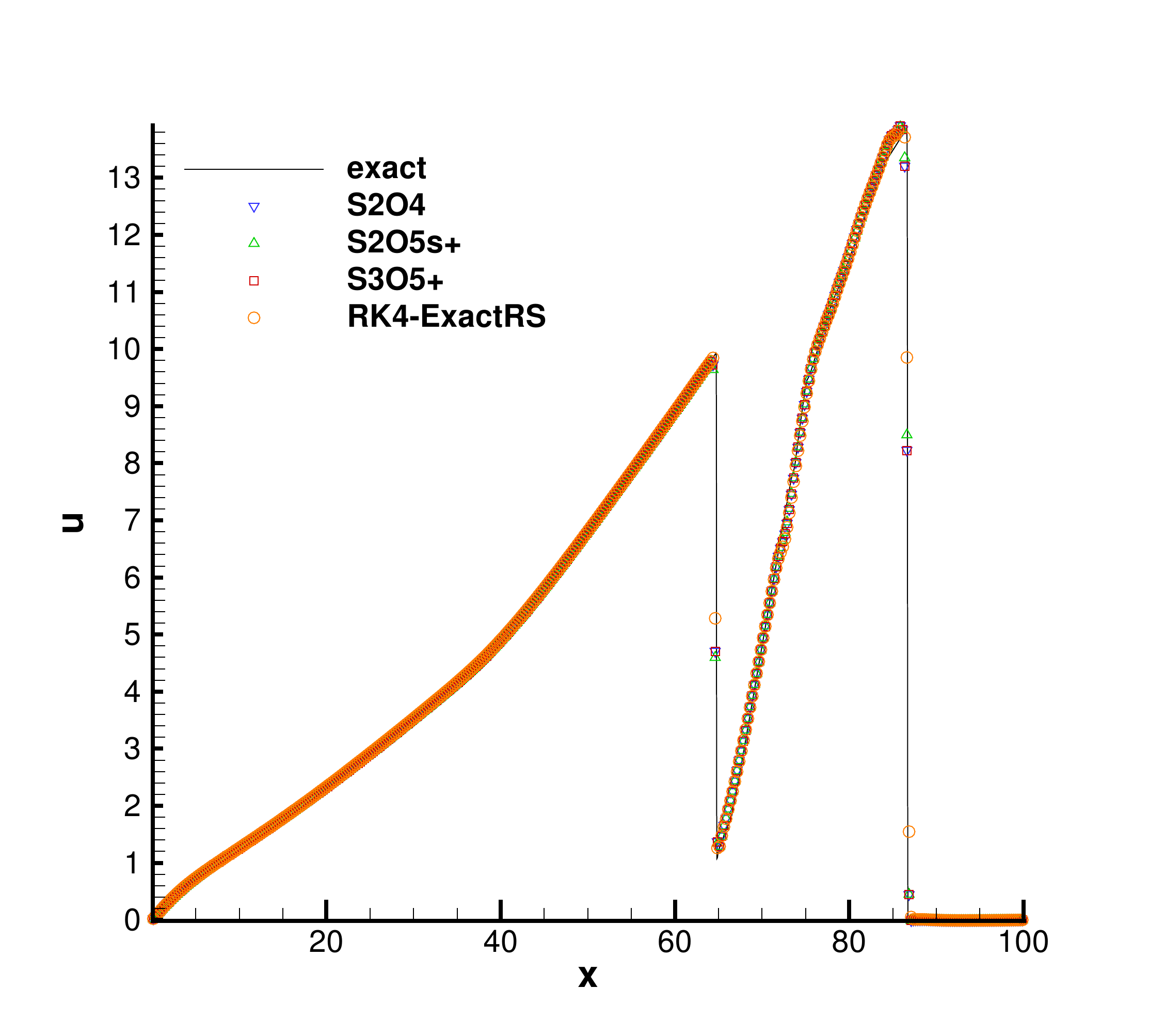}
\caption{\label{blastwave} The density and velocity distributions for 1-D blast wave problem at
$t=3.8$ with $400$ cells. }
\end{figure}

The initial conditions for the blast wave problem are given as follows
\begin{equation*}
(\rho,u,p)=\left\{\begin{aligned}
&(1, 0, 1000), 0\leq x<10,\\
&(1, 0, 0.01), 0.1\leq x<90,\\
&(1, 0, 100),  0.9\leq x\leq 100.
\end{aligned} \right.
\end{equation*}
The computational domain is covered with $400$ uniform mesh points and reflection boundary conditions are applied at both ends.
The CFL number takes $0.5$. The density and velocity distributions at $t=3.8$ are presented in Fig. \ref{blastwave}.
All schemes give almost identical results.
Based on the above observations, it seems that the Sod and blast wave cases may not be the appropriate tests to distinguish
different kinds of high order schemes.

\bigskip
\noindent{\sl{(c) Titarev and Toro problem}}

\begin{figure}[!h]
\centering
\includegraphics[width=0.44\textwidth]{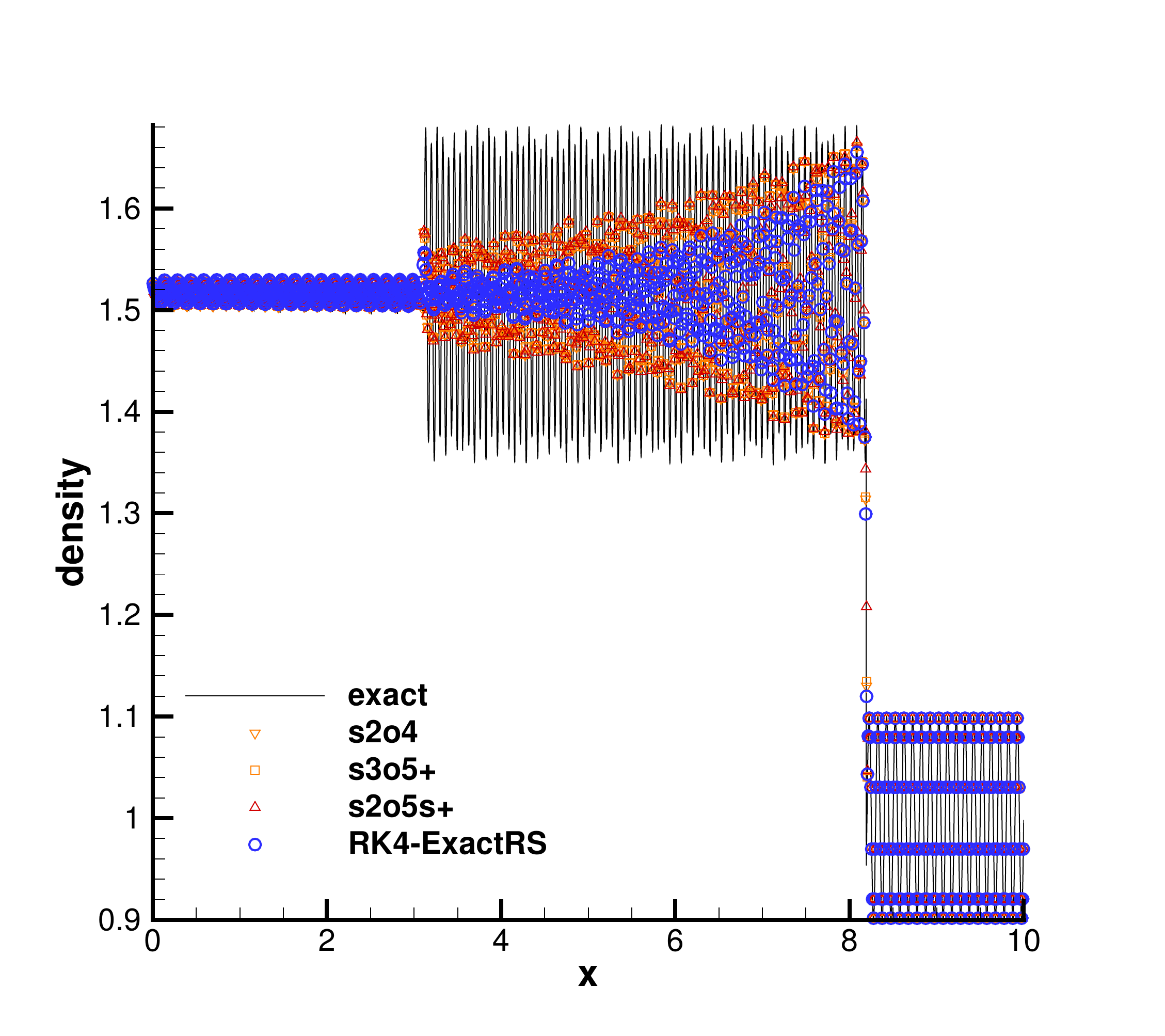}
\includegraphics[width=0.44\textwidth]{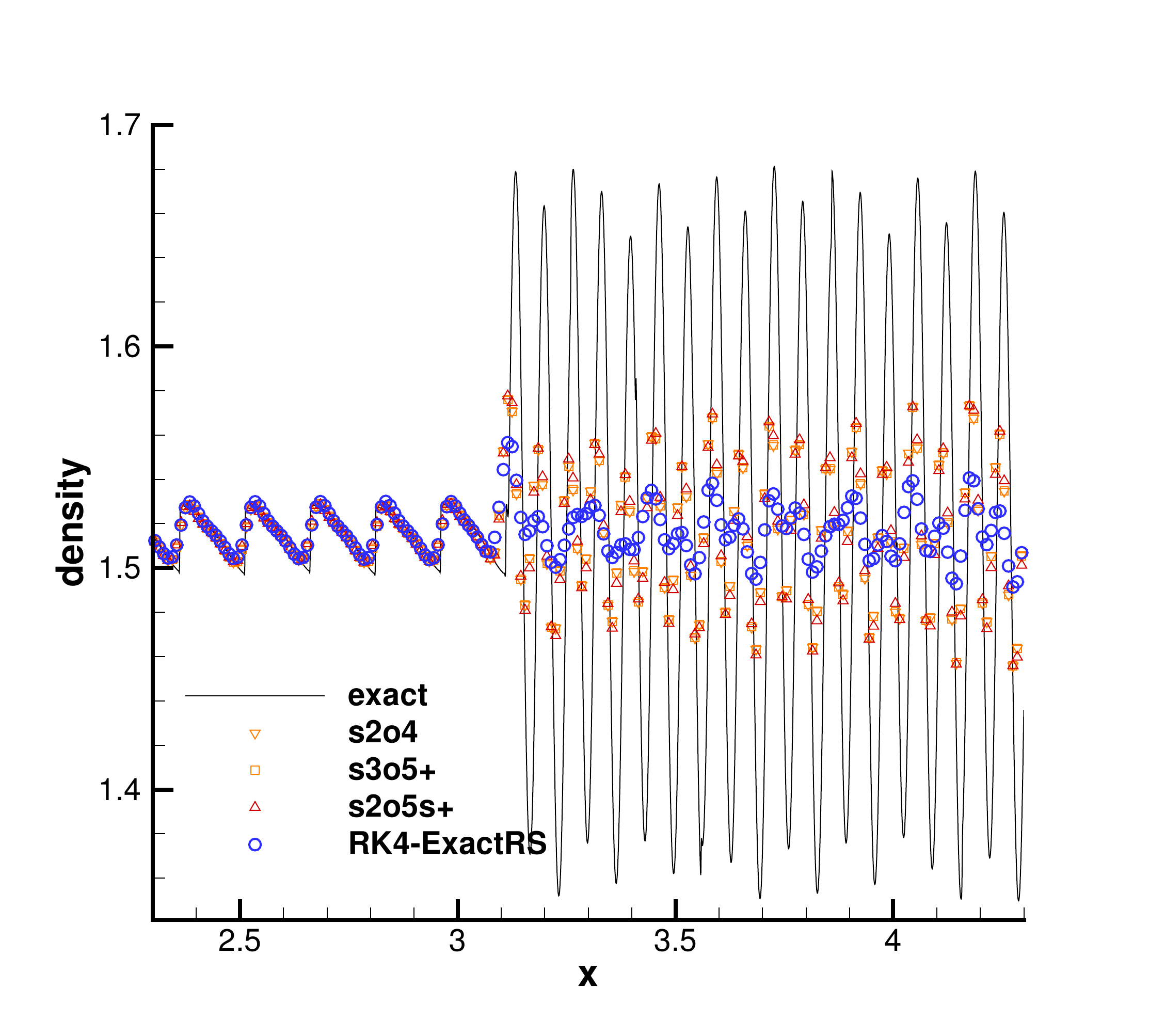}
\caption{\label{ttoro} The density distributions for 1-D Titarev-Toro problem at
$t=5.0$ with $1000$ cells. Left figure: solution in the whole domain. Right figure: local enlargement.}
\end{figure}

Titarev and Toro \cite{ttoro} problem is about high frequency oscillating sinusoidal waves propagation with shock interaction,
which is a great challenge for spatial reconstruction \cite{wenoz+,gks-benchmark} and flux solver \cite{ttoro}.
The initial conditions are given as follows
\begin{equation*}
(\rho,u,p)=\left\{\begin{aligned}
&(1.515695, 0.523346, 1.805), 0\leq x<0.5,\\
&(1+0.1sin(20 \pi x), 0, 1), 0.5\le x<10.
\end{aligned} \right.
\end{equation*}
A uniform mesh with $1000$ mesh points are used in the computational domain
and CFL = 0.5 is used for all schemes.
The result at $t=5$ is shown in Fig. \ref{ttoro}.
Different from previous test cases, there are clear differences between GKS and Riemann solutions in the middle region,
after shock interaction with smooth acoustic wave.
All multi-stage schemes present less dissipative results than those from the high order RK4 Godunov method with exact Riemann solver.
With the same initial reconstruction, the differences must come from the different type of temporal discretization and flux functions.
This test case indicates the usefulness of the high order GKS on acoustic wave computations.


\subsection{Two dimensional inviscid test cases}

\bigskip
\noindent{\sl{(a) Isentropic vortex propagation with $100$ periods}}

The isentropic vortex propagation is tested a 2-D domain for the smooth inviscid flow.
The initial condition is given by
\begin{equation}
\begin{split}
(U,V)=(1,1)+\frac{\kappa}{2\pi}e^{0.5(1-r^2)}(-\overline{y},\overline{x}), \\
T=1-\frac{(\gamma-1){\kappa}^2}{8\gamma{\pi}^2}e^{1-{\gamma}^2},\\
S=1,
\end{split}
\label{eqn_isen1}
\end{equation}
where the density $\rho$ and pressure $p$ are calculated from the temperature $T$ and the entropy $S$ by
\begin{equation}
T=\frac{p}{\rho},S=\frac{p}{\rho^{\gamma}},
\label{eqn_isen2}
\end{equation}
where $(\overline{x},\overline{y})=(x-5,y-5),~r^2=\overline{x}^2+\overline{y}^2$, and the vortex strength $\kappa=5$. The computational domain is $[0,10]\times[0,10]$. Periodic boundary condition is applied to all boundaries.

\begin{figure}[!h]
\centering
\includegraphics[width=0.7\textwidth]{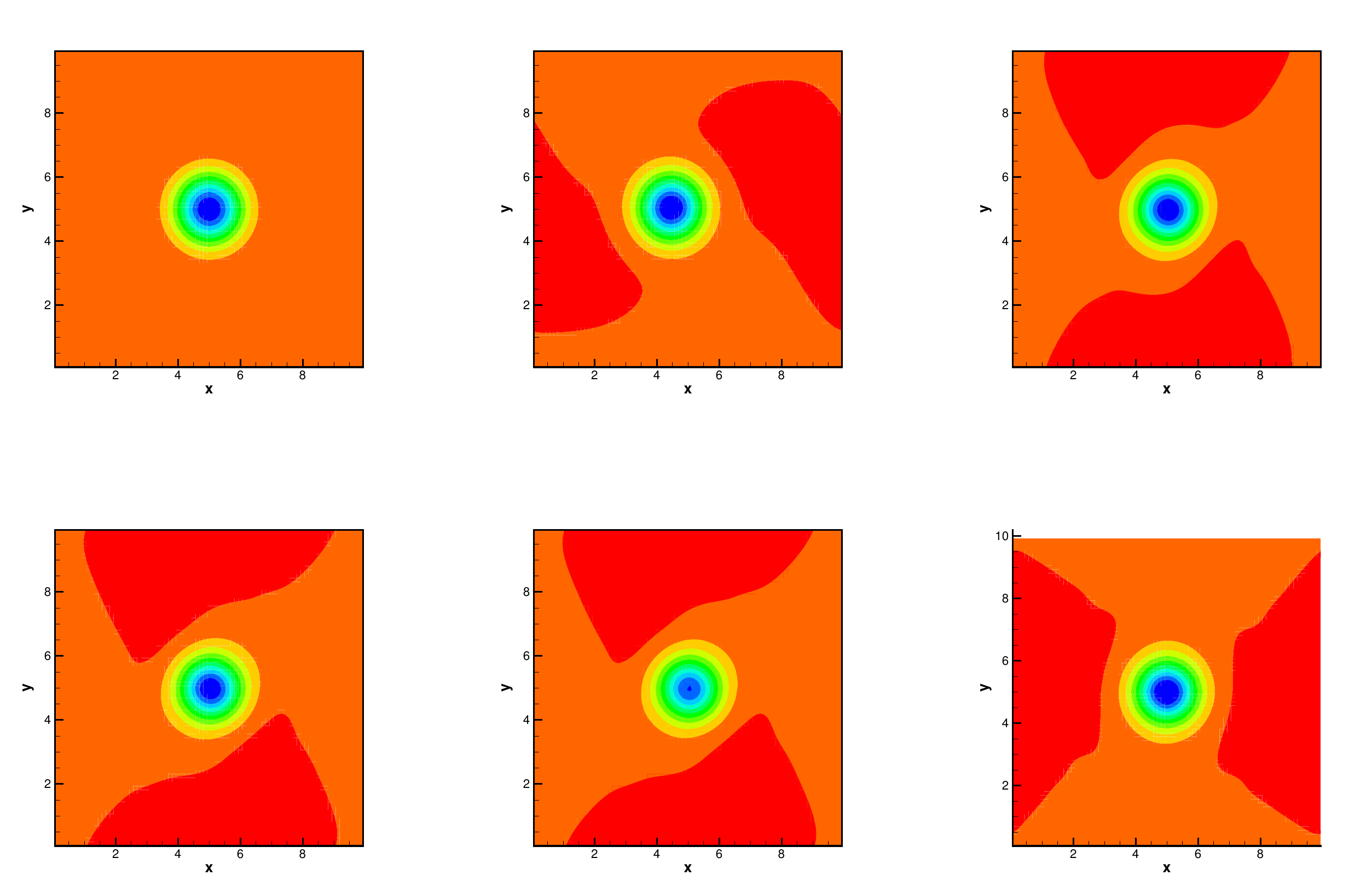}
\caption{\label{isen_long1} The density contours of isentropic vortex propagation after $100$ periods. CFL=0.4, mesh $80\times 80$.
Up row from left to right: exact solution, S1O3c, and S2O4. Down row from left ro right: S2O5s+, S3O5+, RK5-ExactRS.
 }
\end{figure}

\begin{figure}[!h]
\centering
\includegraphics[width=0.44\textwidth]{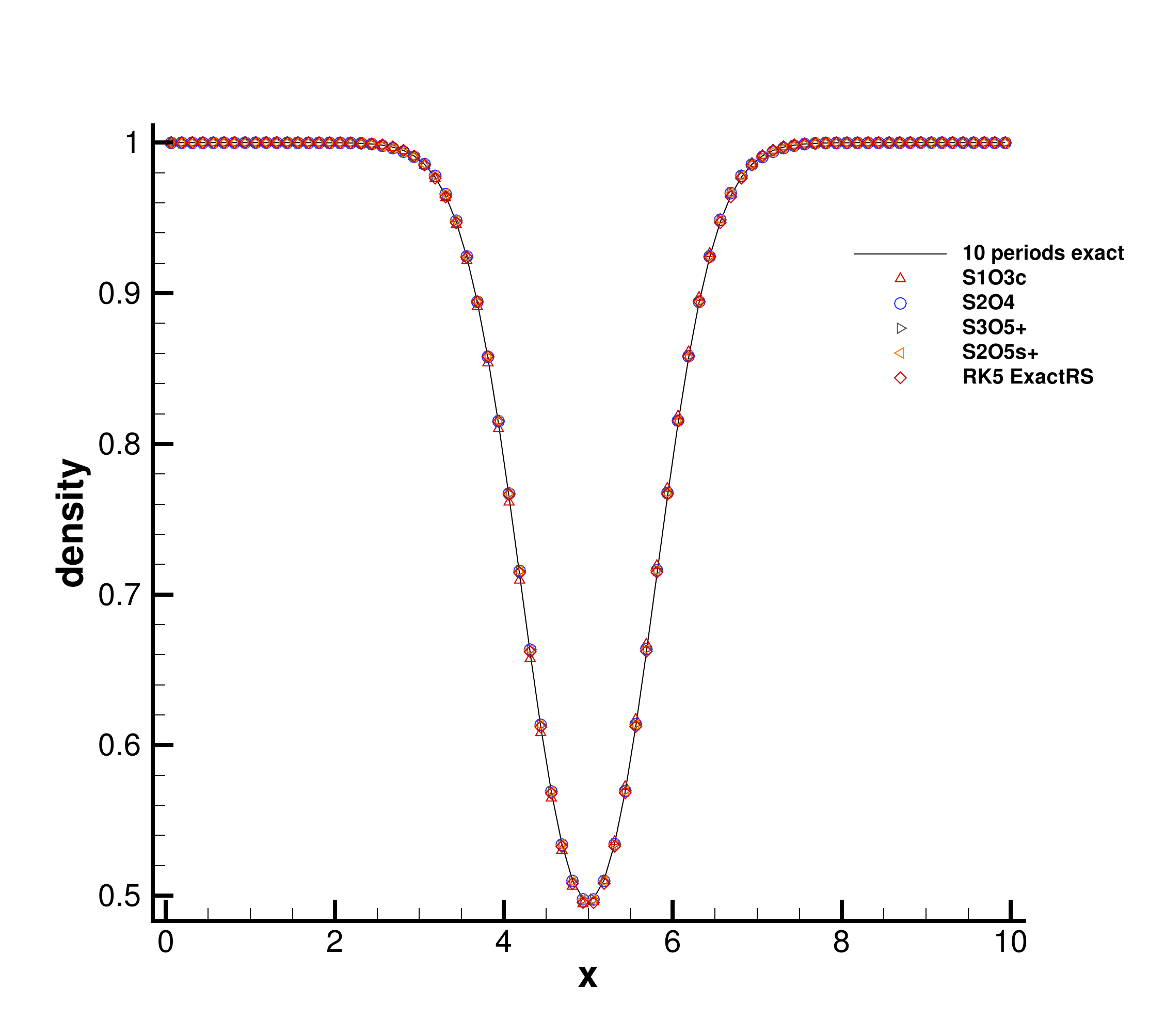}
\includegraphics[width=0.44\textwidth]{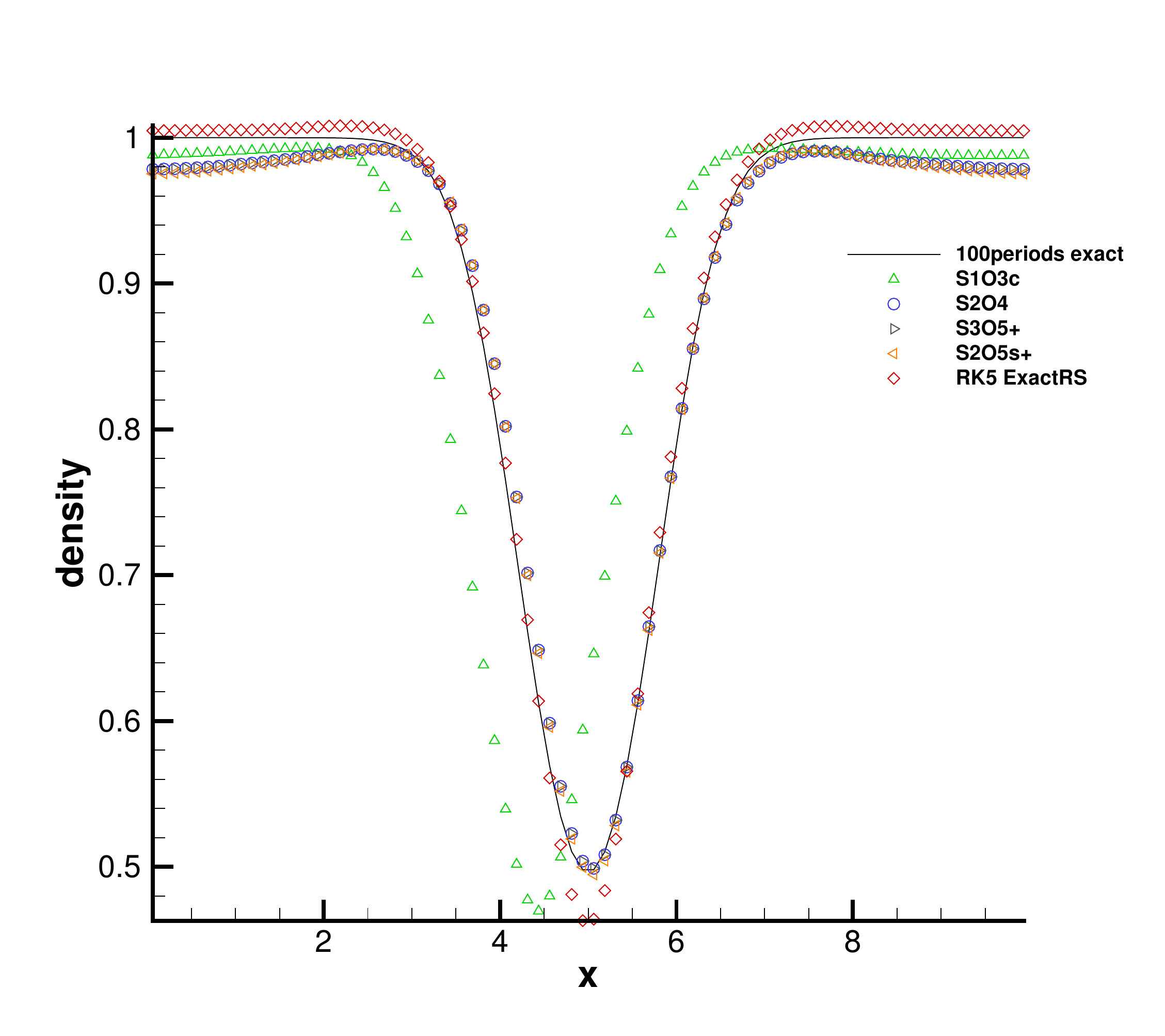}
\caption{\label{isen_long2} Density distributions for isentropic vortex propagation along $y=0$. Left: $10$ periods. Right: $100$ periods.}
\end{figure}

To show performance of different time marching schemes,
this case is tested under $1$ period, $10$ periods, and $100$ periods,
with $t=10, 100$, and $100$ accordingly. Again, the same spatial reconstruction is used for all schemes.
For $t=10$ and $t=100$, the error in density is less than $10^{-4}$ for all schemes,
which could hardly be used to distinguish the performance of different high order schemes.
However, with the output time $t=1000$ for $100$ periods vortex propagation,
the one step 3rd-order S1O3c scheme shows the dispersion and dissipation error, see in Fig.\ref{isen_long1},
while other higher order schemes still keep the vortex center in the computational domain.
It shows that the higher order time accurate scheme is important for capturing long time wave propagating behavior.
Another observation is the anti-diffusive effect in the S1O3 and RK5 Godunov method with exact Riemann solver.
A more quantitative comparison of the density distributions along $y=0.0$ are shown in Fig.\ref{isen_long2}.
It demonstrates that the close coupling of space and time evolution is
important for capturing long time wave propagation.

\bigskip
\noindent{\sl{(b) Two dimensional Riemann problems}}

For high speed compressible flow, two distinguishable flow patterns are the shock-vortex interaction  and
free shear layer \cite{shear-book}.
Here, two dimensional Riemann problems are used for the study the complicated wave structures \cite{2dRM}.
Although both test cases are for the inviscid flow,
the inherent numerical viscosity in schemes can trigger shear instability.
Another advantage for using 2-D Riemann problem is the rectangular domain and simple boundary condition.

\noindent{\sl{(b1) Interaction of planar shocks}}

\begin{figure}[!h]
\centering
  \subfigure[S3O5+]{
    \label{shock-s3o5} 
    \includegraphics[width=0.48\textwidth]{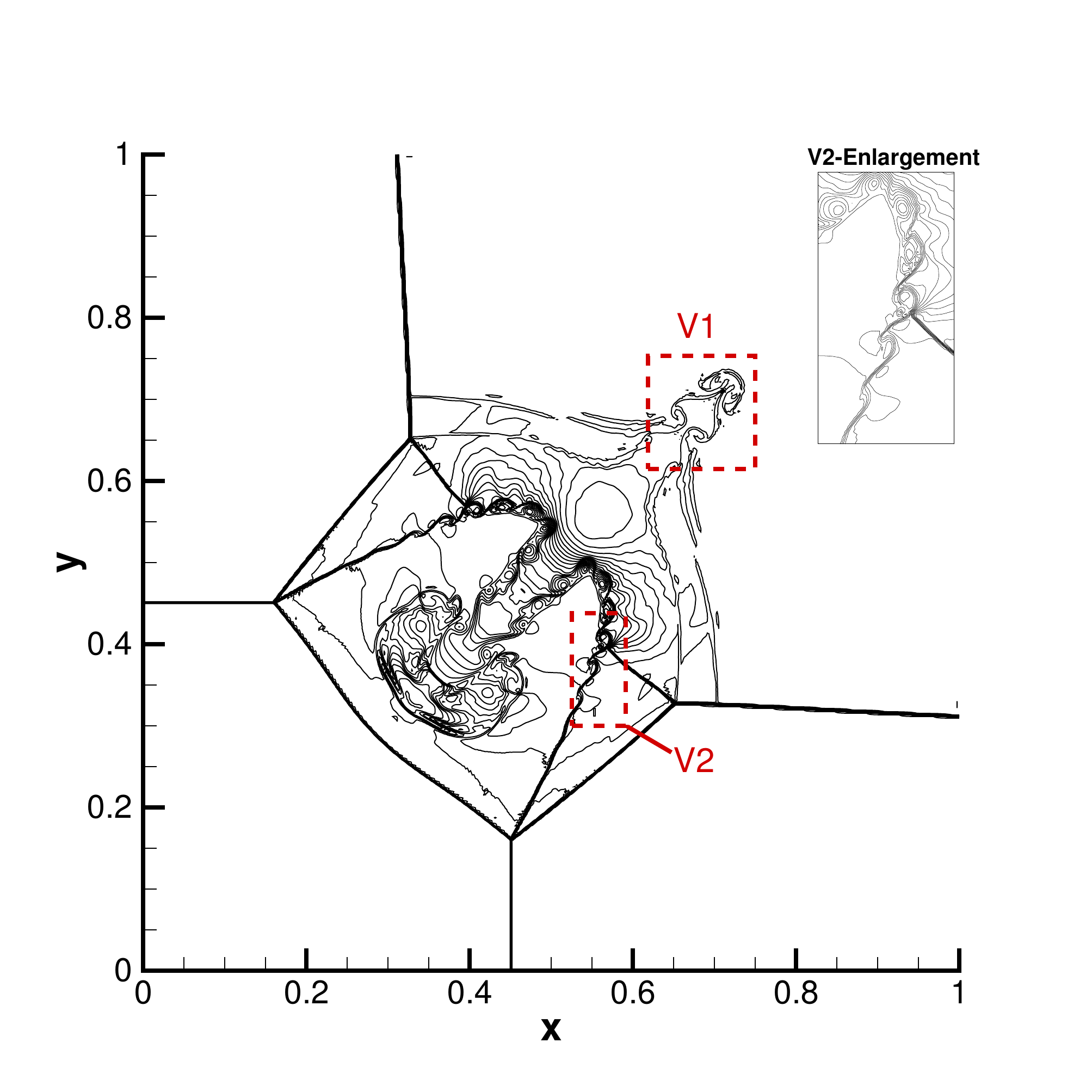}}
   \subfigure[S2O5s+]{
    \label{shock-s2o5} 
    \includegraphics[width=0.48\textwidth]{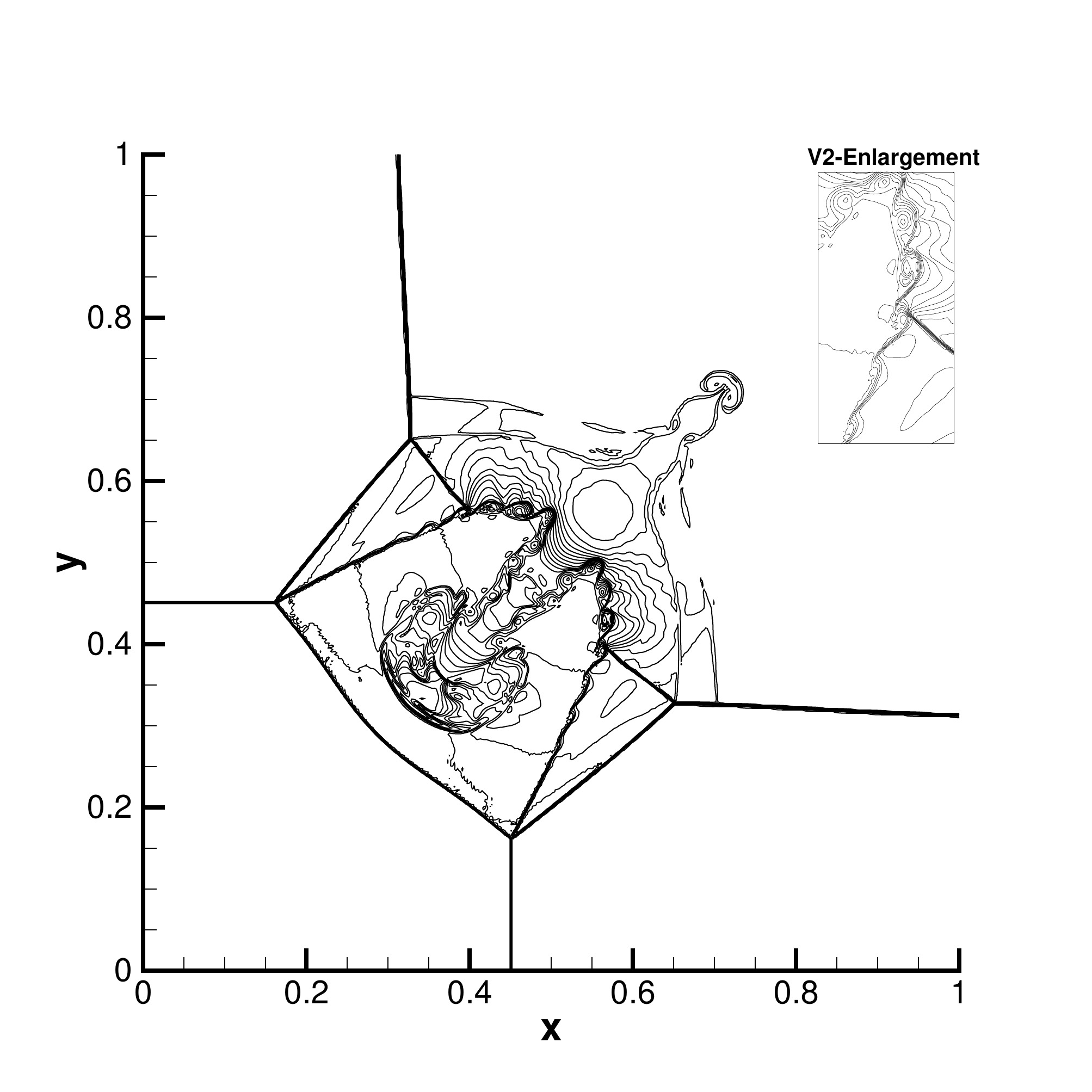}}
    \\
    \subfigure[RK4-HLLC]{
    \label{shock-hllc} 
    \includegraphics[width=0.48\textwidth]{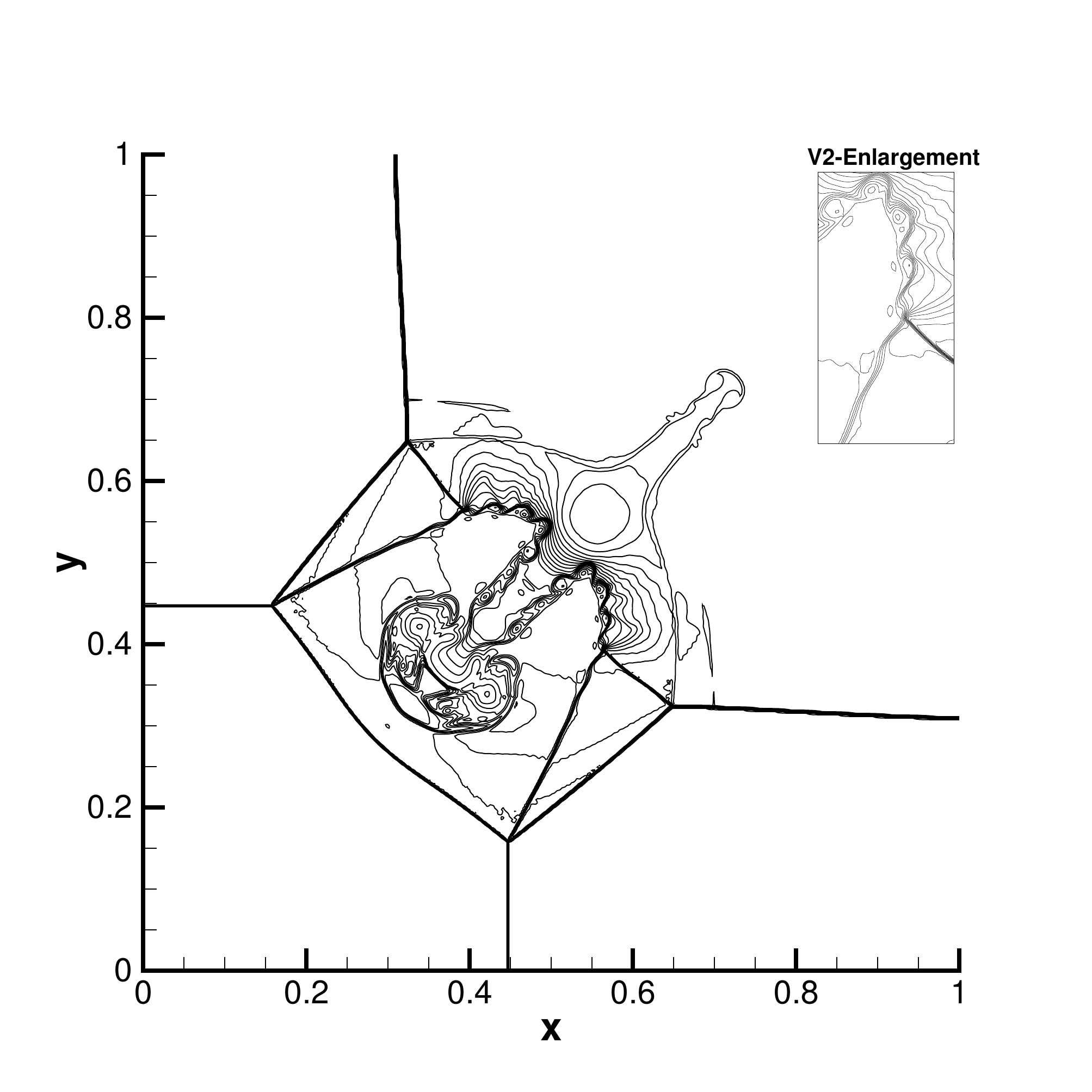}}
    \subfigure[RK4-ExactRS]{
    \label{shock-exact} 
    \includegraphics[width=0.48\textwidth]{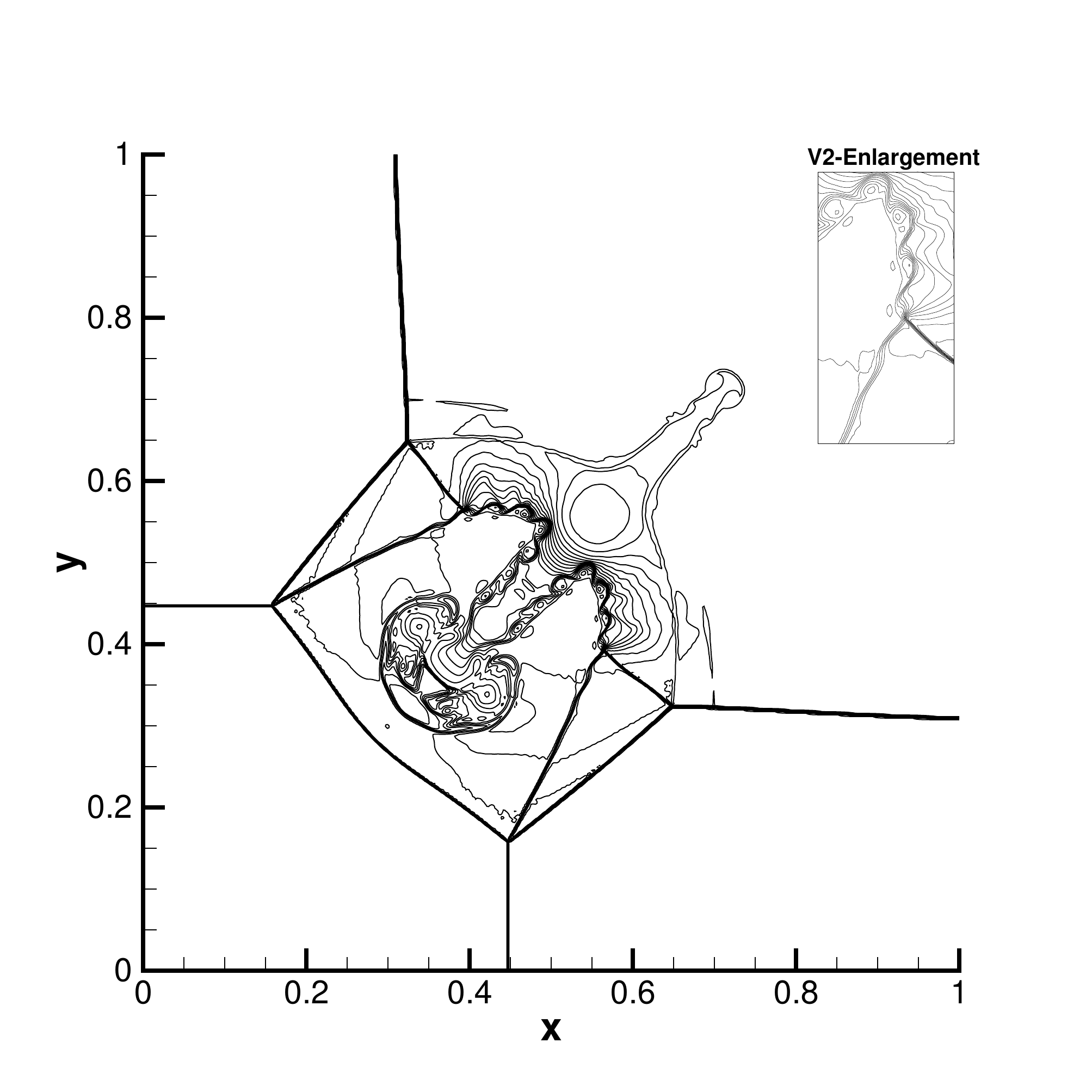}}
    \\
\caption{\label{rm2d-shock-500} The density contour of Configuration 3 case in \cite{2dRM}. $500\times 500$ mesh points are used in all calculations.}
\end{figure}

Configuration 3 in \cite{2dRM} involves the shock-shock interaction and shock-vortex interaction.
The initial condition in a square domain $[0,1]\times[0,1]$ is given by
\begin{equation*}
(\rho,u,v,p)=\left\{\begin{aligned}
&(0.138, 1.206,1.206, 0.029),& x<0.7,y<0.7,\\
&(0.5323, 0,1.206, 0.3),& x\geq 0.7,y<0.7,\\
&(1.5, 0,0, 1.5),& x\geq 0.7,y\geq 0.7,\\
&(0.5323,1.206,0, 0.3),& x<0.7, y\geq 0.7.
\end{aligned} \right.
\end{equation*}
At the output time $t=0.6$,
the same $23$ density contours are plotted in Fig. \ref{rm2d-shock-500} from different schemes.
All schemes could capture the shock sharply.
The main differences among all schemes are the strength of the the shear layers, such as the V1 and V2 regions of  Fig. \ref{shock-s3o5}.
Both S3O5+ and S2O5s+ schemes  could resolve the vortex pairs in V1 region better than those from
the RK4 Riemann solvers based schemes.
At the same time, gas kinetic schemes show stronger instabilities of the vortex sheets in V2 region.

\noindent{\sl{(b2) Interaction of planar contact discontinuities}}

\begin{figure}

  \centering
  \subfigure[S2O4]{
    \label{2ndt16} 
    \includegraphics[width=0.25\textwidth]{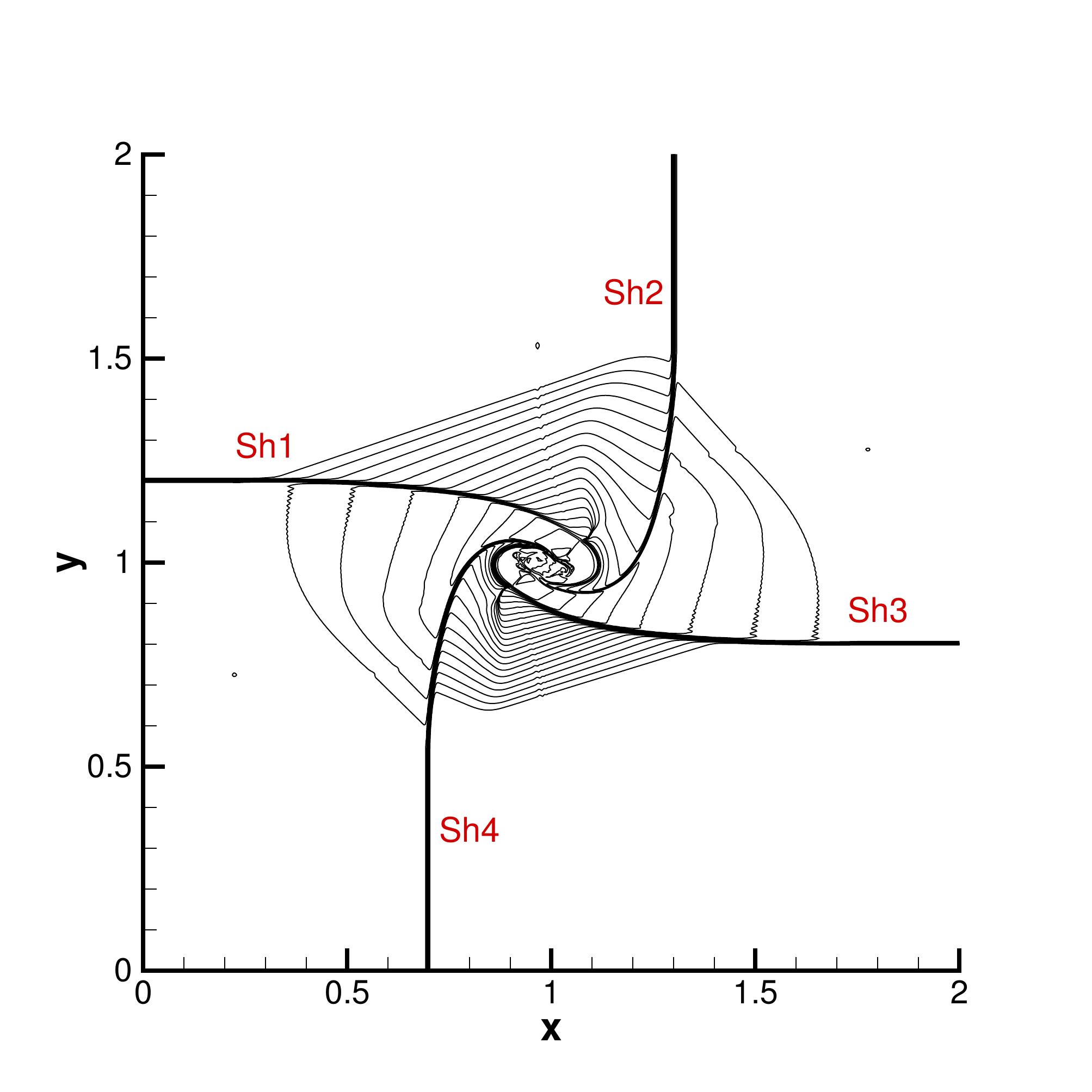}
    \includegraphics[width=0.25\textwidth]{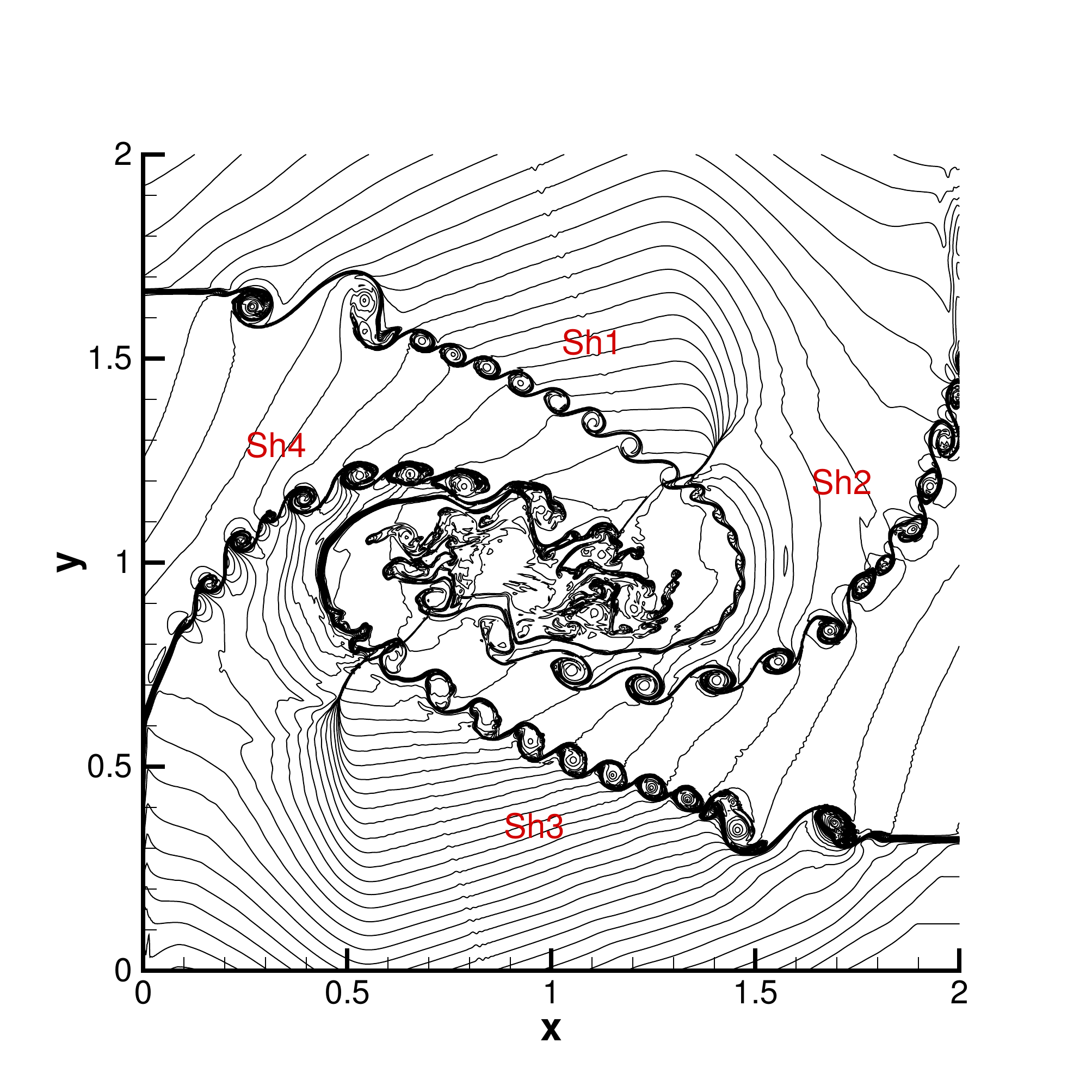}
    \includegraphics[width=0.25\textwidth]{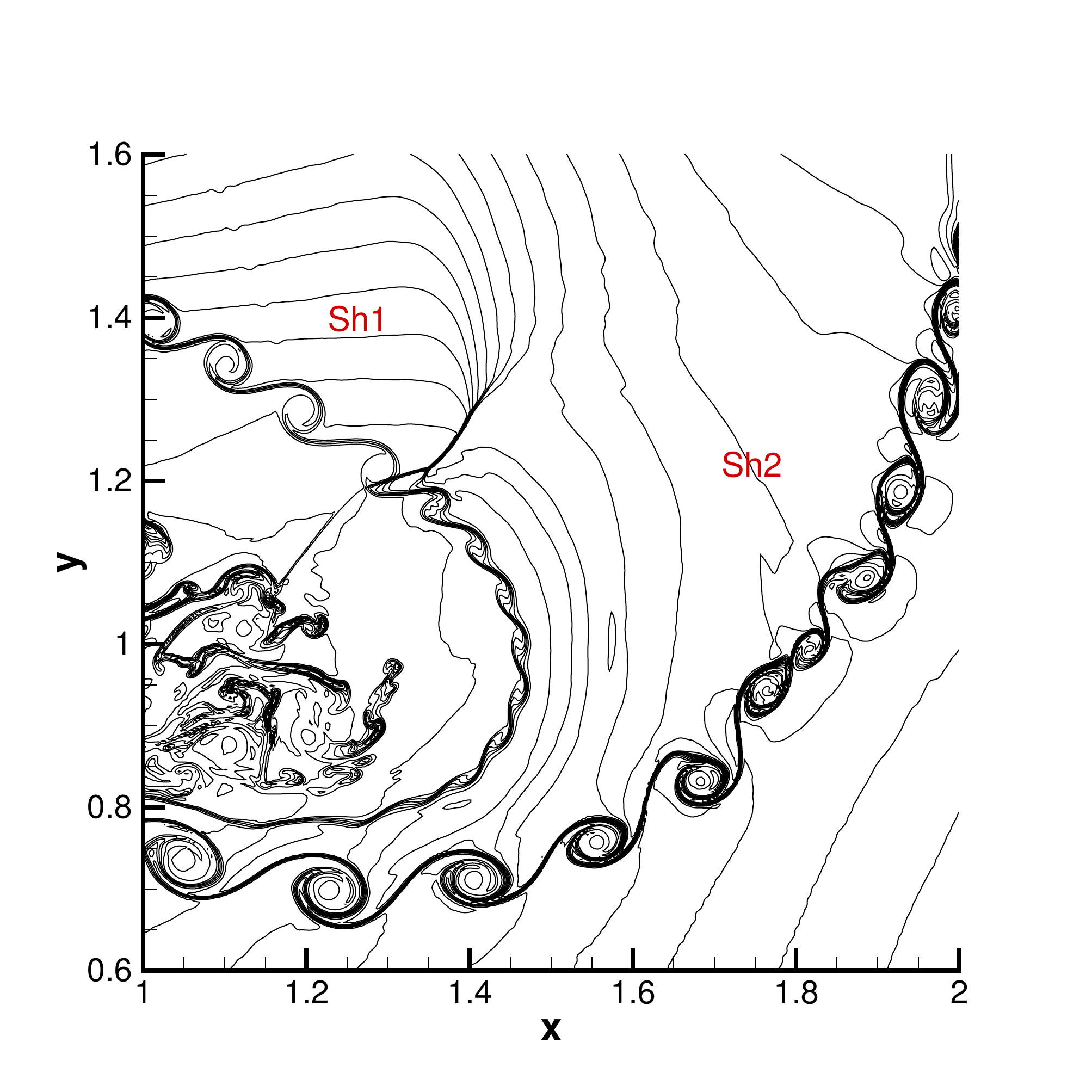}}
    \\
  \subfigure[S3O5+]{
    \label{2ndt16} 
    \includegraphics[width=0.25\textwidth]{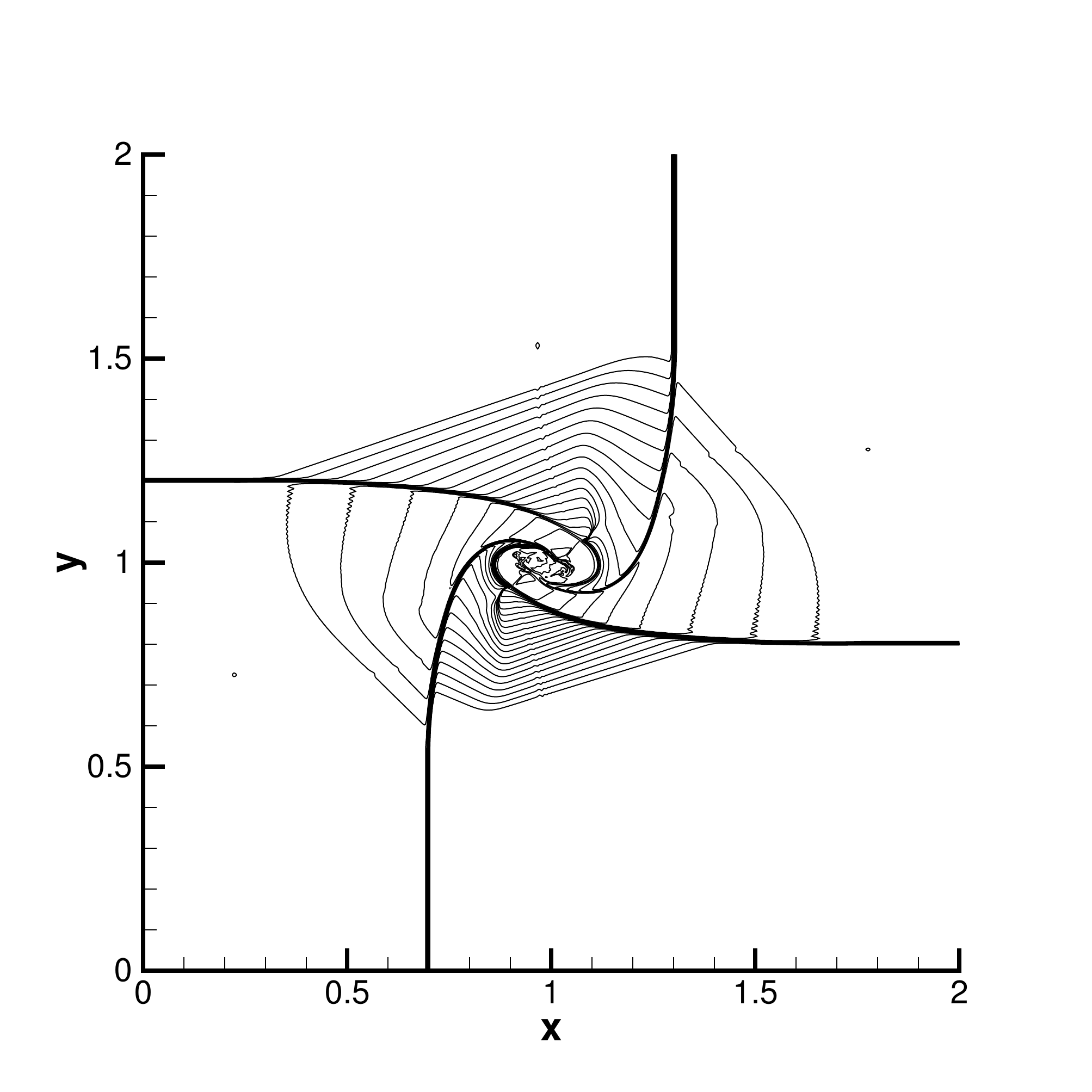}
    \includegraphics[width=0.25\textwidth]{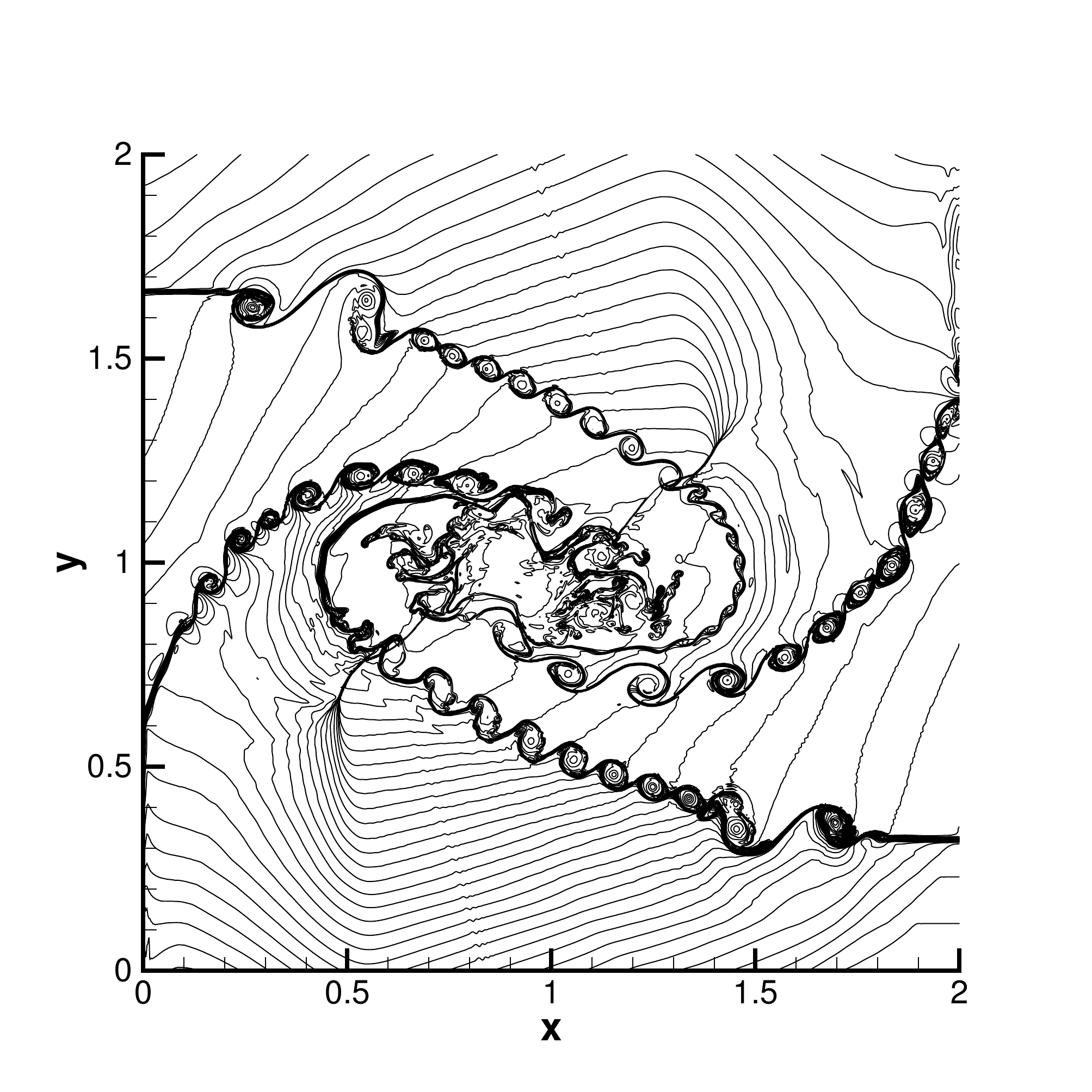}
    \includegraphics[width=0.25\textwidth]{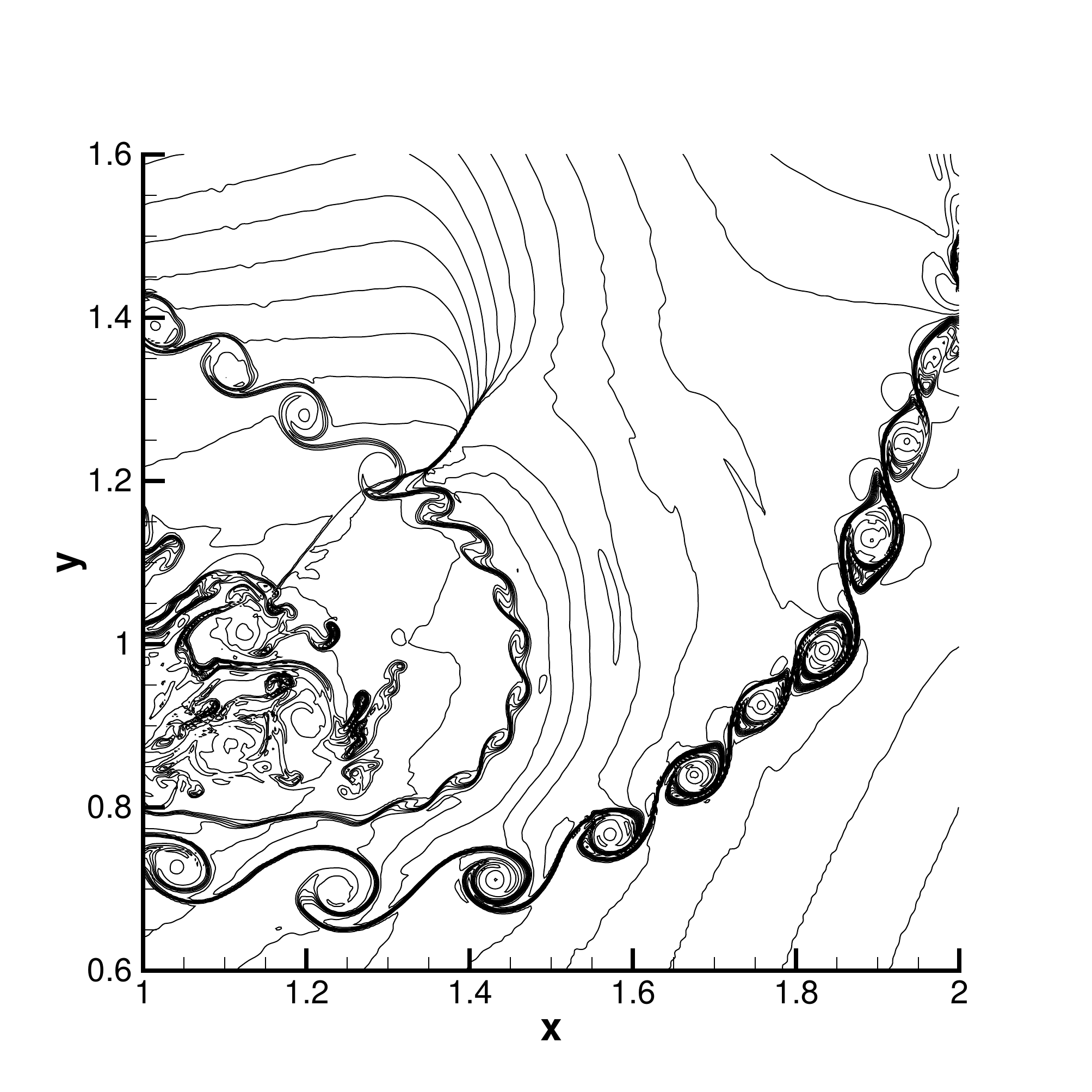}}
    \\
   \subfigure[S2O5s+]{
    \label{2ndt16} 
    \includegraphics[width=0.25\textwidth]{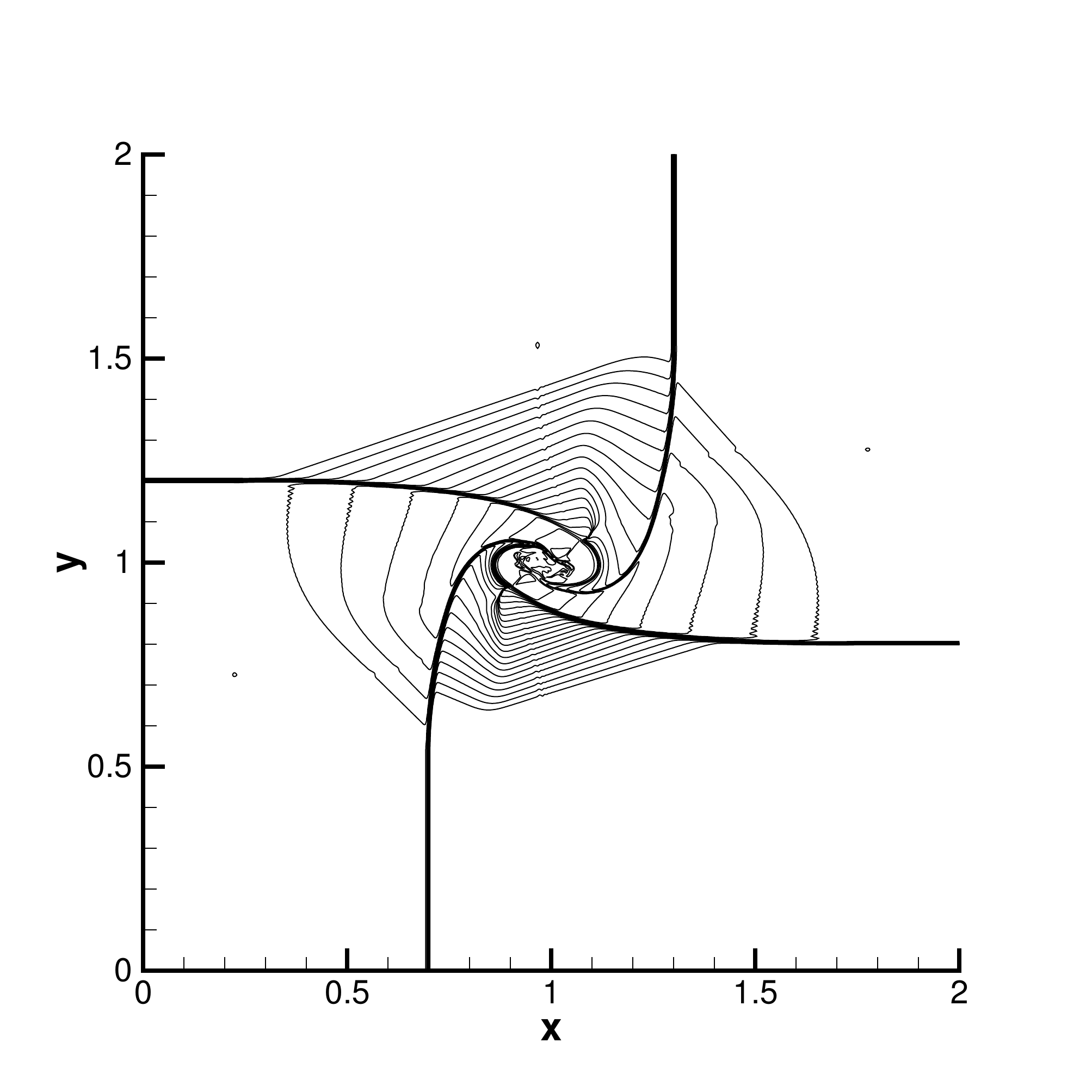}
    \includegraphics[width=0.25\textwidth]{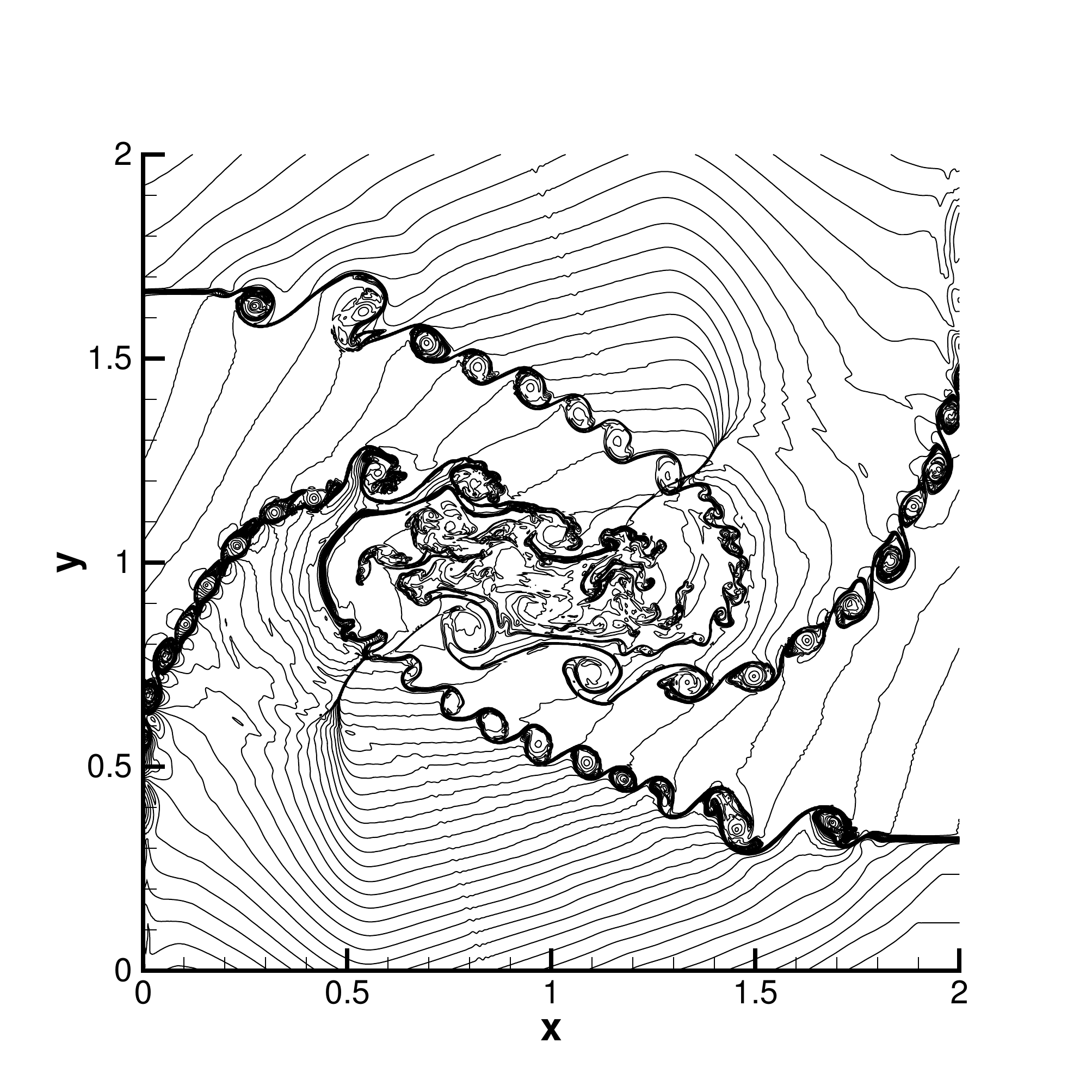}
    \includegraphics[width=0.25\textwidth]{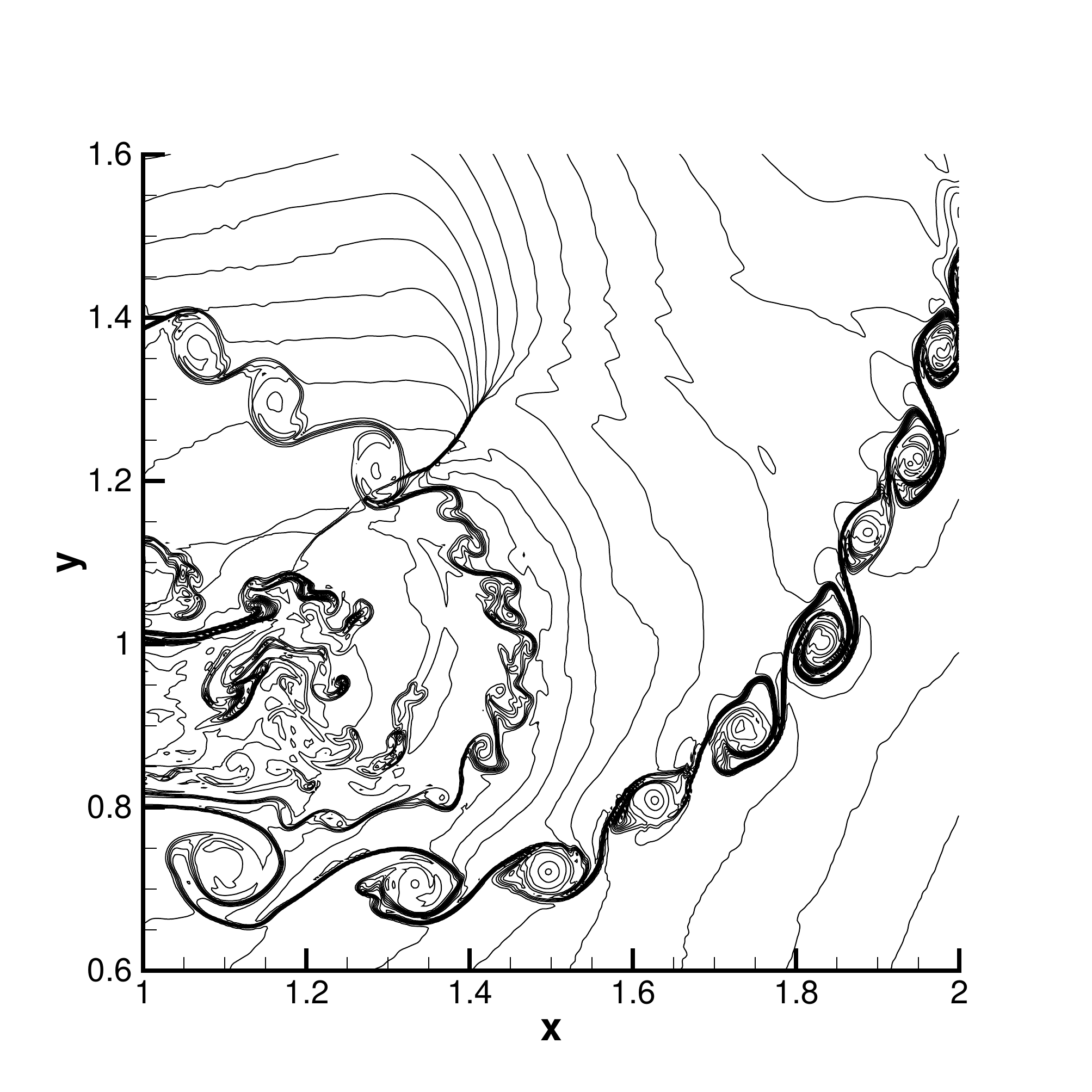}}
    \\
  \caption{The density contours of Configuration 6 case in \cite{2dRM} by using multi-stage GKS.
  Left: $t=0.4$. Middle: $t=1.6$. Right: local enlargement density contours of middle figure.}
  \label{2drm-longshear-1} 
\end{figure}

\begin{figure}
  \centering
    \subfigure[RK4 HLLC]{
    \label{2ndt16} 
    \includegraphics[width=0.25\textwidth]{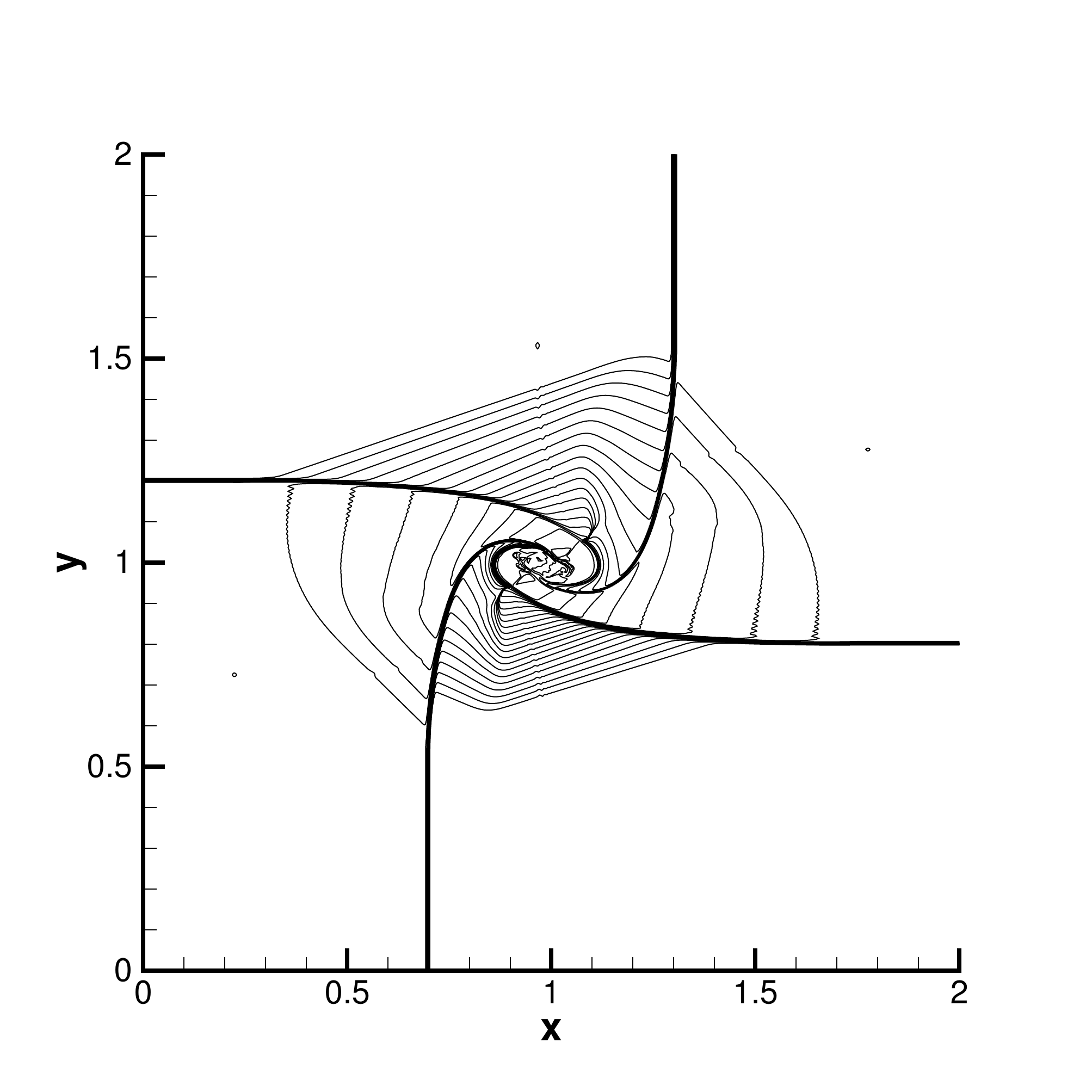}
    \includegraphics[width=0.25\textwidth]{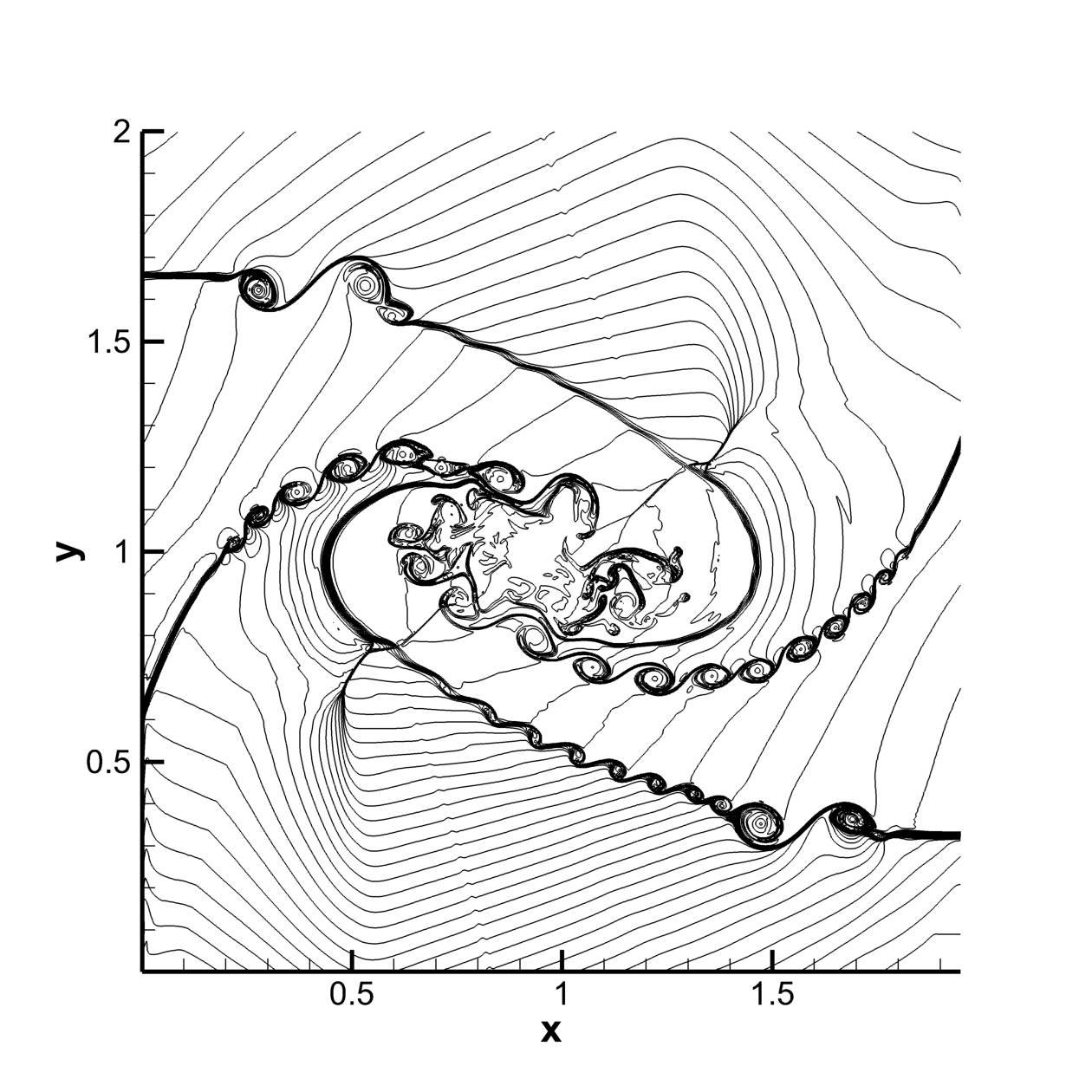}
    \includegraphics[width=0.25\textwidth]{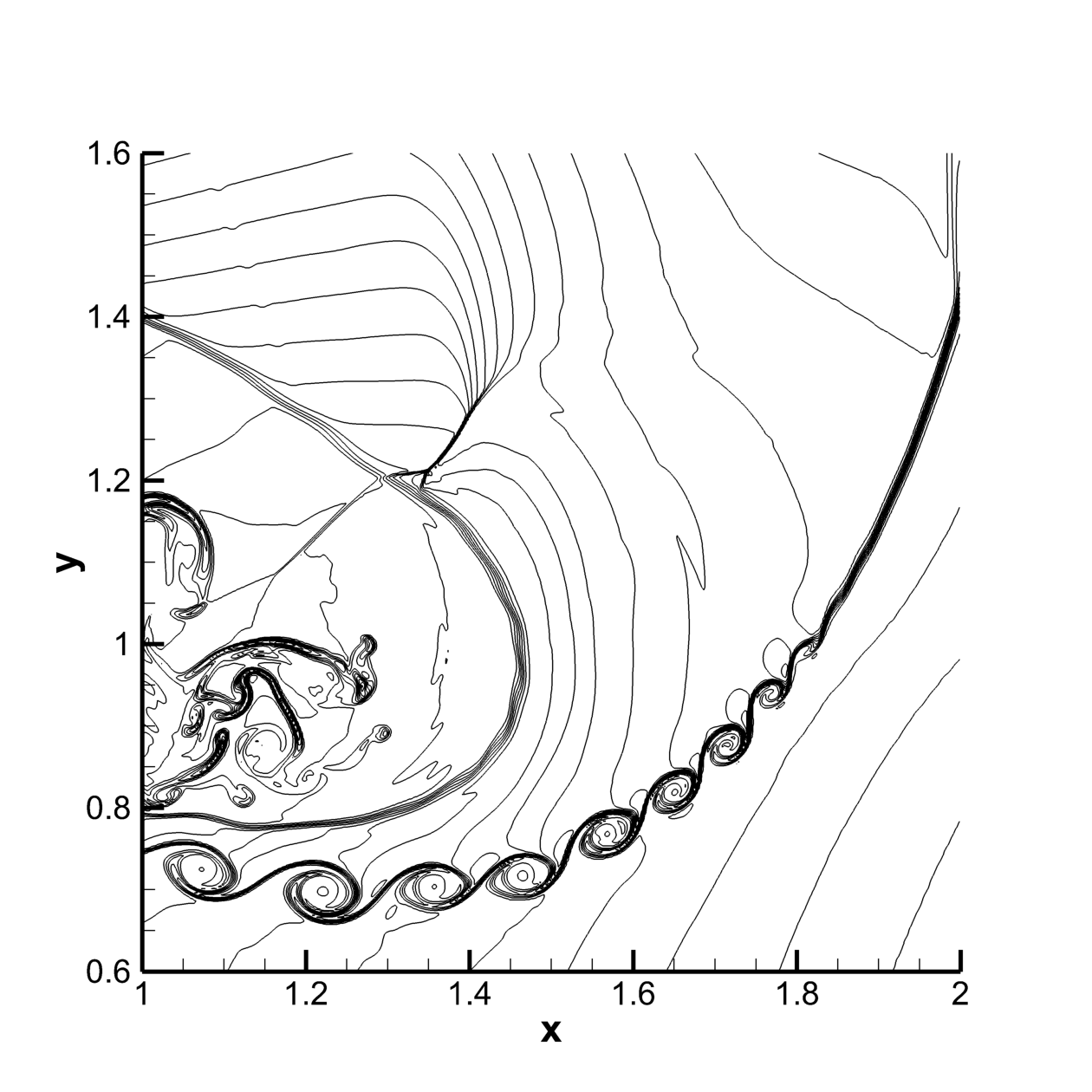}}
    \\
    \subfigure[RK4 ExactRS]{
    \label{2ndt16} 
    \includegraphics[width=0.25\textwidth]{hllcRK4t04}
    \includegraphics[width=0.25\textwidth]{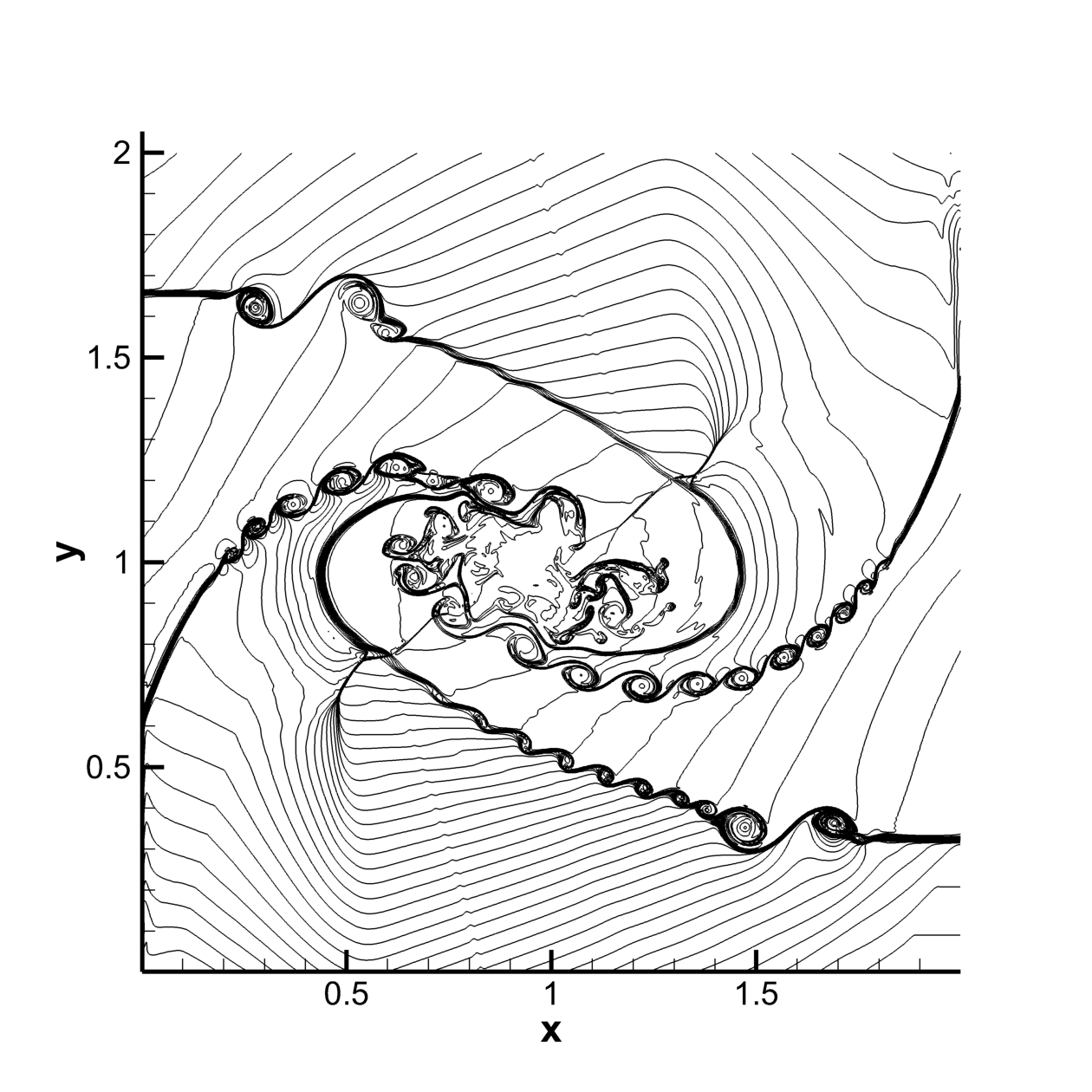}
    \includegraphics[width=0.25\textwidth]{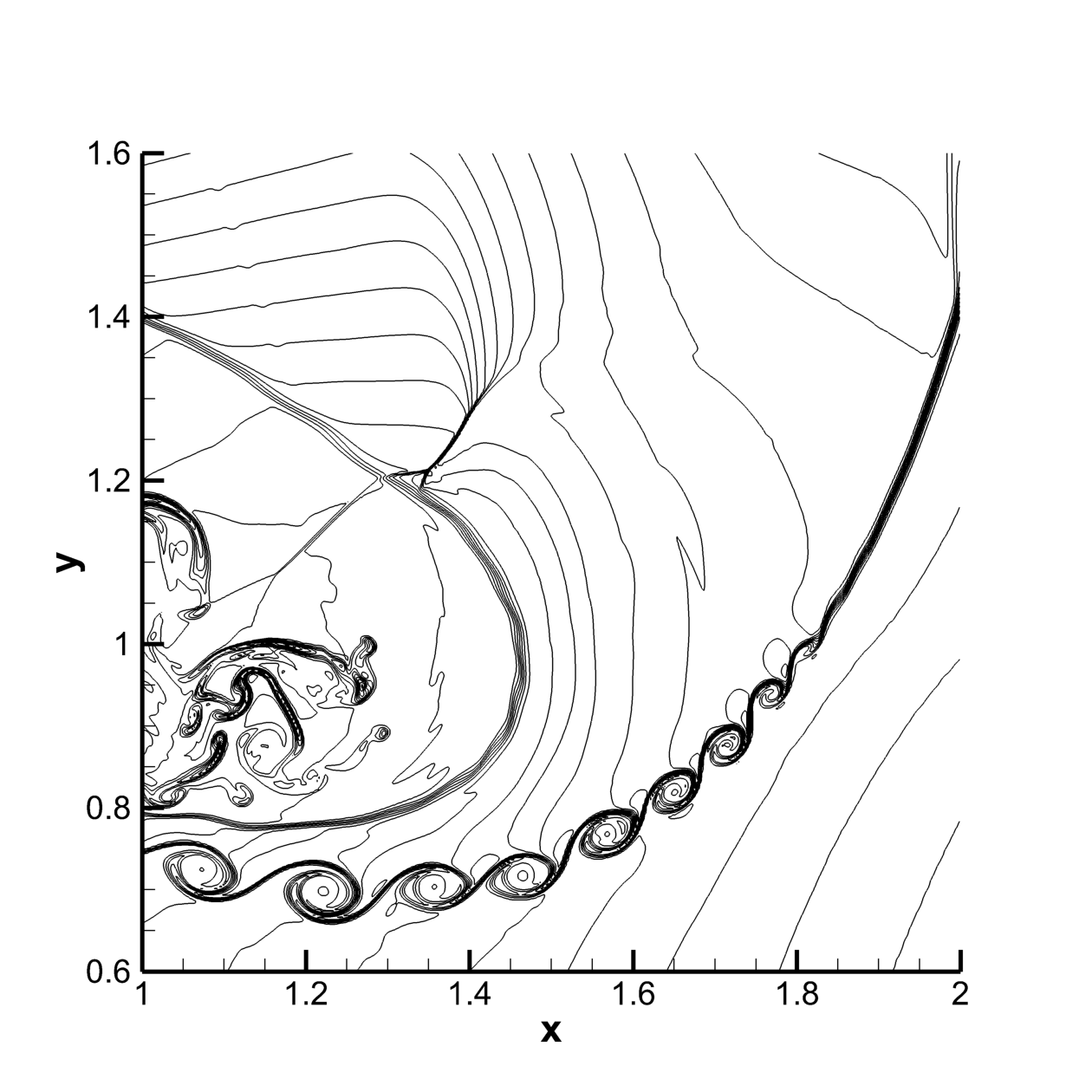}}
    \\
  \caption{The density contours of  Configuration 6 case in \cite{2dRM} by using RK4 Godunov  methods.
  Left: $t=0.4$. Middle: $t=1.6$. Right: local enlargement of middle figure.}
  \label{2drm-longshear-2} 
\end{figure}

The Configuration 6 in \cite{2dRM} is tested.
The initial condition in a square domain $[0,2]\times [0,2]$ is given by
\begin{equation*}
(\rho,u,v,p)=\left\{\begin{aligned}
&(1, -0.75,0.5, 1),& x<1,y<1,\\
&(3, -0.75,-0.5, 1),& x\geq 1,y<1,\\
&(1, 0.75,-0.5, 1),& x\geq 1,y\geq 1,\\
&(2,0.75,0.5, 1),& x< 1,y\geq 1,
\end{aligned} \right.
\end{equation*}
where four zones have different density and velocity, but the same pressure.
Four shear layers will be formed by these planar contact discontinuity interactions.
Similar to the isentropic vortex case, in order to reveal the influence of different time marching schemes
a large domain $[0,2]\times [0,2]$ covered by $800 \times 800$ mesh points is used with the output times $t=0.4$ and $t=1.6$.
The CFL condition is $0.5$ in all calculations.
As shown in the Fig. \ref{2drm-longshear-1}, at $t=0.4$ the results from all schemes seem identical.
However, at time $t=1.6$, the flow structures become much more complicated.
From the local enlargement of the central shear layer,
the S2O4 and S3O5 schemes present more smaller vortexes, and the S2O5 scheme shows slightly less but a little bit larger size.
In comparison with Riemann flux based RK4 Godunov method, as shown in Fig. \ref{2drm-longshear-2},
the traditional RK methods show more dissipative results than those from MSMD GKS.
Different from GKS solutions, almost no instabilities for Sh1 wave are triggered for traditional RK4 methods.
For Sh2 wave, both the number and the size of the vortexes from RK4 Godunov methods are inferior in comparison with MSMD GKS results.

\subsection{Computational Efficiency }

Based on the above 2D Riemann problem test case of interaction of planar contact discontinuities, the computational
efficiency from different schemes will be evaluated.
For S2O4, S3O5, S2O5 methods, the fluxes at three Gaussian points along a cell interface are needed, which are the same
as RK5 Godunov scheme with Riemann flux.
The 2D Riemann problem (Configuration 6 in \cite{2dRM}) with different mesh sizes are tested for the comparison of computation cost,
which is shown in Table. \ref{compT-1}.
The CPU times are recorded after running $10$ time steps for each scheme with a single processor of Intel Xeon E5 2670 $@$2.60GHz.
Based on the table, the computational time of S3O5 GKS is about $1.5$ times of S2O4 GKS due to the differences in number of stages.
Another observation is that the computational speed for S2O5 and S3O5 is almost the same.
However, for S2O5 scheme, only two stages of fluid variables are stored, which is less memory used than S3O5 scheme, but more computational time
spent on the 3rd-order flux function.

The computational time for the traditional RK methods based on Riemann solvers is also obtained in Table. \ref{5th6stage}.
The RK5 method with HLLC flux seems faster than the S2O4 scheme. This is due to the simple HLLC flux function.
The RK5 method with exact Riemann solver is almost two times more expensive than that of 5th-order GKS.
But, even for the current 2D inviscid flow test case, the GKS always solves the Navier-Stokes (NS) equations.
If RK5 Godunov type scheme is extended to solve the NS equations, the computational cost will at least be doubled,
which will become inefficient in comparison with MSMD GKS.
So, for the high order schemes and for the viscous flow computations, the design of high order time accurate flux solver
is worthwhile.

\begin{table}[!h]
\small
\begin{center}
\def\temptablewidth{1\textwidth}
{\rule{\temptablewidth}{1pt}}
\begin{tabular*}{\temptablewidth}{@{\extracolsep{\fill}}c|c|c|c|c|c}

\diagbox{Mesh size}{CPU time}{Schemes} & S204& S2O5s+ &S3O5+&RK5 HLLC&RK5 Exact-RS  \\
\hline
100*100&7.640 & 8.057 &10.959&6.028&20.223\\
200*200&26.309 & 45.551&46.801&24.022&80.880\\
300*300&58.524 &94.102 &94.523&53.833&171.647\\
400*400&102.985& 138.153&158.364&94.293&292.566\\
\end{tabular*}
{\rule{\temptablewidth}{0.1pt}}
\end{center}
\vspace{-4mm} \caption{\label{compT-1} Computational time (in seconds) of different schemes for the 2D Riemann problem.}
\end{table}

\subsection{Two dimensional viscous test cases}

\bigskip
\noindent{\sl{(a) Viscous shock tube problem under low Reynolds numbers}}

\begin{figure}[!h]
\centering
    \subfigure[S204]{
    \label{vishocktube-s2o4} 
    \includegraphics[width=0.44\textwidth]{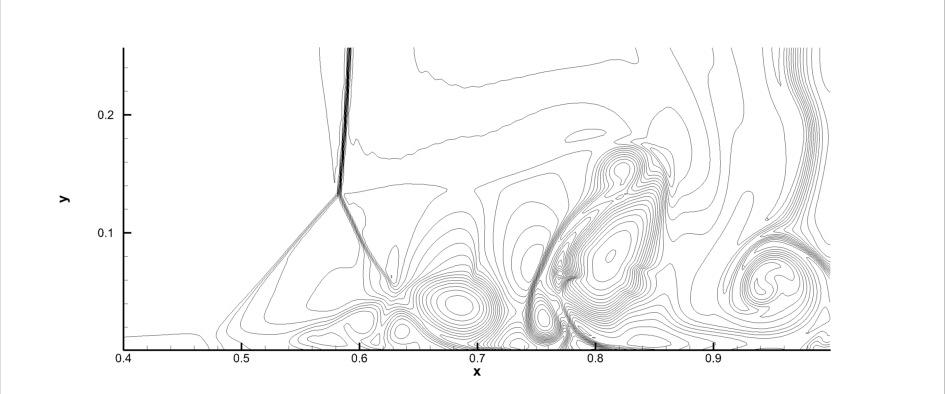}
    \includegraphics[width=0.44\textwidth]{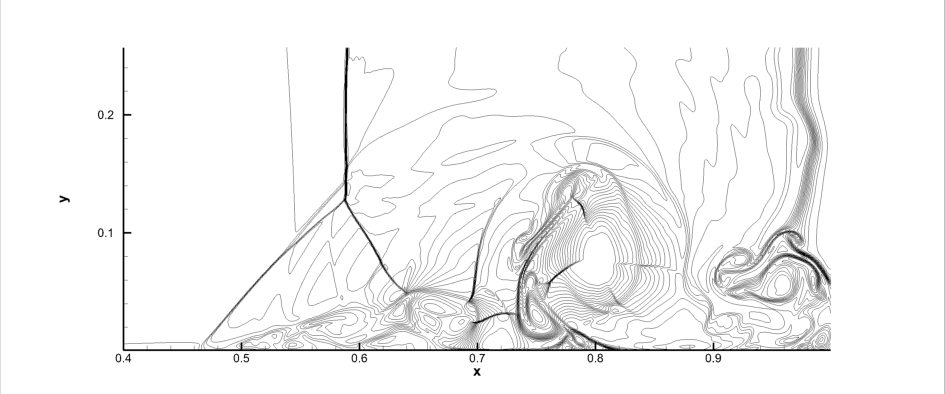}}
    \subfigure[S3O5+]{
    \label{vishocktube-s3o5ssp} 
    \includegraphics[width=0.44\textwidth]{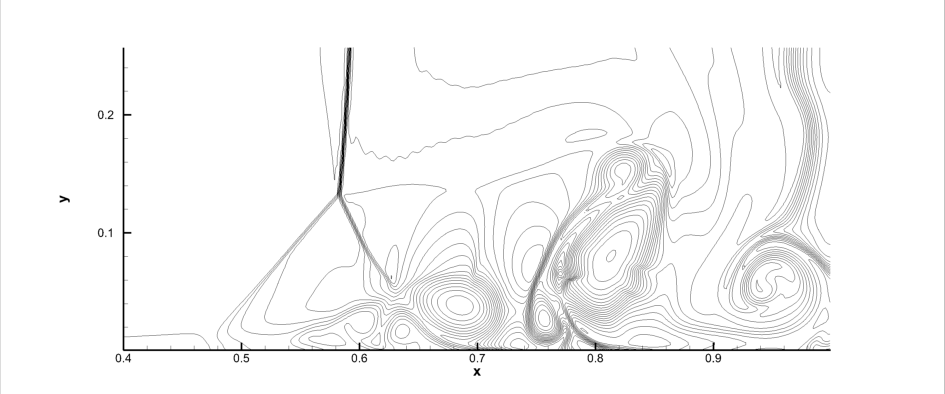}
    \includegraphics[width=0.44\textwidth]{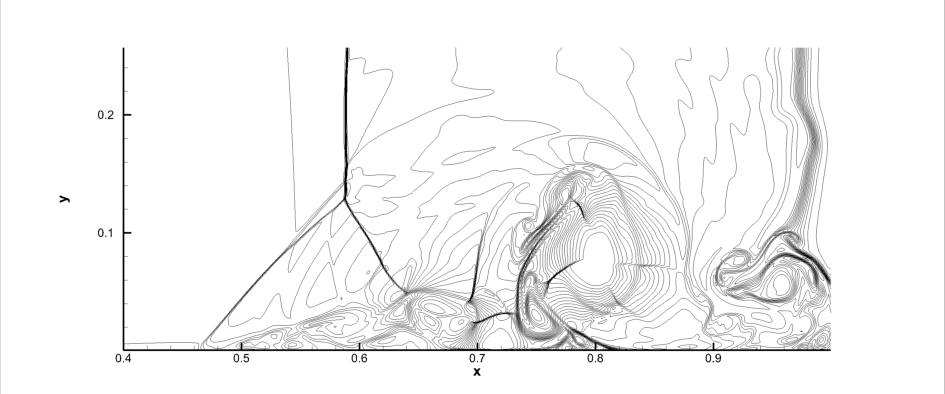}}
    \subfigure[S2O5s+]{
    \label{vishocktube-s2o5} 
    \includegraphics[width=0.44\textwidth]{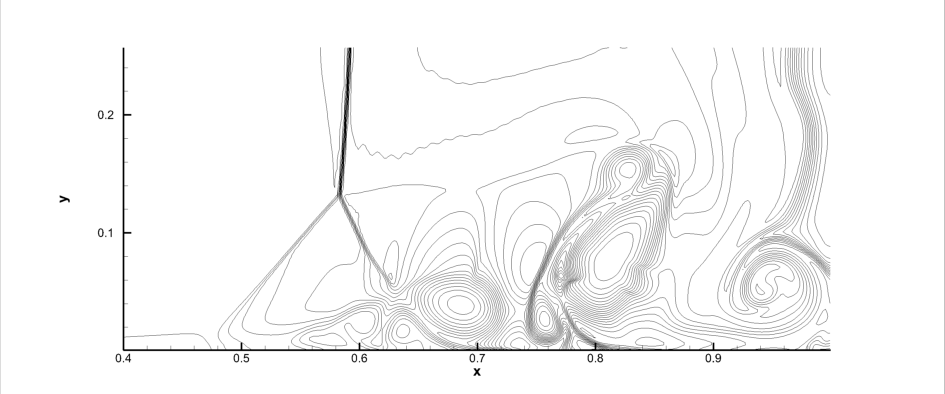}
    \includegraphics[width=0.44\textwidth]{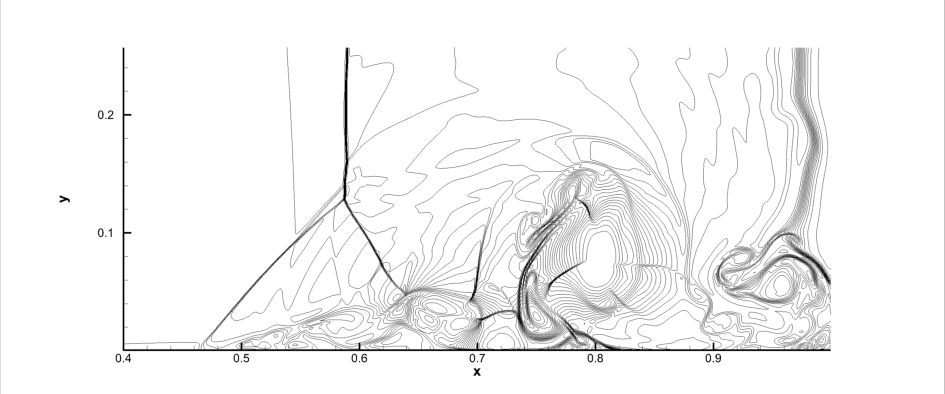}}
\caption{\label{vistube} The density contours of viscous shock tube problems at $t=1$ from  different MSMD GKS.
Left: $Re=200$, $500\times250$ mesh points. Right: $Re=1000$, $1000\times500$ mesh points.  }
\end{figure}

For N-S solvers, the viscous shock tube problem is a nice test case for the validity of the scheme due to
the complicated flow structure and the shock boundary layer interaction \cite{daru}.
The geometry is a two-dimensional unit box $[0,1]\times [0,1]$, with no-slip adiabatic walls.
Two different constant conditions are given on both sides of $x=0.5$,
 \begin{equation*}
(\rho,u,v,p)=\left\{\begin{aligned}
&(120,0,0, 120/\gamma), 0<x<0.5,\\
&(1.2,0,0,1.2/\gamma),  0.5\leq x<1,
\end{aligned} \right.
\end{equation*}
where $\gamma=1.4$ and the Prandtl number $Pr=0.73$.
The computational domain is chosen as $[0,1]\times [0,0.5]$ due to the symmetry of the problem.
The upper boundary is set as a symmetric boundary condition and the others boundaries are no-slip adiabatic condition.
Two Reynolds number $Re=200$ and $1000$ are selected in the simulations.
The results at $t=1$ are presented.
For such a low Reynolds number low,
the time step is determined by
$$ \Delta t = C_{CFL} \mbox{Min} ( \frac{ \Delta x}{|\sqrt{U^2+V^2}|+a}, \frac{ (\Delta x)^2}{4\nu}) ,$$
where $C_{CFL}$ is the CFL number, $a$ is the sound speed, and $\nu = \mu /\rho$ is the kinematic viscosity coefficient.
The results from different GKS schemes are shown in Fig. \ref{vistube}.
For both $Re=200$ and $1000$, almost identical results are obtained from different schemes.
The height of primary vortex for $Re =200$ is shown in Table. \ref{vistube_table} \cite{vistube}.

\begin{table}[!h]
\small
\begin{center}
\begin{tabular}{c|c|c|c}
\Xhline{1.2pt}
Schemes&AUSMPW+&M-AUSMPW+&S2O4\\
\hline
Height&0.163&0.168&0.173\\
\Xhline{1.2pt}
Schemes&S2O5s+&S3O5+&~\\
\hline
Height&0.174&0.173&~\\
\Xhline{1.2pt}
\end{tabular}
\vspace{-1mm} \caption{\label{vistube_table} Heights of primary vortex from different schemes for $Re =200$ and $\Delta x= \Delta y=1/500$.}
\end{center}
\end{table}

\bigskip
\noindent{\sl{(b) A planar jet under high Reynolds number}}

\begin{figure}
  \centering
    \subfigure[Schematic of the computational domain]{
    \label{planarjet-initial} 
    \includegraphics[width=0.44\textwidth]{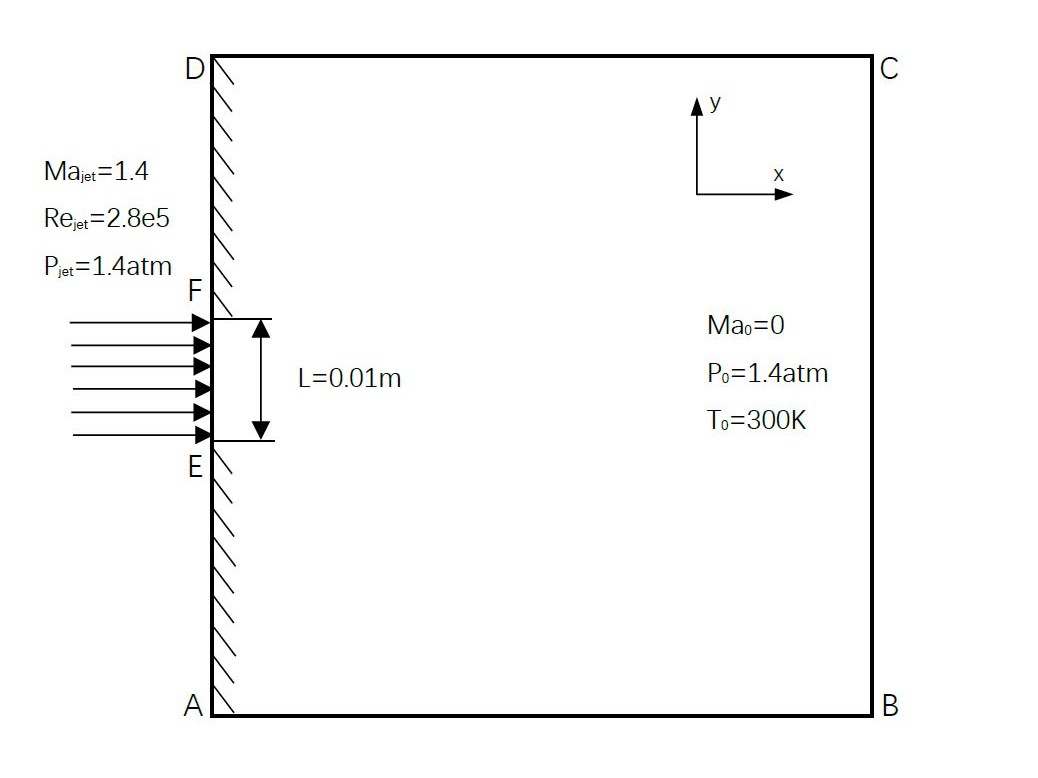}}
    \subfigure[Case 3 result at t=0.15ms]{
    \label{planarjet-2d-1} 
    \includegraphics[width=0.44\textwidth]{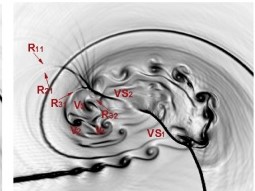}}
  \caption{(a) The sketch of a planar jet case proposed in \cite{planar-jet}. (b) The 2-D result without entrance perturbation in \cite{planar-jet}, at t=0.15ms. The numerical schlieren-type images are plotted for visualization.    }
  \label{planer_jet1} 
\end{figure}

Free supersonic jet flow is widely studied.
A simplified 2-D planar jet, which was proposed by Zhang et al \cite{planar-jet}, is used here to test high order  GKS.
A Mach $1.4$ jet is injected through a width $L=0.01m$ entrance into a square computational domain with the size $10L \times 10L$.
The sketch of the geometry is plotted in Fig. \ref{planarjet-initial}.
The Reynolds number is set as $Re_{\infty}={U_{jet}L}/{\nu}=2.8 \times 10^5$, and the dynamic viscosity coefficient
$\nu=1.73 \times 10^{-5}m^2 /s$. The same mesh size of $1200 \times 1200$ as that in \cite{planar-jet} is used here.

\begin{figure}
  \centering
    \subfigure[S204]
    {
    \label{planarjet-s2o4} 
    \includegraphics[height=0.35\textwidth]{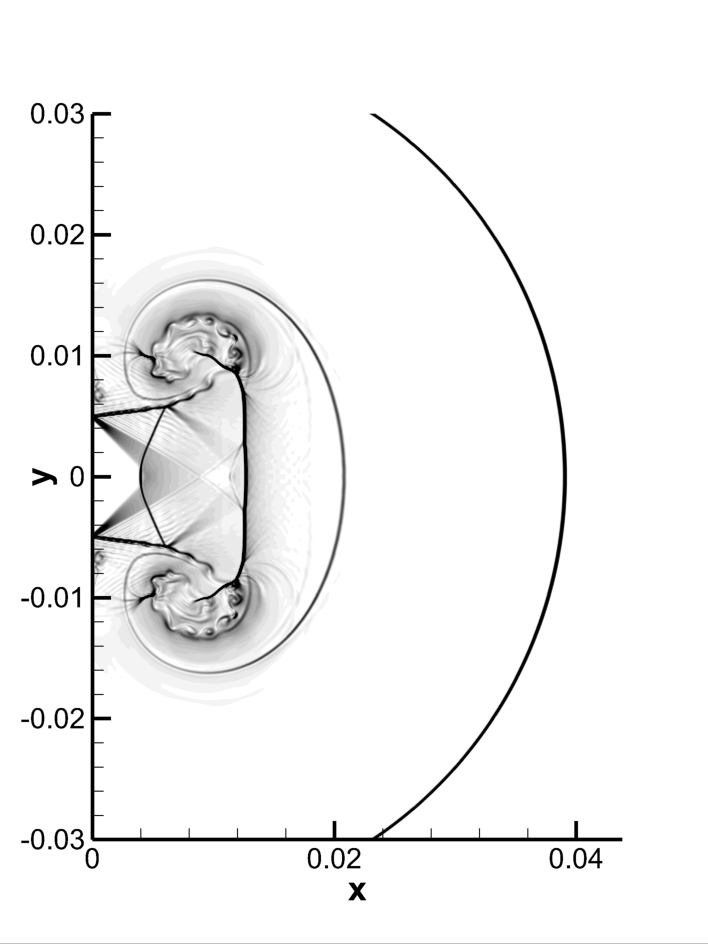}
    \includegraphics[height=0.35\textwidth]{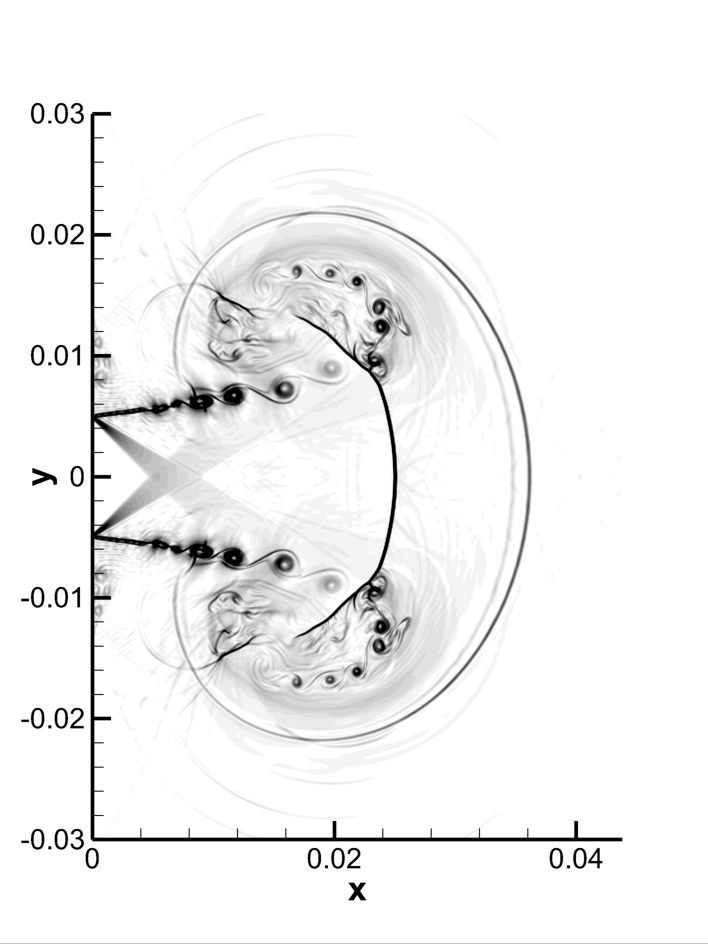}
    \includegraphics[height=0.35\textwidth]{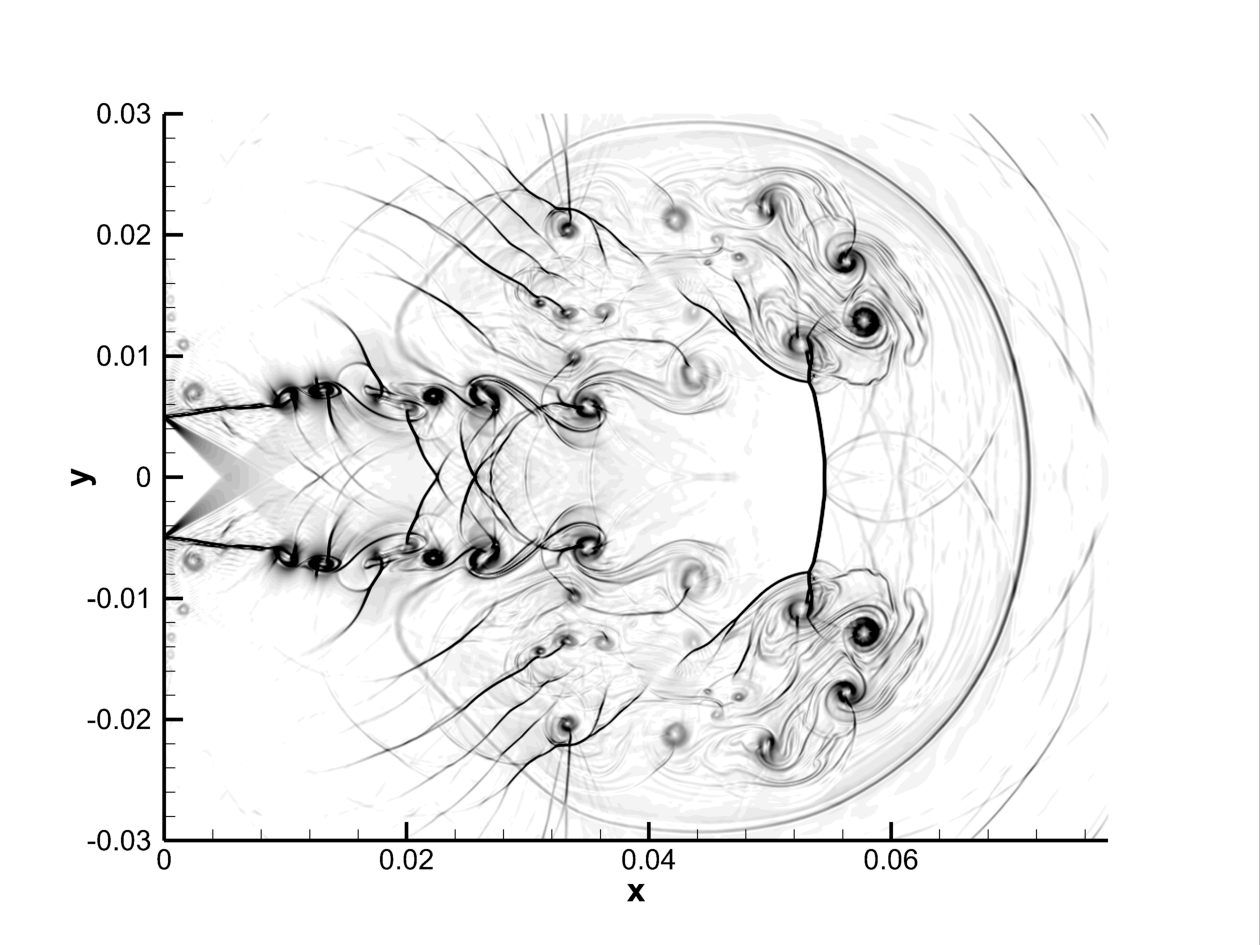}
    }
    \subfigure[S3O5+]
    {
    \label{planarjet-s3o5} 
    \includegraphics[height=0.35\textwidth]{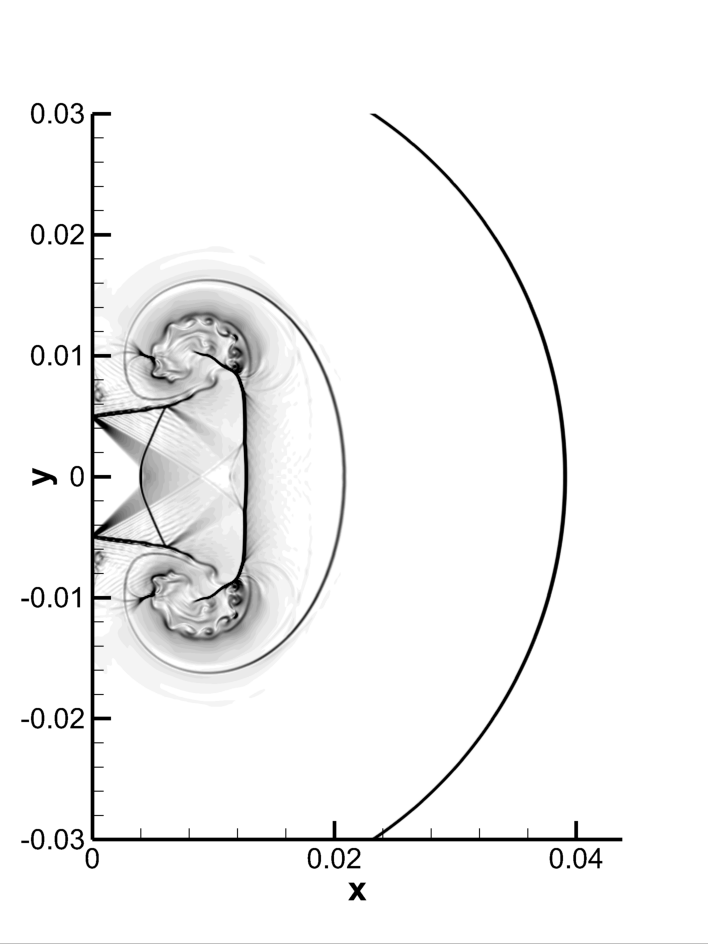}
    \includegraphics[height=0.35\textwidth]{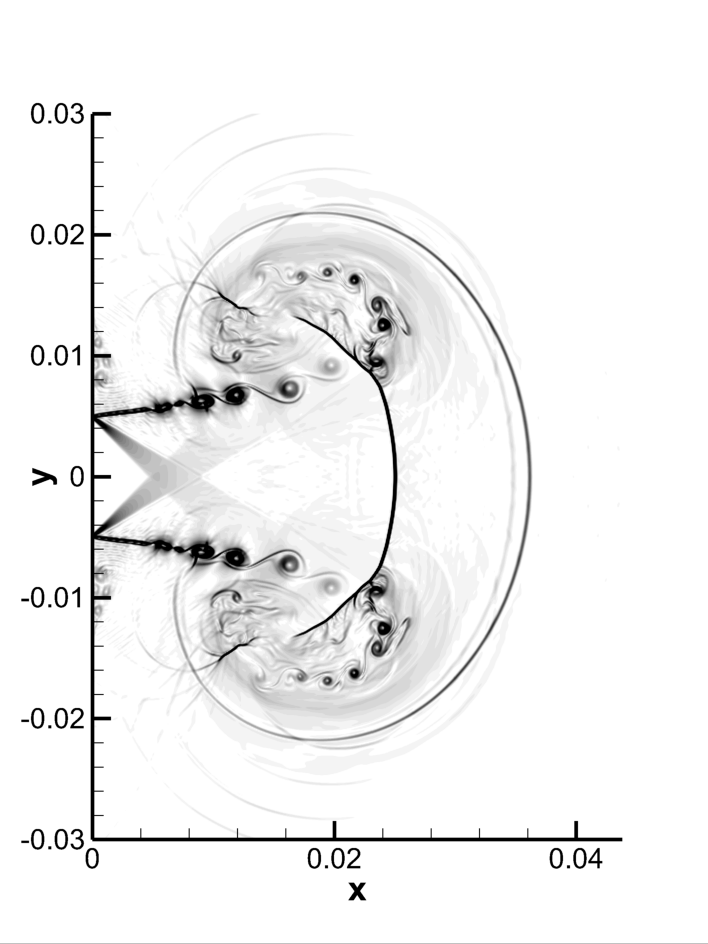}
    \includegraphics[height=0.35\textwidth]{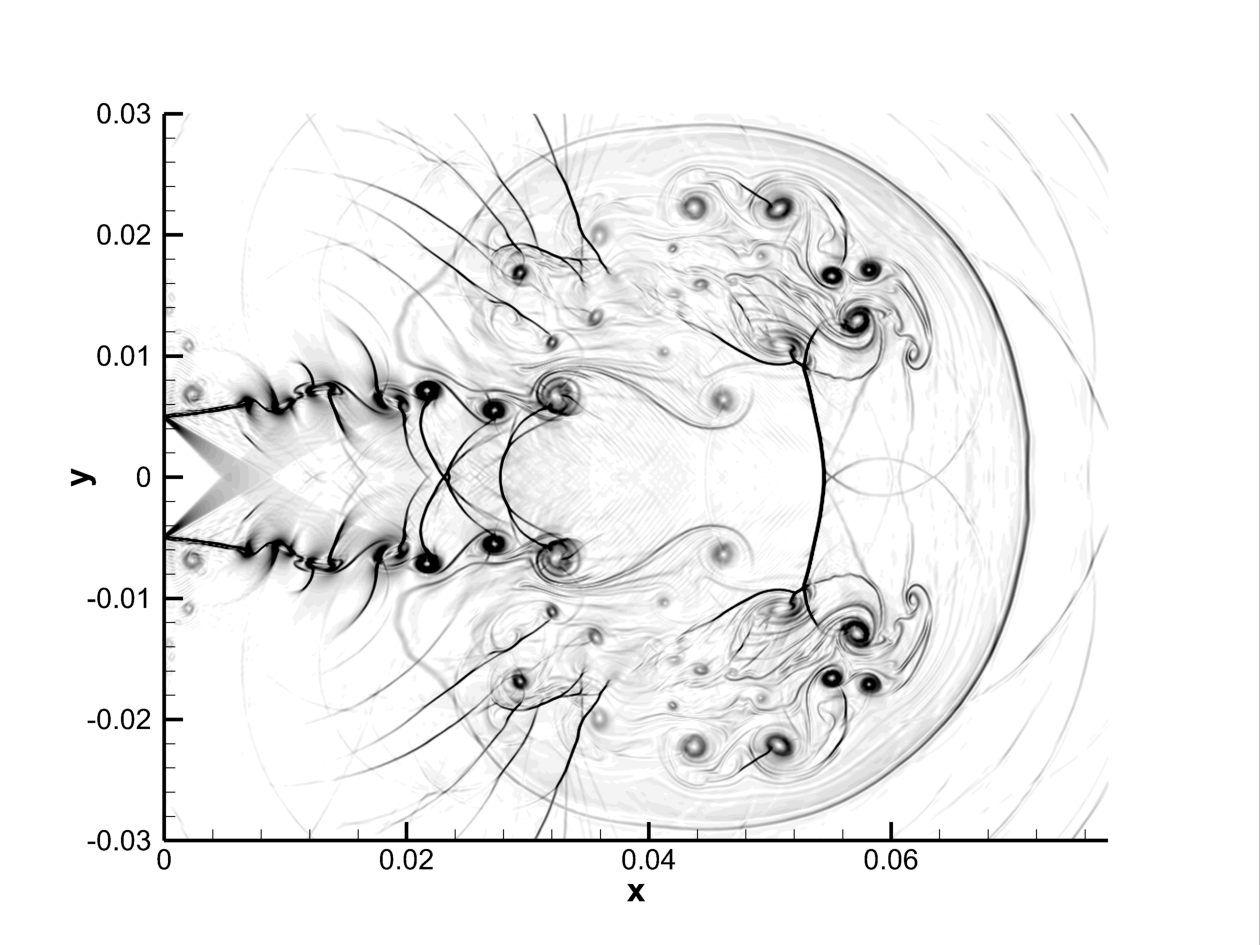}
    }
    \subfigure[S2O5s+]
    {
    \label{planarjet-s2o5} 
    \includegraphics[height=0.35\textwidth]{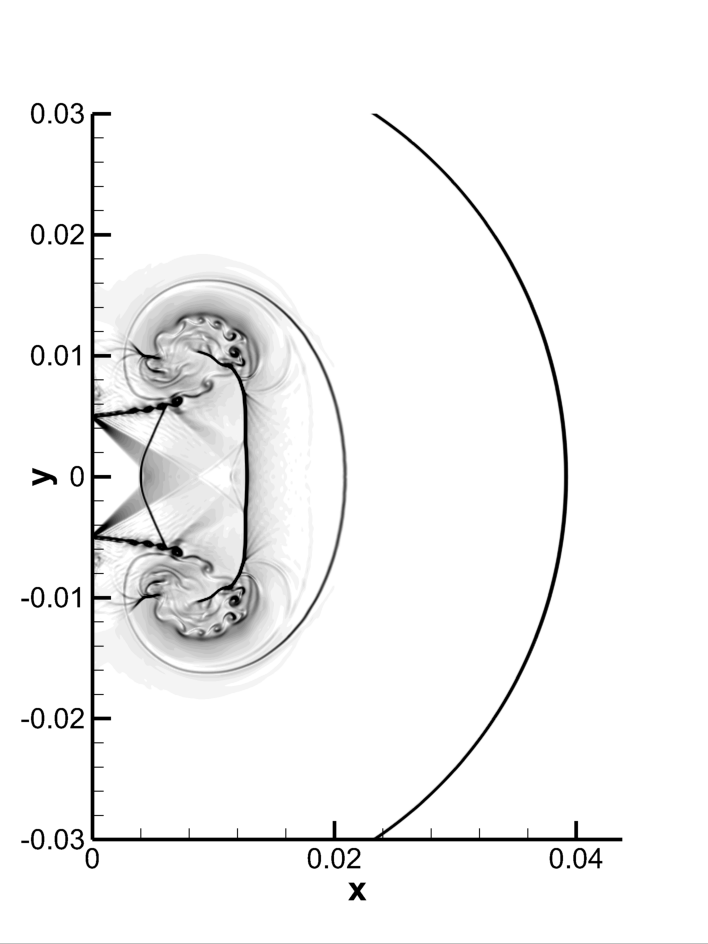}
    \includegraphics[height=0.35\textwidth]{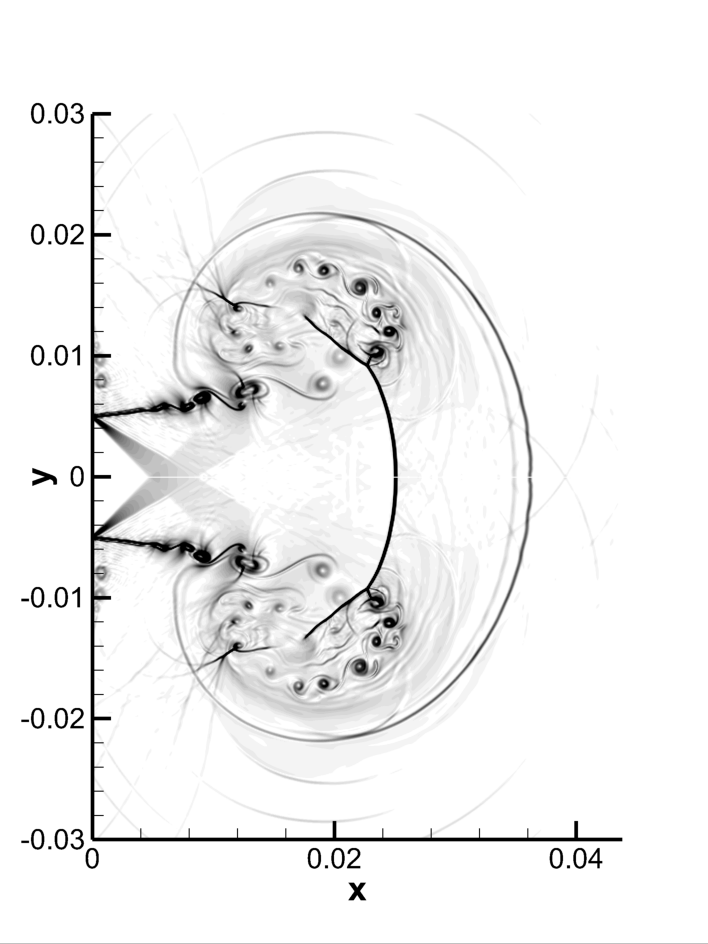}
    \includegraphics[height=0.35\textwidth]{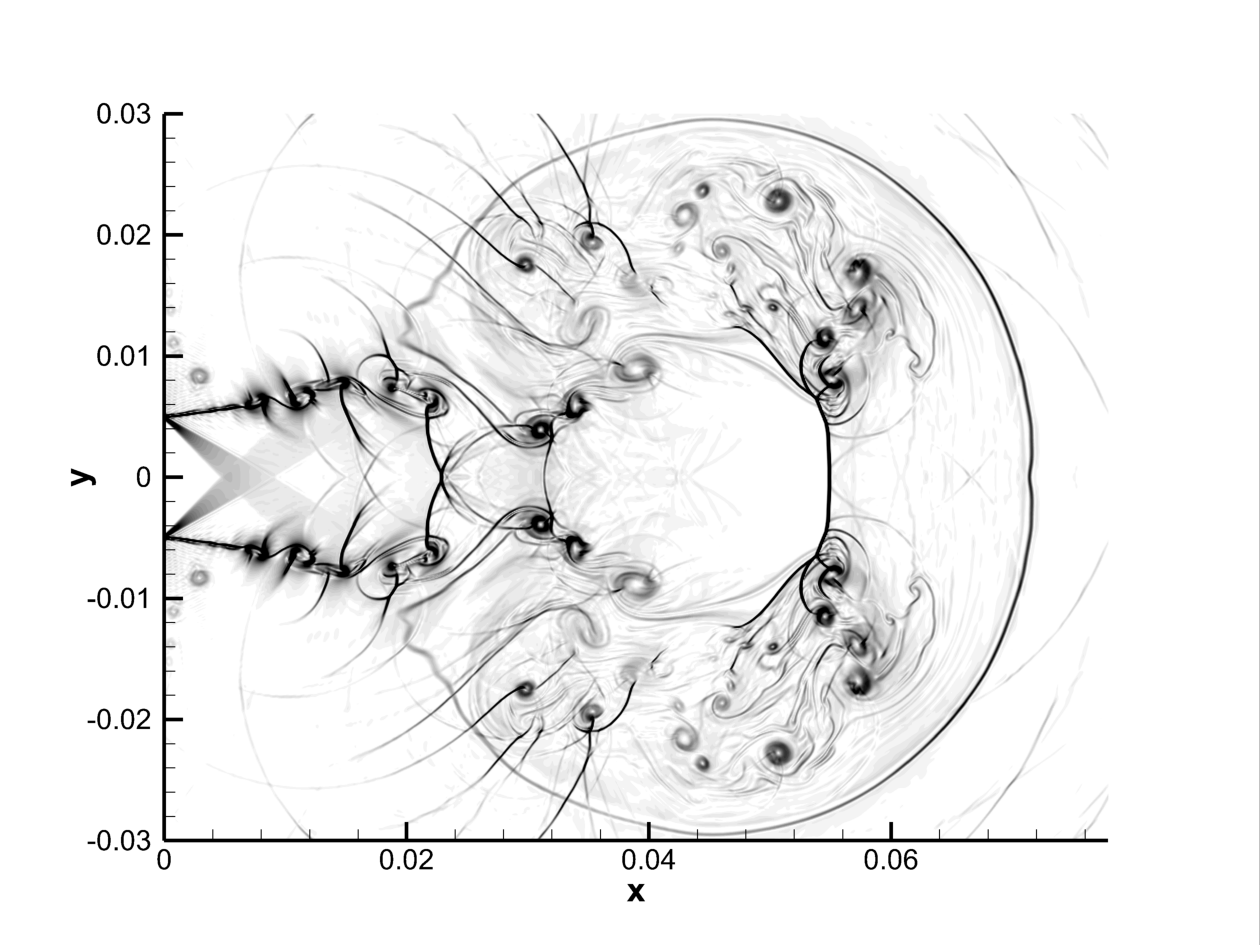}
    }

  \caption{The 2-D simulation of starting structures of Mach $1.4$ planar jet under different MSMD GKS. Mesh size $1200\times1200$. The schlieren images at $t=0.08ms, 0.15ms, 0.32ms$ are given from left to right to present the jet evolution.    }
  \label{planer_jet2} 
\end{figure}

Fig. \ref{planer_jet2} shows the evolution process of the jet at three different output times $t=0.08ms$, $0.15ms$,
and $0.32ms$. At $t=0.08ms$, the three schemes of S2O4, S3O5+, and S2O5s+,
give almost identical results. At $t=0.15ms$, the jet structures from different schemes are similar, with  small variations,
such as the vortex sheets along the main shear layers due to the K-H instability.
Current results can be compared with the one in Fig. \ref{planarjet-2d-1}, while the
high order MSMD GKS seems present more clear flow structure, especially for the capturing of small size vortices.
As the jet further develops up to $t=0.32ms$, significant differences in the shear layer and main vortex pairs
could be observed from different schemes.
The S205s+ scheme shows more distinct flow pattern in the center of vortex pairs.
Fig. \ref{planer_jet2} provides abundant flow structures, such as shear layer instability, shock vortex interaction, and
a wide range of vortex strength and sizes, which clearly demonstrate the power of high order schemes.

\section{Conclusion}

In this paper, a family of high order gas kinetic schemes with multi-stage multi-derivative techniques have been proposed,
especially the two stages and three stages fifth-order schemes S2O5 and S3O5.
Due to the use of time derivatives of the flux function, such as the second and third-order GKS time accurate flux functions,
for the same temporal accuracy the current schemes can reduce the number of middle stages
in comparison with the traditional Runge-Kutta methods, such as the RK5 method with the time independent Riemann flux.
Therefore, the current MSMD GKS becomes more efficient than the RK methods, especially for the Navier-Stokes solutions.
The high order MSMD GKS provides accurate numerical solutions for the compressible flows
with the same robustness as the second-order methods in the flow simulations with strong shock interactions.
The jet simulation provides the state-of-art numerical results from high order schemes.

The current MSMD gas kinetic schemes use the WENO type reconstruction for initial condition and spatial accuracy, which has a large
stencils.
Even though the MSMD GKS can be easily extended to unstructured mesh for the NS solutions,
it is still preferred to develop  high order compact gas kinetic schemes for the engineering applications with
complicated geometry. As a continuation of the compact third-order GKS \cite{unstructured-compact-gks}, the developments of the
fourth-order and fifth-order compact schemes with the implementation of MSMD technique are on the development.

\section*{Acknowledgement}
 The current work  is supported by Hong Kong Research Grant Council (16211014, 16207715), HKUST research fund
(PROVOST13SC01, IRS16SC42, SBI14SC11), and  National Science Foundation of China (91330203,91530319).
\bibliographystyle{plain}%
\bibliography{bib-mmhgks}
\end{document}